\documentclass[a4paper]{article}
\usepackage{a4wide}

\usepackage[T1]{fontenc}
\usepackage{graphicx}
\usepackage{amssymb}
\usepackage{amsmath}
\usepackage[hyphens]{url}
\usepackage{xcolor}
\usepackage{soul}
\usepackage{longtable}
\usepackage{multirow}
\usepackage{booktabs}
\usepackage{array}
\usepackage{cite}
\usepackage{wrapfig}

\usepackage{pstricks}
\usepackage{pst-node}
\usepackage{pst-3d}
\usepackage{pst-tree}
\usepackage{pst-all}

\newlength{\pbs}
\newlength{\corner}
\newcommand{\NodeBB}[1]{\Tr[linecolor=white,linewidth=2pt,
edge={\ncline[linewidth=1pt,linestyle=solid]}]{\psframebox[linearc=\corner,
cornersize=absolute,linecolor=black,linewidth=2pt]{\parbox[c]{\pbs}
{\centering\shortstack{#1}}}}
\psset{linecolor=black}}

\newcommand{\NNodeBB}[2]{\Tr[name=#1,linecolor=white,linewidth=2pt,
edge={\ncline[linewidth=1pt,linestyle=solid]}]{\psframebox[linecolor=black,
linearc=\corner,cornersize=absolute,
linewidth=2pt]{\parbox[c]{\pbs}{\centering\shortstack{#2}}}}
\psset{linecolor=black}}

\newcommand{\BBCC}{\mathop{\vcenter{\offinterlineskip 
\hbox{\psframebox[cornersize=absolute,linecolor=black,linewidth=1pt]{}}}}}

\def\Arca#1#2#3#4{
\psarc[linecolor=black](#2){#1}
  {!\psGetNodeCenter{#2} \psGetNodeCenter{#3} 
    #2.y #3.y sub abs #2.x #3.x sub abs atan 180 add}
  {!\psGetNodeCenter{#4} 
    #2.y #4.y sub abs #2.x #4.x sub abs atan neg 360 add}
}

\newcommand{\legendtlt}{
\begin{tabular}{|cl|}
\hline
\multicolumn{2}{|l|}{\textbf{Legend}}\\
\hspace{2mm}$\BBCC^{\phantom{A}}$ & objectives \\[0.1cm]
\hspace{2mm}\psline[linewidth=0.2mm](0,.3)(-.3,0)
\psline[linewidth=0.2mm](0,.3)(-.1,0)
\psline[linewidth=0.2mm](0,.3)(.1,0)
& disjunctive refinement\\[0.1cm]
\hspace{2mm}\psline[linewidth=0.2mm](0,.3)(-.3,0)
\psline[linewidth=0.2mm](0,.3)(-.1,0)
\psline[linewidth=0.2mm](0,.3)(.1,0)
\psarc[linewidth=0.2mm](0,.3){2mm}{225}{288}
& conjunctive refinement \strut\\
\hline
\end{tabular}
}

\DeclareMathOperator{\Prob}{Pr}

\newcommand{\emptytable}{\psscalebox{0.30}{
\begin{tabular}{|c|c|c|}
\hline
&&\\
\hline
&&\\
&&\\
\hline
\end{tabular}}
}
\newcommand{\legendbayes}{
\begin{tabular}{|cl|}
\hline
\multicolumn{2}{|l|}{\textbf{Legend}}\\
\multirow{2}{0.04\textwidth}{\hspace*{.25cm}$\BBCC$} & threat 
sources and\\[-1mm]
&  internal states \\
\multirow{2}{0.04\textwidth}{
\begin{pspicture}(-1,-0.3)(-.7,.5)
\psline[linewidth=0.2mm]{<-}(-0.7,.25)(-.8,-.15)
\rput(-1,0){{\tiny 0.65}}
\end{pspicture}
}
& probability of\\[-1mm]
& successful exploit\\
\multirow{2}{0.04\textwidth}[-1.5mm]{\hspace*{1.5mm}\emptytable
}
& local conditional\\[-1mm]
& probability distribution \strut\\
\hline
\end{tabular}
}

\newcommand{\probtabA}{
\begin{tabular}{|c|c|c|c|}
\hline
$\strut B\strut$ & $C$ & $\Prob(A)$ & $\Prob(\lnot A)$\\
\hline
$1$ & $1$ & $1.00$ & $0.00$ \\
$1$ & $0$ & $0.65$ & $0.35$ \\
$1$ & $1$ & $1.00$ & $0.00$ \\
$0$ & $0$ & $0.00$ & $\strut0.00\strut$ \\
\hline
\multicolumn{4}{c}{$\Prob(A) = 0.61$}
\end{tabular}
}

\newcommand{\probtabB}{
\begin{tabular}{|c|c|c|}
\hline
$\strut D \strut$ & $\Prob(B)$ & $\Prob(\lnot B)$\\
\hline
$1$ & $0.85$ & $0.15$ \\
$0$ & $0.00$ & $\strut 1.00 \strut $\\
\hline
\multicolumn{3}{c}{$\Prob(B) = 0.60$}
\end{tabular}
}

\newcommand{\probtabC}{
\begin{tabular}{|c|c|c|}
\hline
$\strut D \strut$ & $\Prob(C)$ & $\Prob(\lnot C)$\\
\hline
$1$ & $0.70$ & $0.30$ \\
$0$ & $0.00$ & $\strut 1.00 \strut$ \\
\hline
\multicolumn{3}{c}{$\Prob(C) = 0.49$}
\end{tabular}
}

\newcommand{\probtabD}{
\begin{tabular}{|c|c|}
\hline
$\strut \Prob(D) \strut$ & $\Prob(\lnot D)$\\
\hline
$0.70$ & $\strut 0.30 \strut$\\
\hline
\multicolumn{2}{c}{$\Prob(D) = 0.70$}
\end{tabular}
}

\usepackage[belowskip=\abovecaptionskip]{caption}
\usepackage[colorlinks]{hyperref}

\definecolor{lightblue}{rgb}{0.5,0.5,1.0}
\definecolor{darkred}{rgb}{0.5,0,0}
\definecolor{darkgreen}{rgb}{0,0.5,0}
\definecolor{darkblue}{rgb}{0,0,0.5}

\hypersetup{colorlinks=true
,linkcolor=darkred
,filecolor=darkgreen
,urlcolor=darkred
,citecolor=darkblue}

\usepackage[hyphenbreaks]{breakurl}
\usepackage{totcount}

\author{
Barbara Kordy$^1$,
Ludovic Pi\`{e}tre-Cambac\'{e}d\`{e}s$^2$,
Patrick Schweitzer$^1$
\\
$^1$University of Luxembourg,~$^2$EDF, France
\date{}
}

\title{DAG-Based Attack and Defense Modeling:\\
Don't Miss the Forest for the Attack Trees.\footnotemark[1]}

\begin{document}
\maketitle

\begin{abstract}
This paper presents the current state of the art on \emph{attack and defense
modeling approaches that are based on directed acyclic graphs (DAGs)}. DAGs
allow for a hierarchical decomposition of complex scenarios into simple, easily
understandable and quantifiable actions. Methods based on threat trees and
Bayesian networks are two well-known approaches to security modeling. However
there exist more than $30$  DAG-based methodologies, each having different
features and goals. The objective of this survey is to present a complete
overview of  graphical attack and defense modeling techniques based on DAGs.
This consists  of summarizing the existing methodologies, comparing their
features  and proposing a taxonomy of the  described formalisms. This article
also  supports the selection of an adequate modeling technique depending on
user requirements. 
\end{abstract}

\renewcommand*{\thefootnote}{\fnsymbol{footnote}}
\footnotetext[1]{
The research leading to these results has received 
funding from the Fonds National de la 
Recherche Luxembourg under the grants C$08$/IS/$26$ and PHD-$09$-$167$ and 
the European Commission's Seventh Framework Programme
(FP$7$/$2007-2013$) under grant agreement number~$318003$ (TREsPASS).
}
\renewcommand*{\thefootnote}{\arabic{footnote}}

\section{Introduction} 

Graphical security models provide a useful method to represent and analyze 
security scenarios that examine vulnerabilities of systems and organizations.
The great advantage of graph-based approaches lies in combining user friendly,
intuitive, visual  features with formal  semantics and algorithms that allow for
qualitative and quantitative analysis. Over the course of the last two decades,
graphical approaches attracted the attention of numerous security and formal
methods experts and are quickly becoming a stand-alone research area  with
dedicated national and international research 
projects~\cite{Website_SHIELDS,trespass,aniketos,ATREES,VISPER}. Graphical 
models constitute a valuable support tool to facilitate threat assessment and
risk management of real-life systems.  Thus, they have also become popular in
the industrial sector. Notable application domains of graphical models include
security analysis of supervisory control and data acquisition (SCADA)
systems~\cite{ByFrMi,TeLiGo,TaJo}, voting systems~\cite{LaDiEpHa,BuTr},
vehicular communication systems~\cite{HeApFuRoRuWe,AiBoDoFeGeKrLe}, Internet
related attacks~\cite{TiLaFiHa,LiZaRuLi}, secure software
engineering~\cite{JuElBaRa}, and socio-technical
attacks~\cite{BaKoMeSc,EoPaPaCh,ReVeOlCu}.

In this paper we focus on graphical methods for analysis of attack and defense 
scenarios. We understand attack and defense scenarios in a general sense: they
encompass any malicious action of an attacker who wants to harm or damage 
another party or its assets as well as any defense or countermeasure that could
be used to prevent or mitigate such malicious actions. In~$1991$,
Weiss~\cite{Weis} introduced threat logic trees as the first graphical attack
modeling technique. The obvious similarity of threat logic trees to fault
trees~\cite{VeGoRoHa} suggests that graph-based security modeling has its roots
in safety modeling. Weiss' approach can be seen as the origin of numerous
subsequent models, including attack trees~\cite{SaSaScWa,Schn} 
which are nowadays one of the most popular graphical security models. 

Today, more than~$30$ different approaches for analysis of attack and defense
scenarios exist. Most of them extend the original model of threat logic trees in
one or several dimensions which include defensive components, timed and ordered
actions, dynamic aspects and different types of quantification. Moreover,
methods for computation of various security related parameters, such as the
cost, the impact or likelihood of an attack, the efficiency of necessary
protection measures, or the environmental damage of an attack, have been
developed or adapted. 

\begin{figure}[ht]
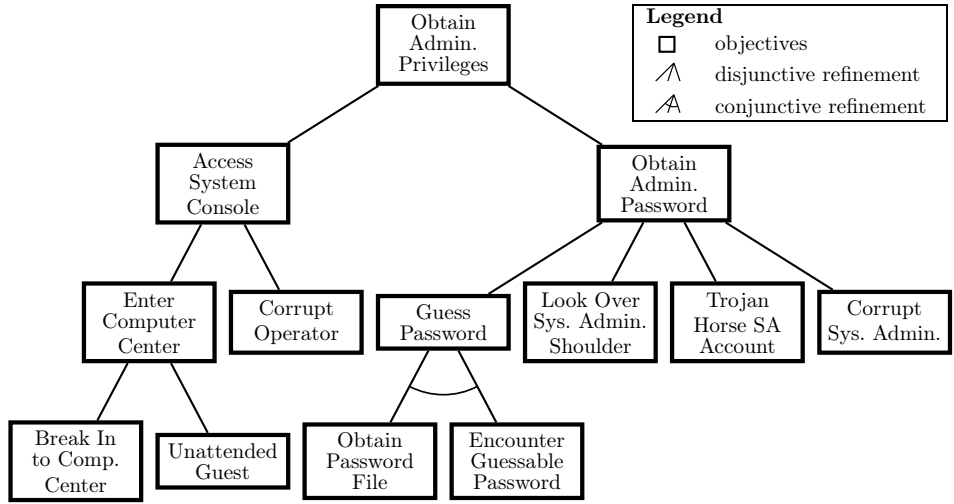

\centering
\settowidth{\pbs}{Sys. Admin.}
\psscalebox{0.8}{
\pstree[levelsep=2.3cm,treesep=0.2cm,nodesep=0cm]{\NNodeBB{Root}{Obtain\\Admin.
\\Privileges}}{
	\pstree{\NodeBB{Access\\System\\Console}}{
		\pstree{\NodeBB{Enter\\Computer\\Center}}{
			\NodeBB{Break In\\to Comp.\\Center}
			\NodeBB{Unattended\\Guest}
		}
		\NodeBB{Corrupt\\Operator}
	}
	\pstree{\NodeBB{Obtain\\Admin.\\Password}}{
		\pstree{\NNodeBB{A}{Guess\\Password}}{
			\NNodeBB{B}{Obtain\\Password\\File}
			\NNodeBB{C}{Encounter\\Guessable\\Password}
		}
		\NodeBB{Look Over\\Sys. Admin.\\Shoulder}
		\NodeBB{Trojan\\Horse SA\\Account}
		\NodeBB{Corrupt\\Sys. Admin.}
	}
}
\Arca{1.2cm}{A}{B}{C}
\rput(-2.9,-0.3){\legendtlt}
}
\caption{A threat logic tree taken from 
\protect\cite{Weis}: Obtaining administrator privileges on a UNIX system.}
\label{fig:treat_logic_tree}
\end{figure}

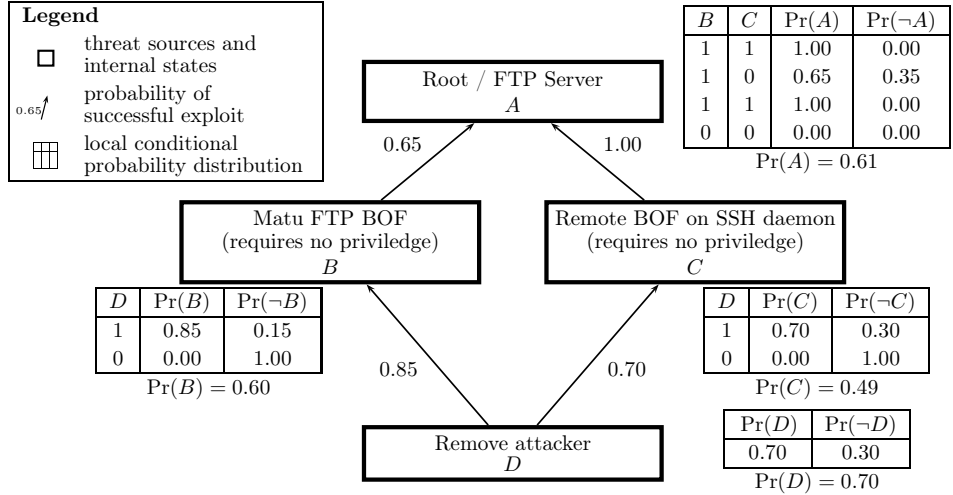
\begin{figure}[ht]
\centering
\psscalebox{0.8}{
\settowidth{\pbs}{Remote BOF on SSH daemon}
\renewcommand*\arraystretch{1.1}
\begin{pspicture}(-8.25,-3.5)(7.25,4.35)
\rput(0,3){\NNodeBB{A}{Root / FTP Server \\ $A$}}
\rput(-3,0.5){\NNodeBB{B}{Matu FTP BOF\\ (requires no priviledge)\\ $B$}}
\rput(3,0.5){\NNodeBB{C}{Remote BOF on SSH daemon\\ (requires no priviledge)\\ 
$C$}}
\rput(0,-3){\NNodeBB{D}{Remove attacker\\ $D$}}
\ncline[nodesep=0cm]{->}{B}{A}
\naput{0.65}
\ncline[nodesep=0cm]{->}{C}{A}
\nbput{1.00}
\ncline[nodesep=0cm]{->}{D}{B}
\naput{0.85}
\ncline[nodesep=0cm]{->}{D}{C}
\nbput{0.70}
\rput(-5.7,3){\legendbayes}

\rput(5,3){\probtabA}
\rput(-5,-1.2){\probtabB}
\rput(5,-1.2){\probtabC}
\rput(5,-3){\probtabD}
\end{pspicture}
\renewcommand*\arraystretch{1.0}
}
\caption{Bayesian attack graph taken 
from~\protect\cite{PoDeRa}: A test network with local conditional 
probability distributions (tables) and updated unconditional 
probabilities 
(below each table).} 
\label{fig:bayesian_network}
\end{figure}

This survey concentrates on formalisms based on directed acyclic graphs (DAGs), 
rather than on arbitrary graphs.  Described approaches can be divided into two
main classes: formalisms derived from or extending threat trees,  and
formalisms based on Bayesian networks. The model creation in all threat
tree-based methodologies  starts with the identification of a  feared event
represented as the root node. Then, the event's causes or consequences,
depending on the specific approach, are deduced and depicted as refining nodes. 
The refinement process is illustrated in  Figure~\ref{fig:treat_logic_tree},
which recreates the first threat tree model proposed by Weiss~\cite{Weis}.
The DAG structure allows to use refinements  with a customizable level of
detail. The root of a DAG is refined as long as the refining children provide
useful and adequate  information about the modeled scenario. Refinements paired
with the acyclic  structure allow for modularization which in turn allows 
different experts to work in parallel on the same model. This is highly
appreciated in case of large-scale, complex models, where analysis of different
parts requires different types of expertise. A big advantage of the DAG-based
approaches is that they are fairly scalable.  They do not suffer from the state
space explosion problem, which is common  for models based on general graphs
with cycles.  In the case of trees, most of the analysis algorithms are linear
with respect  to the number of nodes of the model. Due to multiple incoming
edges, this property is no longer true for DAGs and the complexity of analysis
methods might in theory be exponential. However in practice, the largest
exponent in the runtime of DAG-based  approaches is still acceptable, since it
can be kept small due to the underlying cycle-free structure.  This is, for
instance, the case for Bayesian  inference algorithms  used for the analysis of
security models based on Bayesian networks. Figure~\ref{fig:bayesian_network}
depicts a simple Bayesian attack graph borrowed from~\cite{PoDeRa} and
illustrates how to compute the  unconditional probability of a vulnerability
exploitation.

This paper surveys DAG-based graphical formalisms for attack and defense 
modeling. These formalisms provide a systematic, intuitive and practical
representation of a large amount of possible attacks, vulnerabilities and
countermeasures, while at the same time allowing for an efficient formal and
quantitative analysis of security scenarios. 
The contribution of this work is to 
provide a complete overview of the field and systematize  existing 
knowledge. More specifically, the survey 

\begin{itemize}
\item presents the state of the art in the field of DAG-based
graphical attack and defense modeling as of~$2012$;
\item identifies relevant key aspects allowing to compare different formalisms;
\item proposes a taxonomy of the presented approaches, which helps in selecting
an appropriate formalism;
\item lays a foundation for future research in the field, with the goal to
prevent reinvention of already existing features.
\end{itemize}

In Section~\ref{sec:prelims}, we introduce terminology used in the field of
graph-based security modeling and provide a template for the description of the
formalisms. Section~\ref{sec:main_survey} is the main part of the survey and
presents the DAG-based attack and defense modeling approaches published
before~$2013$. In Section~\ref{sec:table}, we provide a concise tabular 
overview of the presented formalisms. We illustrate how to use the tables in
order to select the most relevant modeling technique, depending on the
application requirements. Section~\ref{sec:alternative} briefly mentions
alternative  graphical security models. We close the survey with concluding
section, which  summarizes our findings and proposes future research directions 
in the field. 

\section{Preliminaries}
\label{sec:prelims}

In this section we introduce our terminology and make a link to existing 
definitions and concepts. We then present and define the aspects that we have
taken into account while analyzing different formalisms. We conclude with a
detailed description of how formalisms from Section~\ref{sec:main_survey} are
described. 

\subsection{Keywords and Terminology}
\label{sec:terminology}

When examining different models in the same context, it is imperative to have a 
common language. Over the last~$20$ years, numerous concepts and definitions 
have emerged in the field of graphical security modeling. This section is
intended to introduce the language used in this paper, and to serve as quick 
reference guide over the most commonly occurring concepts. Our goal here is not 
to point out the differences in definitions or other intricate details. 

\paragraph{Attack and defense modeling} 
By techniques for attack and defense modeling we understand formalisms that
serve for representation and analysis of malicious behavior of an
\emph{attacker} and allow to reason about possible defending strategies of the
attacker's opponent, called the \emph{defender}. In our survey we use attacks in
a very broad sense. Attacks can also be thought of as \emph{threats}, 
\emph{obstacles}, and \emph{vulnerabilities}. Contrary, defenses can appear in 
form of \emph{protections}, \emph{mitigations}, \emph{responses} and 
\emph{countermeasures}. They oppose, mitigate or prevent attacks. 

\paragraph{Nodes}
\emph{Nodes}, also called \emph{vertices}, are one of the main components of
graph-based security models. They are used to depict the concept that is being
modeled. Nodes may represent \emph{events}, \emph{goals}, \emph{objectives} and
\emph{actions}. Depending on whether the models are constructed in an inductive
or deductive way, nodes may also express \emph{causes} or \emph{consequences}.

\paragraph{Root node}
In a rooted DAG (and therefore in any tree) the \emph{root} is the single
designated node that does not have any predecessor. From it all other nodes can
be reached via a directed path. This distinguished node usually depicts the
entire concept which is being modeled. In the context of security models, 
various existing names for this special node include \emph{top event}, 
\emph{main goal}, \emph{main consequence}, \emph{main objective} or \emph{main
action}.

\paragraph{Leaf nodes}
In a DAG, nodes that do not have any children are called \emph{leaves}. They
usually display an atomic component of a scenario that is no longer refined.
They are also called \emph{primary events}, \emph{basic components},
\emph{elementary attacks}, \emph{elementary components} or \emph{basic actions}.

\paragraph{Edges}
Edges are the second main component of graph-based security models. They link
nodes with each other, and thusly determine relations between the modeled
concepts. Edges are also called \emph{arcs}, \emph{arrows}, or \emph{lines}. In
some models, edges may have special semantics and may detail a cause-consequence
relation, a specialization or some other information.

\paragraph{Connectors}
Connectors usually specify more preciously how a parent node is connected with
its children. A connector might be a set of edges or a node of a special type. 
Connectors are also called \emph{refinements} or \emph{gates}. Some examples 
include: AND, OR, XOR,~$k$-out-of-$n$, priority AND, triggers, etc.

\paragraph{Priority AND}
A \emph{priority AND} (PAND) is a special kind of AND connector which prescribes
an order in which the nodes are to be treated. The origin of the prescribed
order is usually time or some priority criterion. The PAND is also called an
\emph{ordered-AND}, an \emph{O-AND} or a \emph{sequential AND}. Sometimes the
underlying reason behind the priority is specified as in the case of the
\emph{time-based AND}.

\paragraph{Attributes}
\emph{Attributes} represent aspects or properties that are relevant for
quantitative analysis of security models. Examples of attributes, sometimes also
called \emph{metrics}, include: impact of an attack, costs of necessary
defenses, risk associated with an attack etc. Proposed computation methods range
from versatile approaches that can be applied for evaluation of a wide class of
attributes, to specific algorithms developed for particular measures. An 
example of the former is the formalization of an attribute domain proposed 
in~\cite{MaOo}, which is well suited for calculation of any attribute whose 
underlying algebraic structure is a semi-ring. An example of the latter are the 
specific methods for probability computation proposed in~\cite{Yage}.

\subsection{Examined Aspects}
\label{sec:examined_aspects}

One of the goals of this paper is to provide a classification of existing
formalisms for attack and defense modeling. Thus, all approaches described in
Section~\ref{sec:main_survey} were analyzed based on the same $13$ criteria,
which we refer to as \emph{aspects} and define in this section.

The formalisms are grouped according to the following two main aspects:
\begin{enumerate}
\item \textbf{Attack and/or defense modeling:}
\emph{Attack} modeling techniques are focused on an attacker's actions and
vulnerabilities of systems; \emph{defense} modeling techniques concentrate on 
defensive aspects, such as detection, reaction, responses and prevention. 
\item \textbf{Static or sequential approaches:} 
\emph{Sequential} formalisms take temporal aspects, such as dynamics time 
variations, and dependencies between considered actions, such as order or 
priority, into account; \emph{static} approaches cannot model any of such 
relations.
\end{enumerate}

The above two aspects provide a partition of all considered approaches. 
Furthermore, they correspond to questions that a user selecting a suitable
formalism is most likely to ask, namely \emph{'What do we want to model?'} and
\emph{'How do we want to model?'}. The proposed classification allows a reader
to easily make a primary selection and identify which formalisms best fit his 
needs. 

Besides the two main aspects, each formalism is analyzed according to 
additional criteria, listed in Table~\ref{tab:keywords_table2}. All aspects
taken into account in our work, can be grouped into three categories:
\begin{itemize}
\item Aspects relating to the formalism's \emph{modeling capabilities}, i.e., 
what we can model: attack or defense modeling, sequential or static modeling, 
quantification, main purpose, extensions.
\item Aspects relating to the formalism's \emph{characteristics}, i.e., how we 
can model: structure, connectors, formalization.
\item Aspects related to the formalism's \emph{maturity and usability}: tool
availability, case study, external use, paper count, year.
\end{itemize}

In Table~\ref{tab:keywords_table2}, we define all $13$ aspects in form of
questions and provide possible values that answer the questions.

\setlength{\tabcolsep}{.1cm}

\begin{longtable}[c]{m{0.15\textwidth}m{0.25\textwidth}m{0.141\textwidth}
m{0.40\textwidth}}
\caption{Table summarizing aspects taken into account in formalism 
description.}
\label{tab:keywords_table2}
\\
\toprule
\textbf{Aspect} & 
\textbf{Aspect Description} & 
\textbf{Possible\newline Values} & 
\textbf{Value Explanation}\\
\midrule
\multirow{3}{0.15\textwidth}[-0.2cm]{Attack or defense} 
& 
\multirow{3}{0.25\textwidth}[-0.2cm]{Is the formalism offensively or 
defensively oriented?} 
& 
Attack & 
Only attack modeling \\
\cmidrule{3-4}
& & Defense & Only defense modeling \\
\cmidrule{3-4}
& & Both & 
Integrates attack and defense modeling
\\
\midrule
\multirow{3}{0.15\textwidth}[0.13cm]{Static or sequential} 
& 
\multirow{3}{0.25\textwidth}[0.13cm]{Can the formalism deal with 
depen\-den\-cies and time varying scenarios?} 
& 
Static & 
\multirow{2}{0.36\textwidth}{Does not support any dependencies}\newline 
\\\cmidrule{3-4}
& & 
Sequential &
Supports time and order dependencies
\\
\midrule
\multirow{5}{0.15\textwidth}[0cm]{Quantifi\-cation} 
& 
\multirow{5}{0.25\textwidth}[0cm]{Can numerical values be computed using the 
formalism?} & 
Versatile & 
Supports numerous generic and diverse metrics 
\\\cmidrule{3-4}
& & 
Specific & 
Dedicated, tailored for (a couple of) specific metrics
\\\cmidrule{3-4}
& & 
No & 
Does not support quantification\\
\midrule
\pagebreak
\midrule
\multirow{10}{0.15\textwidth}[-0.6cm]{Main purpose} 
& 
\multirow{10}{0.25\textwidth}[-0.6cm]{Why was the formalism invented?} 
& 
Sec. mod. 
& General security modeling \\
\cmidrule{3-4}
&& Unification 
& 
Unification of existing formalisms 
\\\cmidrule{3-4}
&& 
Quantitative 
& 
Provide better methods for quantitative analysis
\\\cmidrule{3-4}
&& Risk 
& Support risk assessment \\\cmidrule{3-4}
&& Soft. dev. 
& 
Support secure software development 
\\\cmidrule{3-4}
&& Int. det. 
& 
Automated intrusion detection and response analysis 
\\\cmidrule{3-4}
&& Req. eng. 
& 
Support security requirements enginee\-ring 
\\
\midrule
\multirow{12}{0.15\textwidth}[-0.2cm]{Extensions} 
& 
\multirow{12}{0.25\textwidth}[-0.2cm]{What are added features of the 
formalism with respect to the state of the art?} 
& 
Structural & 
New connectors, extended graph structure 
\\\cmidrule{3-4}
& & 
\multirow{2}{0.141\textwidth}{Com\-puta\-tional}\newline & 
How the formalism handles computations (e.g., top down)
\\\cmidrule{3-4}
& & 
Quantitative 
& 
Which computations can be performed (e.g., specific attributes)
\\\cmidrule{3-4}
& & 
Time & 
The formalism can handle time dependencies
\\\cmidrule{3-4}
& & 
Order & 
The formalism can handle order dependencies
\\\cmidrule{3-4}
& & 
\multirow{2}{0.141\textwidth}{New formalism} \newline & 
Entirely new formalism\\
\midrule
\multirow{4}{0.15\textwidth}[-0.15cm]{Structure} 
& 
\multirow{4}{0.25\textwidth}[-0.15cm]{Which graphical structure is the 
formalism based on?} 
&
Tree & 
Tree (possibly with repeated nodes)
\\\cmidrule{3-4}
& & 
DAG & 
Directed acyclic graph \\\cmidrule{3-4}
& & 
Unspecified 
& 
It is not specified whether the models are DAGs or trees
\\
\midrule
\multirow{1}{0.15\textwidth}{Connectors} 
& 
\multirow{3}{0.25\textwidth}[0.415cm]{What type of connectors does the 
formalism use?} 
&
\multirow{2}{0.141\textwidth}{List of connectors} \newline 
& 
AND, OR, trigger, sequential AND, ordered-AND, priority AND,~$k$-out-of-$n$, 
OWA nodes, split gate, countermeasures, counter leaves
\\\midrule
\multirow{5}{0.15\textwidth}{Formali\-zation} 
&
\multirow{5}{0.25\textwidth}{Is the formalism formally defined?} 
& 
Formal & 
Defined using a mathematical framework; with clear syntax and semantics 
\\\cmidrule{3-4}
& & 
Semi-formal & 
Parts of the definitions are given\newline verbally, parts are
precise\\\cmidrule{3-4}
& &
Informal & 
Models only verbally described\\
\midrule
\multirow{4}{0.15\textwidth}[0cm]{Tool availability} 
&
\multirow{4}{0.25\textwidth}[0cm]{Does a software tool supporting the 
formalism exist?} & 
Commercial &
A commercial
software tool exists \\\cmidrule{3-4}
& &
Prototype
& A prototype tool exists\\\cmidrule{3-4}
& &
No
& No implementation exists\\
\midrule
\pagebreak
\midrule
\multirow{5}{0.15\textwidth}[0cm]{Case study} 
& 
\multirow{5}{0.25\textwidth}[0cm]{Do papers or reports describing case studies 
exist?} & 
Real(istic) & 
Real or realistic case study has been documented
\\\cmidrule{3-4}
& & 
\multirow{2}{0.141\textwidth}{Toy case study} \newline & 
\multirow{2}{0.36\textwidth}{Toy case study has been described} \newline 
\\\cmidrule{3-4}
& & 
No & 
No documented case study exist\\
\midrule
\multirow{8}{0.15\textwidth}[0cm]{External use} 
&
\multirow{8}{0.25\textwidth}[0cm]{Do papers or reports having a disjoint set 
of authors from the formalism inventors exist?} 
&
Indepen\-dent & 
People and institutions who did not invent the formalism have used it 
\\\cmidrule{3-4}
& & 
Collabora\-tion 
&
The formalism has been used by external researchers and institutions in 
collaboration with its inventors
\\\cmidrule{3-4}
& & 
No & 
The formalism has only been used by its inventors or within the institution 
where it was invented 
\\
\midrule
Paper count & 
\multirow{2}{0.25\textwidth}{How many papers on the formalism 
exist?}\newline & 
Number & 
Number of papers that have been identified\footnote{
Different versions of the same paper (e.g., an official publication and a 
corresponding technical report) have been counted as the same publication.}
\\
\midrule
Year & 
\multirow{3}{0.25\textwidth}{What year was the formalism first 
published?} \newline $\phantom{x}$ \newline & Year & Before~$2013$\\
\bottomrule
\end{longtable}

\subsection{Template of the Formalism Descriptions}
\label{sec:template_of_formalisms}
The description of each formalism presented in 
Section~\ref{sec:main_survey} complies with the following template.

\paragraph{General presentation}
The first paragraph mentions the name of the formalism and its authors, as well
as lists main papers. The year when the approach was proposed is given. Here
we also present the main purpose for which the technique was introduced. 
If nothing is indicated about the formalism structure, it means that it is a
generic DAG. If the structure is more specifically a tree, then it is indicated
either in the formalism's name or in the first paragraph of the description.

\paragraph{Main features}
In the second paragraph, we briefly explain 
the main features of the formalism, in particular
what its added features are with respect to the state of 
the art at the time of its invention.
Moreover, we state whether the modeling technique is 
formalized, i.e., whether it complies with proper mathematical definitions.

\paragraph{Quantification}
Next, we focus on quantitative aspects of the considered methodology.
We explain whether the formalism
is tailored for a couple of specific parameters or metrics, or 
whether a general framework has been introduced to deal with
computations. 
In the first case, we list relevant attributes, in the second case, 
we briefly explain the new algorithms or calculation
procedures. 

\paragraph{Practical aspects}
When relevant, we mention industrialized or prototype software tools
supporting the described approach. We also indicate when real
or realistic scenarios have been modeled and analyzed with the help of the
described approach. In this paragraph, we also refer to
large research projects and Ph.D. theses applying the methodology. This
paragraph is optional.

\paragraph{Additional remarks}
We finish the formalism description by relating it to follow-up methodologies.
If it is the case, we point out the formalism's limitations that have been
identified by its authors or other researchers from the field. In this part we 
also point out various other peculiarities related to the formalism. This 
paragraph is optional.

\section{Description of the Formalisms}
\label{sec:main_survey}
This section constitutes the main part of this survey. It describes numerous 
DAG-based approaches for graphical attack and defense modeling according to the 
template outlined in Section~\ref{sec:template_of_formalisms}. 
Models gathered within each subsection are ordered chronologically, 
with respect to the year of their introduction. 

\subsection{Static Modeling of Attacks}
\label{sec:offensive_static}

\subsubsection{Attack Trees} 
\label{sec:attack_trees}

Inspired by research in the reliability area, Weiss~\cite{Weis} in~$1991$ and
Amoroso~\cite{Amor} in~$1994$ proposed to adopt a tree-based concept of visual
system reliability engineering to security. Today, \emph{threat
trees}~\cite{Amor,SwSn,HoLe,MaDoHeKoXu,DepD}, \emph{threat logic
trees}~\cite{Weis}, \emph{cyber threat trees}~\cite{OnTuThNaSzMa},
\emph{fault trees} for attack modeling~\cite{StSc}, and the \emph{attack
specification language}~\cite{TiLaFiHa} can be subsumed under \emph{attack
trees}, which are AND-OR tree structures used in graphical security modeling.
The name attack trees was first mentioned by Salter~et~al. in
1998~\cite{SaSaScWa} but is often only attributed to Schneier and cited
as~\cite{Schn,Schn2}.

In the attack tree formalism, an attacker's main goal (or a main security
threat) is specified and depicted as the root of a tree. The goal is then
disjunctively or conjunctively refined into sub-goals. The refinement is 
repeated recursively, until the reached sub-goals represent basic  actions.
Basic actions correspond to atomic components, which can  easily be understood
and quantified. Disjunctive refinements represent  different alternative ways of
how a goal can be achieved, whereas conjunctive refinements depict different
steps an attacker needs to take in order to achieve a goal~\cite{QiLe}.
In~$2005$, Mauw and Oostdijk formalize attack trees  by defining their semantics
and specifying tree transformations consistent with  their
framework~\cite{MaOo}. Kienzle and Wulf present an extensive general procedure
for tree construction~\cite{KiWu} while other researchers are engaged  in
describing how to generate attack tree templates using \emph{attack 
patterns}~\cite{MoElLi,LiMo}.

Quantification of security with the help of attack trees is a very active
topic of research~\cite{WhPhWaPa}. A first simple procedure for quantification
of attack trees was proposed by Weiss~\cite{Weis} and is based on a bottom-up
algorithm. In this algorithm, values are provided for all leaf nodes and the
tree is traversed from the leaves towards the root
in order to compute values of the refined nodes.  
Depending on the type of
refinement, different functional operators are used to combine the values of
the children. This procedure allows to analyze simple aspects, such as the 
costs of an attack, the time of an attack or the necessary skill level
\cite{Weis,Amor,SaSaScWa,Schn,ByFrMi,HiUnJaSaLu,FuChWaLeTaAnLi,MaOo,%
BiDaPe,Yage,EdDaRaMi,SaDuPa,HeApFuRoRuWe,LiLiFeHe,AbCeKa,BaPe,TaJo,WhPhWaPa}.
Whenever more complicated attributes, such as probability of occurrence, 
probability of success, risk or similarity measures are analyzed, additional 
assumptions, for example mutual independence of all leaf nodes, are
necessary, or methods different from the bottom-up 
procedure have to be used
\cite{Schn,ByFrMi,BuLaPrSaWi,EdDaRaMi,Yage,JuWi3,HeApFuRoRuWe,LiLiFeHe,%
Buon2,AbCeKa,BuFeMe,OnTuThNaSzMa,BuFeMe2,MaThFe,WaWhPhPa,%
ReSeKoStHo,RoKiTr2,ZhYu}.
Propagation of fuzzy numbers that model fuzzy preference relations has 
initially been proposed in~\cite{BoFeGi} and extended in~\cite{BuFe}. Using 
Choquet integrals it is possible to take interactions between nodes into 
account.

Commercial software for attack tree modeling, such as
SecurITree~\cite{Program1} from Amenaza or AttackTree+~\cite{Program2} from
Isograph provides a large database of attack tree templates. Academic tools,
including SeaMonster~\cite{Program6} developed within the SHIELDS
project~\cite{Website_SHIELDS} offer visualization and library support. Attack
trees may occur in the SQUARE methodology~\cite{MeHoSt}. The entire methodology 
and therefore visualization of attack trees are supported by the SQUARE
tool~\cite{Website_Square_Tool}. AttackDog~\cite{Program3} was developed as a
prototype software tool for managing and evaluating attack trees with voting
systems in mind but is believed to be much more widely applicable to evaluating
security risks in systems~\cite{Accurate_Annual_Report}. Numerous case studies
\cite{MoElLi,TiLaFiHa,ByFrMi,CoCoFr,EvHeKyPiWa,HiUnJaSaLu,%
BuPaUnPaWaSa,FuChWaLeTaAnLi,MeHoSt,AiBoDoFeGeKrLe,%
KhSe,TeLiGo,ClSiTyHa,GrJo,Mars,NiXiYoSi,PaLeChKiLeKw,ReVeOlCu,SaDuPa,%
HeApFuRoRuWe,LiZaRuLi,MoMaCaJi,CaKiKi,FeBaMoJi,LaPoDa,LaPoMiDeDa,TeMaLi,%
EoPaPaCh,LaDiEpHa,MoCaMa,SaWoXu,WaLeRo,SuSv,ZhYu}
account for the applicability of the attack tree methodology. Attack trees are
used in large international research
projects~\cite{Website_EVITA,Website_SHIELDS,trespass,Nucl}.
They have been focus of various Ph.D. and 
Master theses~\cite{Kien,Pumf,Mobe,Fost,Sche,Opel,Karp,Edge,Espe,Hogg,Magi,%
Harr,Jurg,Piet,Roy,Niel,Ostl,Sameer,Zonouz,Buon,Koot,Pose}.

Since attack trees only focus on static modeling and only take an attacker's
behavior into account, numerous extensions that include dynamic modeling and a
defender's behavior, exist. Except for formalisms involving Bayesian inference
techniques, all other DAG-based formalisms refer back to the attack tree 
methodology.  They point out a need for modeling defenses, dynamics, and ordered
actions,  as well as  propose computation procedures for probability or highly
specified key figures. Neither the name attack trees, nor the initial
formalization of Mauw and Oostdijk is universally accepted. Some researchers
consider attack trees,  threat trees or fault trees to essentially be the 
same~\cite{Lams,MoYa,SoEkNo,Andefirst,Ingo,StSc} while other researchers point 
out specific differences~\cite{LiLiFeHe,MiMu}. As common ground all  mentioned
methodologies use an AND-OR tree structure but are divided on what  the tree can
actually model (attacks, vulnerabilities, threats, failures, etc.) 

\subsubsection{Augmented Vulnerability Trees} 
\label{sec:vulnerability_trees}

\emph{Vulnerability trees}~\cite{ViJo} have been proposed by Vidalis and Jones
in~$2003$ to support the decision making process in threat assessment.
Vulnerability trees are meant to represent  hierarchical interdependence between
different  vulnerabilities of a system. In~$2008$, Patel, Graham and
Ralston~\cite{PaGrRa} extended this model to \emph{augmented vulnerability
trees} which combine the concepts of vulnerability trees, fault tree analysis,
attack trees, and cause-consequence diagrams. The aim of augmented vulnerability
trees is to express the financial risk that computer-based information systems
face, in terms of a numeric value, called ``degree of security''.

The root of a vulnerability tree is an event that represents a vulnerability;
the branches correspond to different ways of exploiting it. The leaves of the
tree symbolize steps that an attacker may perform in order to get to the parent
event. The model, which is not formally defined, uses only AND and OR connectors
depicted as logical gates. Vulnerability trees are very similar to attack trees,
they differ in how the root event is defined (vulnerability event vs. an
attacker's goal). A step-wise methodology consisting of a sequence of six steps
is proposed in~\cite{PaGrRa} to create an augmented vulnerability tree
and analyze security related indexes.

The authors of~\cite{ViJo} propose a number of attributes on vulnerability
trees, including: complexity value (the smaller number of steps that an attacker
has to employ in order to achieve his goal), educational complexity
(qualifications that an attacker has to acquire in order to exploit a given
vulnerability), and time necessary to exploit a vulnerability. However, the
paper~\cite{ViJo} does not detail how to compute these attributes.
In~\cite{PaGrRa}, the model is augmented with two indexes: the threat-impact
index and the cyber-vulnerability index. The first index, represented by a value
from~$[0,100]$, expresses the financial impact of a probable cyber threat. The
lower the index the smaller the impact from a successful cyber attack. The
second index, also expressed by a value from~$[0,100]$, represents system flaws
or undesirable events that would help an intruder to launch attacks. The lower
this index, the more secure the system is.

In~\cite{TaJo}, the augmented vulnerability tree approach has been used to
evaluate risks posed to a SCADA system exposed to the mobile and the Internet 
environment.

\subsubsection{Augmented Attack Trees} 
\label{sec:augmented_attack_trees}

In~$2005$, Ray and Poolsappasit\footnote{In early papers spelled
Poolsapassit~\cite{RaPo,PoRa}} first published about \emph{augmented
attack trees} to provide a probabilistic measure of how far an
attacker has progressed towards compromising the system~\cite{RaPo}. This
tree-based approach was taken up by H. Wang~et~al. in~$2006$ and extended to
allow more flexibility in the probabilistic values provided for the leaf
nodes~\cite{WaLiZh}. When again publishing in~$2007$, Poolsappasit and Ray used
a different definition of augmented attack trees to be able to perform a
forensic analysis of log files~\cite{PoRa}. Using the second definition of
augmented attack trees, J.~Wang~et~al. performed an analysis of SQL injection
attacks~\cite{WaPhWhPa} and DDoS attacks~\cite{WaPhWhPa3}. They also extended
augmented attack trees further to measure the quality of detectability of an
attack~\cite{WaPhWhPa2}. Co-authors of Dewri, namely Poolsappasit and Ray,
formalized attack trees as AND-OR structure where every node is interpreted to
answer a specific binary question~\cite{DePoRaWh,DeRaPoWh}. This
formalization is then again extended to augmented attack trees by adding to
every node an indicator variable and an additional value with the help of
which the residual damage is computed. On the enhanced structure they are able
to optimize how to efficiently trade-off between spent money and residual
damage.

The various ways of defining augmented attack trees are based on attack trees
(Section~\ref{sec:attack_trees}). In the first definition, attack trees are
augmented by node labels that quantify the number of compromised subgoals on
the most advanced attack path as well as the least-effort needed to compromise
the subgoal on the most advanced path to be able to compute the probability of
attack~\cite{RaPo}. H.~Wang~et~al. generalized this definition from integer
values to general weights. Both approaches include tree pruning and tree
trimming algorithms to eliminate irrelevant nodes with respect to intended
operations (behavior) of a user~\cite{WaLiZh}. In the second definition,
attack trees are augmented by descriptive edge labels and attack signatures.
Each edge defines an atomic attack which is described by the label and
represents a state transition from a child node to the corresponding parent.
An attack signature is a sequence of groups of incidents, from which a sequence
of incidents can be formed, which executed constitutes an atomic attack.
The sequences are then exploited to filter log files for relevant intrusion
incidences~\cite{PoRa} and used to describe state transitions in SQL injection
attacks using regular expressions~\cite{WaPhWhPa}. Moreover they are exploited
to model state transition in DDoS attacks~\cite{WaPhWhPa3} and adapted to
provide a measure for quality of service detection, called quality of
detectability~\cite{WaPhWhPa2}. In an extension of the third
definition~\cite{DeRaPoWh} the system administrator's dilemma is thoroughly
examined. The purpose of this extension is to be able to compute a bounded
minimization of the cost of the security measures while also keeping the
residual damage at a minimum.

Augmented attack trees were designed with a specific quantitative purpose in
mind. The first formalization of augmented attack trees was introduced to
compute the probability of a system being successfully attacked. Additionally
to increasing the descriptive capabilities of the methodology, the second
definition is accompanied by several algorithms that help compute the quality
of detectability in~\cite{WaPhWhPa2}. As mentioned before, the third
definition targets solving the system administrator's dilemma. This is
achieved by using a simplistic cost model a multi-objective optimization
algorithm which guides the optimization process of which security hardening
measures best to employ.

The authors of the first formalism state that attempts by system
administrators to protect the system will not change the outcome of their
analysis. A similar shortcoming is suggested for the second formalization.

\subsubsection{OWA Trees} 
\label{sec:OWA_trees}

In~$2005$, Yager proposed to extend the AND and OR nodes used in attack trees by
replacing them with ordered weighted averaging (OWA) nodes. The resulting
formalism is called \emph{OWA trees}~\cite{Yage} and it forms a general
methodology for qualitative and quantitative modeling of attacks.

Regular attack trees make use of two (extreme) operators only: AND (to be used
when \emph{all} actions need to be fulfilled in order to achieve a given goal)
and OR (to be used when the fulfillment of \emph{at least one} action is
sufficient to reach a desired result). OWA operators represent quantifiers such
as \emph{most}, \emph{some}, \emph{half of}, etc. Thus, OWA trees are well
suited to model uncertainty and to reason about situations where the number of
actions that need to be satisfied is unknown. OWA trees are static in the sense
that they do not take interdependencies between nodes into account. They have
been formally defined in~\cite{Yage} using the notion of an \emph{OWA weighting
vector}. Since AND and OR nodes can be seen as special cases of OWA nodes,
mathematically, attack trees form a subclass of OWA trees. Therefore, algorithms
proposed for OWA trees are also suitable for the analysis of attack trees.

In~\cite{Yage}, Yager provides sound techniques for the evaluation of success
probability and cost attributes on OWA trees. For the probability attribute, he
identifies two approaches that can be explained using two different types of
attackers. The first approach assumes that the attacker is able to try all
available actions until he finds one that succeeds. Since in most situations
such an assumption is unrealistic, the author proposes a second model, where an
attacker simply chooses the action with the highest probability of success.
Furthermore,~\cite{Yage} presents two algorithms for computing the success
probability attribute: one assumes independent actions which leads to a simpler
calculation procedure, the other can deal with dependent actions. Finally, the
author discusses how to join the two attributes together, in order to correctly
compute the cheapest and most probable attack.

In~\cite{BoFeGi}, Bortot, Fedrizzi and Giove proposed the use of Choquet
integrals in order to reason about OWA trees involving dependent actions.

\subsubsection{Parallel Model for Multi-Parameter Attack Trees} 
\label{sec:multi-parameter_attack_trees}

In~$2006$, Buldas, Laud, Priisalu, Saarepera and Willemson initiated a series of
papers on rational choice of economically relevant security measures using
attack trees. The proposed model is called \emph{multi-parameter attack trees}
and was first introduced in~\cite{BuLaPrSaWi}. Between~$2006$ and $2010$,
researchers from different research institutes in Estonia proposed six follow-up
papers~\cite{BuTr,JuWi,JuWi3,WiJu,JuWi2,Niit}, extending and improving the
original model proposed in~\cite{BuLaPrSaWi}.

Most of the approaches for quantitative analysis using attack trees, prior
to~\cite{BuLaPrSaWi}, focus on one specific attribute, e.g., cost or feasibility
of an attack. In reality, interactions between different parameters play an
important role. The aim of the mentioned series of papers was to study how tree
computations must be done when several interdependent parameters are considered.
The model of multi-parameter attack trees assumes that the attacker behavior is
rational. This means that attacks are considered unlikely if their costs are
greater than the related benefits and that the attacker always chooses the most
profitable way of attacking. The parallel model for multi-parameter attack trees
has been studied in~\cite{BuLaPrSaWi,BuTr,JuWi,JuWi3,JuWi2,Jurg}. This model
assumes that all elementary attacks take place simultaneously, thus the attacker
does not base his decisions on success or failure of some of the elementary
attacks.

Multi-parameter attack trees concentrate on the attribute called expected
attacker's outcome. This outcome represents a monetary gain of the attacker and
depends on the following parameters: gains of the attacker in case the attack
succeeds, costs of the attack, success probability of the attack, probability of
getting caught and expected penalties in case of being caught. First, a game
theoretical model for estimation of the expected attacker's outcome was proposed
by Buldas~et~al.~\cite{BuLaPrSaWi}, where values of all parameters are
considered to be precise point estimates. In~\cite{JuWi}, J\"{u}rgenson and
Willemson extend the computation methods proposed in~\cite{BuLaPrSaWi} to the
case of interval estimations. Later it turned out that the computational model
from~\cite{BuLaPrSaWi} was imprecise and inconsistent with the mathematical
foundations of attack trees introduced in~\cite{MaOo}. Hence, an improved
approach for the parallel attack tree model was proposed by J\"{u}rgenson and
Willemson~\cite{JuWi3}. Since this new approach requires exponential running
time to determine possible expected outcome of the attacker, an optimization
solution, based on a genetic algorithm for fast approximate computations, has
been proposed by the same authors in~\cite{JuWi2}.

In~\cite{BuTr}, Buldas and M\"{a}gi applied the approach developed
in~\cite{BuLaPrSaWi} to evaluate the security of two real e-voting schemes: the
Estonian E-voting System in use at the time (EstEVS) and the Secure Electronic
Registration and Voting Experiment (SERVE) performed in the USA in~$2004$. A
detailed description of this case study is given in the Master thesis of
M\"{a}gi~\cite{Magi}. A prototype computer tool supporting the security analysis
using the multi-parameter attack trees has been implemented~\cite{Program4} and
described in~\cite{AndrusenkoMaster}. 

In Section~\ref{sec:multi-parameter_attack_trees-serial}, we describe
the serial model for multi-parameter attack trees,
which extends the parallel model with an order on the set of elementary
components.

\subsubsection{Extended Fault Trees} 
\label{sec:extended_fault_trees}

\emph{Extended fault trees} were presented by Fovino~et~al. at the ESREL
conference in~$2007$~\cite{MaFoCi} and published in an extended version as a
journal paper~\cite{FoMaCi} issued in~$2009$. The formalism aims at combining
malicious deliberate acts, which are generally captured by attack trees
(Section~\ref{sec:attack_trees}), and random failures, which are often
associated with classical fault trees (Section~\ref{sec:attack_trees}).

Extended fault trees and attack trees are structurally similar. The main
difference between the two formalisms is in the type of basic events  that can
be modeled. In EFT basic events can represent both non-malicious, accidental
failures  as well as attack steps or security events. Basic events of attack
trees usually correspond to malicious attacker's actions only. Logical AND and
OR gates are explicitly represented in the same way as in classical fault trees.
A step-by-step model construction process is described in~\cite{FoMaCi},
defining how existing fault-trees can be extended with attack-related components
to form extended fault tree models. The modeling technique complies with proper
mathematical foundations,  directly issued from fault trees as defined in the
safety and reliability area.

Quantification capabilities are focused on the computation of the probability
of occurrence of the top-event (root node). Generic formulas from fault tree
quantitative analysis are recalled in~\cite{FoMaCi}, including treatment of 
independent or mutually exclusive events. However, no concrete examples of 
quantification are provided.

A simple example, analyzing the different failure and attack scenarios leading
to the release of a toxic substance by a chemical plant, is described
in~\cite{FoMaCi}. No particular tool has been developed to support extended
fault trees, however, all classical fault tree tools may be used directly.

One of the limitations explicitly stressed by the inventors of extended fault
trees is that they do not take into account time dynamics. 

\subsection{Sequential Modeling of Attacks}
\label{sec:offensive_sequential}

\subsubsection{Cryptographic DAGs} 
\label{sec:cryptographic_dag}

Meadows described \emph{cryptographic DAGs} in~$1996$ (proceedings published
in~$1998$), in order to provide a simple representation of an attack
process~\cite{Mead2}. The purpose of the formalism is limited to visual
description. The attack stages of the overall attack process correspond to the
nodes of a DAG. The difficulty of each stage is shown by a color code.
In~$1996$, the novelty of cryptographic DAGs was to provide a simple
representation technique of sequences and dependencies of attack steps towards a
given attacker's objective.

From a modeling point of view, each stage (represented as a colored box)
contains a textual description of atomic actions needed for the realization
of the stage. Arrows represent dependencies between the boxes. A simple arrow 
indicates that one stage is needed to realize another stage. Two arrows fanned 
out symbolize that one stage enables another one repeatedly. More generally 
speaking, cryptographic DAGs are an informal formalism targeted at high level 
system descriptions.

Cryptographic DAGs do not support any type of quantification.

Cryptographic DAGs have been used in~\cite{Mead2} to demonstrate attacks on
cryptographic protocols (with SSL and Needham-Schroeder scheme as a use-cases),
however this representation technique may be used to model other types
of attacks as well.

This formalism allows the representation of sequences of attack steps, and
dependencies between those steps, but cannot capture ``static'' relations like
``AND'' and ``OR''. Moreover, the clarity and usability of the models depends
heavily on the text inside the boxes, which is not standardized.

\subsubsection{Fault Trees for Security} 
\label{sec:fault_trees_for_sec}

Fault tree analysis was born in~$1961$ and has initially been developed into a
safety, reliability and risk assessment
methodology~\cite{Wats,VeGoRoHa,StVeDuFrMiRa,Elec2}. A short history of
non-security related fault trees was published by Ericson II~\cite{Eric}
in~$1999$. Fault trees have also been adopted to software
analysis~\cite{LeHa,Leve,HeWoSlHoMiLu,HeWoSlHoMiWaWaSt} and were even equated
with attack trees by Steffen and Schumacher~\cite{StSc}. In $2003$, however,
Brooke and Paige adopted \emph{fault tree for security}, extending the classical
AND-OR structure of attack trees (Section~\ref{sec:attack_trees}), to include
well known concepts from safety analysis~\cite{BrPa}.

Based on an AND-OR structure, three additional connectors (priority AND,
exclusive OR and inhibit), specific node types (basic, conditioning,
undeveloped, external and intermediate) as well as transfer symbols (transfer
in, transfer out) to break up larger trees are adopted from fault tree
analysis in its widest sense. Fault trees for security are an aid to the
analysis of security-critical systems, where first an undesired (root) event is
identified. Then, new events are constructed by inserting connectors that
explicitly identify the relationship of the events to each other. Several
rules, like the ``no miracle'' rule, the ``complete the gate'' rule and the
``no gate to gate'' rule are adopted directly from fault trees. Construction
stops when there are no more uncompleted intermediate events. In the end, a
completed fault tree serves as an ``attack handbook'' by providing information
about the interactions by which a security critical system fails.

In~\cite{BrPa}, Brooke and Paige state that in computer security ``it is
difficult to assign useful probabilities to the events''. Consequently
probabilistic quantitative analysis is debatable. Instead the authors
recommend to perform risk analysis which answers how the system fails based on
the primary events (leaf nodes).

While~\cite{BrPa} only provides a toy example, the authors state that any tool
used in fault tree analysis can be used. They refer to~\cite{Website_FTAP} as
a good overview of available programs.

\subsubsection{Bayesian Networks for Security} 
\label{sec:bayesian_networks}

Starting in~$2004$, different researchers proposed, seemingly independently, to
adopt \emph{Bayesian networks}, whose origin lies in artificial intelligence, 
as a security modeling technique~\cite{Pear2,Pear,Neap,JeNi}. Bayesian networks
are also known as \emph{belief network} or \emph{causal network}. In Bayesian
networks, nodes represent events or objects and are associated with
probabilistic variables. Directed edges represent causal dependencies between
nodes. Mathematical algorithms developed for Bayesian networks are suited to
solve probabilistic questions on DAG structures. They are aimed at keeping the
exponent small when the computing algorithm is exponential and reduce to 
polynomial algorithms if the DAG is actually a tree.

According to Qin and Lee, the objective of Bayesian Networks for Security is to
``use probabilistic inference techniques to evaluate the likelihood of attack
goals(s) and predict potential upcoming attacks''~\cite{QiLe}. They proposed the
following procedure that converts an attack tree into a Bayesian network. Every
node in the attack tree is also present in the Bayesian network. An ``OR''
relationship from an attack tree is modeled in the Bayesian network with  edges
pointing from refining nodes that represent causes into the corresponding 
refined nodes that represent consequences. Deviating from regular attack trees,
an ``AND'' relationship is assumed to have an explicit (or implicit) order in
which the actions have to be executed. This allows to model the ``AND'' 
relationship by a directed path, which starts from the first (according to the 
order) child and ends with the parent node. Dantu~et~al. follow a different 
strategy when using Bayesian networks to model security risk management 
starting from behavior-based attack graphs\footnote{The authors do not appear to
make a distinction between attack trees and attack graphs. Since their
methodology is only applicable to cycle-free structures and they do not mention
how to deal with cycles, we assume that the methodology is actually based on
attack DAGs or attack trees.}~\cite{DaKaKo,DaKo,DaKoAkLo,DaKoCa}. When
processing multi-parameter attack trees with estimated parameter values
(Section~\ref{sec:multi-parameter_attack_trees}) J\"{u}rgenson and Willemson use
Qin and Lee's conversion of an attack tree to a Bayesian network~\cite{JuWi}.
An~et~al. propose to add a temporal dimension and to use \emph{dynamic Bayesian
networks} for intrusion detection without specifying how the graph is set
up~\cite{AnJuCe}.  Althebyan and Panda use knowledge graphs and dependency
graphs as basis for the construction of a Bayesian network~\cite{AlPa}. They
analyze a specific type of insider attack and state that their computational
procedures were inspired by Dantu~et~al. Another approach involving Bayesian
networks is described by Xie~et~al. who analyze intrusion detection
systems~\cite{XiLiOuLiLe}. They state that the key to using Bayesian networks is
to ``correctly identify and represent relevant uncertainties'' which governs
their setup of the Bayesian network. 

Bayesian networks are used to analyze security under uncertainty. The DAG
structure is of great value because it allows to use efficient algorithms. On
the one hand there exist efficient inference algorithms that compute a single
query (variable elimination, bucket elimination and importance, which are
actually equivalent according to Pouly and Kohlas~\cite{PoKo}) and on the other
hand there are inference algorithms that compute multiple queries at once
(bucket tree algorithm and Lauritzen-Spiegelhalter algorithm). In fact, the
efficiency of these algorithms can be seen as main reason to the success of
Bayesian networks, since querying general graphs is an NP-hard
problem~\cite{Arnb,Bodl}. Another strength of Bayesian networks is their ability
to update the model, i.e., compute a posteriori distribution, when new 
information is available.

We have not found any dedicated tools for analysis of Baysian networks for 
security. However, numerous tools exist that allow a visual treatment of 
standard Bayesian networks. One such tool is the Graphical Network Interface
(GeNIE) that uses the Structural Modeling, Inference, and Learning Engine
(SMILE)~\cite{Website_Genie}. It was, for example, used in~\cite{NaJoLaFrEk} to
analyze the interoperability of a very small cluster of services and mentioned
as hypothetical use in~\cite{FrSoEkJo}. Another one, called
MulVal~\cite{OuGoAp}, was actually developed for attack graphs 
(Section~\ref{sec:attack_graphs}), but used in~\cite{XiLiOuLiLe} to implement a
Bayesian network model. A third tool, tailored to statistical learning with 
Bayesian networks is bnlearn~\cite{Scut}.

There also exist isolated papers that promote the use of Bayesian networks in
security without any relation to attack trees or attack graphs. Houmb~et~al.,
quantify security risk level from Common Vulnerability Scoring System (CVSS)
estimates of frequency and impact using Bayesian networks~\cite{HoFrEn}. Feng
and Xie also use Bayesian networks and provide an algorithm of how to merge two
sources of information, expert knowledge and information stored in databases,
into one graph~\cite{FeXi}. Note that in this section we have gathered
approaches that rely on Bayesian networks and where their construction starts
from graphs that do not contain any cycles. Graphical models that make use of
Bayesian networks and initially contain cycles are treated in
Section~\ref{sec:bayesian_attack_graphs}, ones that include defenses are treated
in Section~\ref{sec:bayesian_defense_graphs}.

\subsubsection{Bayesian Attack Graphs} 
\label{sec:bayesian_attack_graphs}

\emph{Bayesian Attack Graphs} combine (general) attack graphs
(Section~\ref{sec:attack_graphs}), with computational procedures of Bayesian
networks (Section~\ref{sec:bayesian_networks}). However, since Bayesian
inference procedures only work on cycle-free structures, the formalism includes
instructions on how to remove any occurring cycles. Hence any final Bayesian
attack graph is acyclic. After the elimination of cycles, Bayesian attack graphs
model causal relationships between vulnerabilities in the same way as Bayesian
networks (Section~\ref{sec:bayesian_networks}) Bayesian attack graphs were first
proposed by Liu and Man in order to analyze network vulnerability scenarios with
the help of Bayesian inference methods in $2005$~\cite{LiMa}. Therefore the
formalism advances computational methods in security where uncertainty is
considered.

The formalism of Man and Liu is not the only fusion of attack graphs and
Bayesian networks. Starting in~$2008$ a group of researchers including Frigault,
Noel, Jajodia and Wang published a paper on a modified version of Bayesian
attack graphs. Their goal was to be able to calculate general security metrics
regarding information system networks which also contain probabilistic
dependencies~\cite{NoJaWaSi,FrWa}. Later they extended the formalism, using a
second copy of the model as time slice, to also capture dynamic behavior in so
called \emph{dynamic Bayesian networks}~\cite{FrWaSiJa}. In~$2012$,
Poolsappasit~et~al. revisited the framework to be able to deal with asset
identification, system vulnerability and connectivity analysis as well as
mitigation strategies~\cite{PoDeRa}.

All three approaches eliminate cycles that possibly exist in the underlying
attack graph. A shortcoming of Liu and Man is that they do not provide a
specific procedure on how to achieve this. The group including Frigault refers
to a paper on attack graphs~\cite{WaIsLoSiJa} which removes cycles through an
intricate procedure.  Poolsappasit~et~al. state that they rather analyze ``why
an attack can happen'' and not ``how an attack can happen'' and therefore,
``cycles can be disregarded using the monotonicity constraint'' mentioned
in~\cite{AmWiKa}.

Since Bayesian attack graphs are cycle-free, evaluation on them can make use
of Bayesian inference techniques. For this it is necessary to provide
probabilistic information. The three approaches differ in how they compute
quantitative values. Liu and Man provide edge probabilities~\cite{LiMa},
Frigault~et~al. give conditional probability tables for nodes which are
estimated according to their CVSS score~\cite{FrWa} and Poolsappasit~et~al.
use (local) conditional probability distributions for nodes~\cite{PoDeRa}.
Furthermore, Poolsappasit~et~al. augment Bayesian attack graphs with
additional nodes and values representing hardening measures (defenses). On the
augmented structure they propose a genetic algorithm that solves a
multiobjective optimization problem of how to assess the risk in a network
system and select optimal defenses~\cite{PoDeRa}.

The research group including Wang uses a Topological Vulnerability Analysis
(TVA) tool~\cite{JaNoOb,NoElJaKaOhPr} to create the attack graphs that serve
as basis for constructing Bayesian attack graphs. Poolsappasit~et~al. have
developed an unreferenced in-house tool that allows them to compute with
conditional probability distributions.

Wang~et~al.~\cite{FrWa,FrWaSiJa} state that their work is also based on that
of An~et~al.~\cite{AnJuCe}, who use Bayesian networks without cycles for
modeling risks of violating privacy in a database.

\subsubsection{Compromise Graphs} 
\label{sec:compormise_graphs}

McQueen~et~al. introduced \emph{compromise graphs} in~$2006$~\cite{QuBoFlBe}.
Compromise graphs are based on directed graphs\footnote{The authors do not state
whether these directed graphs are acyclic or not, but the description of
compromise graphs and their examples led us to consider compromise graphs as
DAGs.}, and are used to assess the efficiency of various technical
security measures for a given network architecture. The nodes of a compromise
graph represent the phases of an attack, detailing how a given target can get
compromised. The edges are weighted according to the estimated
time required to complete the corresponding phase for this compromise.
The overall time needed for the attacker to succeed is computed and compared
along different defensive settings, providing a metric to assess and compare
the efficiency of these different defensive settings.

The formalism has a sound mathematical formalization: a time to compromise (TTC)
metric is modeled for each edge as a random process combining three
sub-processes. Each of these processes has a different probability distribution
(mixing exponential, gamma and beta-like distributions). The value for the
process model parameters are based on the known vulnerabilities of the
considered component and the estimated skill of the attacker. A complete
description and justification of such a stochastic modeling is provided by the
same authors in a previous paper~\cite{McQueen2005}. In compromise graphs, five
types of stages, corresponding to the vertices of the graph, are modeled:
recognition, breaching the perimeter, penetration, escalation of privilege,
damage.

Compromise graphs are used to evaluate the efficiency of security measures, 
such as system hardening, firewalls or enhanced authentication. This is achieved
by comparing the shortest paths (in terms of TTC) of compromise graphs with and
without such measures in place. 

The approach is illustrated in~\cite{QuBoFlBe} by modeling attacks on a
SCADA system.

Byres and Leversage adopt a very similar approach in~\cite{LeBy,LeBy2}, called
state-time estimation algorithm (STEA), directly inspired by McQueen~et~al.
They combine a slightly modified TTC calculation approach with a decomposition
of the attack according to the architectural areas of the targeted system.

\subsubsection{Enhanced Attack Trees} 
\label{sec:enhanced_attack_trees}

\emph{Enhanced attack trees} have been introduced by \c{C}amtepe and Yener to to
support an intrusion detection engine by modeling complex attacks with time
dependencies.  This model was first described in a technical
report~\cite{Camtepe2006} in~$2006$. One year later, an official conference
proceedings~\cite{CaYe} appeared.

In addition to classical OR and AND gates, enhanced attack trees rely on the use
of a new gate, the ``ordered-AND'', which allows to capture sequential behavior
and constraints on the order of attack steps.  The model of enhanced attack
trees has sound mathematical foundations. Additionally to the formalism
description,~\cite{CaYe} devises  a new technique for detection of attacks.  The
new technique is based  on automata theory and it allows to verify completeness
of  enhanced attack tree models with respect to the observed attacks.

The quantification capabilities described in~\cite{CaYe} are  directly related 
to intrusion detection (probability of a given attack occurring based on a set
of observed events). A confidence attribute measured in percent is defined for
subgoals as ``the chance of reaching the final goal of the attacker when a
subgoal is accomplished''. It is computed as the ratio of all accomplished
events until a subgoal is realized, over all events of the modeled scenario.
This attribute aims at supporting an early warning system, supporting
decision-making and reaction before actual damages occur. Moreover,~\cite{CaYe}
introduces an original parameter  called ``time to live'' which allows to
express that some steps are  only available in a given time window.

In~\cite{MiKaYa}, Mishra~et~al. also make use of ordered-AND operators,
referring to~\cite{CaYe}. The authors visually describe Stuxnet and
similar attacks, but do not use \c{C}amtepe and Yener's rigorous formalization
to analyze the models.

\subsubsection{Vulnerability Cause Graphs} 
\label{sec:vulnerability_cause_graphs}

\emph{Vulnerability cause graphs} were invented in~$2006$ by Ardi, Byers and
Shahmehri as a key element of a methodology that supports security activities
throughout the entire software development lifecycle~\cite{ArBySh}.

The formalism can be seen as a root cause analysis for security-related software
failures, because it relates vulnerabilities with their causes. In a
vulnerability cause graph, every node except for one, has an outgoing directed
edge. The single node without a successor is called the exit node and represents
the considered vulnerability. All other nodes represent causes. The
predecessor-successor (parent-child) relationship shows how certain conditions
(nodes) might cause other conditions (nodes) to be a concern. In an improved
version of vulnerability cause graphs~\cite{ByArShDu}, nodes can be simple,
compound or conjunctions. Simple nodes represent conditions that may lead to a 
vulnerability. Compound nodes facilitate reuse, maintenance and readability of 
the models. Conjunctions represent groups of two or more other nodes. Contrary,
disjunctions occur if a node has two or more predecessors. In this case, the
original nodes might have to be considered if either of its predecessors might
have to be considered. Finally, if the causes have to follow a certain order,
they are modeled as sequences of nodes. To construct a vulnerability cause
graph, the exit node is considered as a starting point and refined with causes.

In vulnerability cause graphs, nodes can be annotated as ``blocked'' if the
underlying causes are mitigated. The ``blocked'' flag allows the user to compute
whether the underlying vulnerability (exit node) is also mitigated.
Vulnerability cause graphs are also equipped with a notion of graph
transformations that do not change whether the vulnerability is mitigated or
not. The transformations include conversions of conjunctions, reordering of
sequences, combination of nodes, conversion to compound nodes as well as derived
transformations.

In~\cite{ByArShDu} the vulnerability CVE-$2003$-$0161$, in~\cite{BySh3} the 
vulnerability CVE-$2005$-$2558$, and  in~\cite{MaCaMoArBySh} the vulnerability
CVE-$2005$-$3192$ is analyzed with the help of VCGs. Furthermore,~\cite{ChHa}
contains an additional three case studies on common software vulnerabilities
which have been performed using VCGs. The SHIELDS project~\cite{Website_SHIELDS}
has developed a software tool GOAT~\cite{GOAT} to be used in conjunction with
vulnerability cause graphs.

Vulnerability cause graphs were developed as part of a comprehensive
methodology to reduce software vulnerabilities that arise in ad hoc software
development. They are the starting point to build security activity graphs
(Section~\ref{sec:security_activity_graphs}). By introducing compound nodes,
the inventors of the formalism have created a model that allows different 
layers of abstraction, which in turn introduced a problematic design
decision of how many layers of abstraction are needed.

\subsubsection{Dynamic Fault Trees} 
\label{sec:dynamic_fault_trees}

In~$2009$, Khand~\cite{Khan} adapted several dynamic fault 
tree~\cite{Dugan1990,DuBaBo} gates to attack trees, in order to add a dynamic
dimension to classical attack trees. The aim of the formalism are similar
to those of attack trees (Section~\ref{sec:attack_trees}).

Dynamic fault trees~\cite{Dugan1990,DuBaBo} were invented by Dugan~et~al. in
the early~$1990$s to overcome limitations of static fault trees. They aim at
combining the dynamic capacities of Markovian models with the ``look and feel''
of fault trees. To achieve this, four dynamic gates are used: the
``priority-AND'' (PAND), the ``sequence gate'' (SEQ), the ``functional 
dependency gate'' (FDEP) and the ``cold spare gates'' (CSP). Khand reuses 
directly the three first gates (although renaming FDEP gates by CSUB, for 
Conditional Subordination, gates), leaving out the CSP gates. The PAND gate 
reaches a success state if all of its input are realized in a pre-assigned 
order (from left to right in the graphical notation). The SEQ gate allows to 
model a dependency between events, such that these events can only be realized 
in a particular order (from left to right in the graphical notation). Once all 
the input events are realized, the gate is verified. The CSUB gate models the 
need of the realization of a trigger event to allow a possible realization of 
others events. Dynamic fault tree combines dynamic gates with classical logical 
gates (AND, OR). Dynamic gates are formally defined with truth tables 
in~\cite{Khan}, and by Markov processes in the general definitions of dynamic 
fault trees from the safety literature~\cite{Dugan1990,DuBaBo} (although the 
description is still incomplete~\cite{Boui}).

There is no quantification aspects developed in~\cite{Khan}.

The paper by Khand does not specify which tool to use in order to treat the
models, but several tools exist for dynamic fault trees in the reliability
area, e.g., Galileo~\cite{DuSuCo}.

In safety studies, quantifications associated with dynamic fault trees are
usually made using Markovian analysis techniques; those might be used here also
although nothing is said about computation aspects.

\subsubsection{Serial Model for Multi-Parameter Attack Trees} 
\label{sec:multi-parameter_attack_trees-serial}

In~$2010$, the parallel model for multi-parameter attack trees
(Section~\ref{sec:multi-parameter_attack_trees}) has been extended by adding a
temporal order on the set of elementary attacks~\cite{WiJu}. This new
methodology is called \emph{serial model for multi-parameter attack trees} and
was studied further in~\cite{Jurg,Niit} and~\cite{BuSt}.

The model described in~\cite{Jurg} and~\cite{Niit} assumes that an adversary
performs the attacks in a given prescribed order. In~\cite{BuSt}, the authors
introduce so called fully-adaptive adversary model, where an attacker is allowed
to try atomic attacks in an arbitrary order which is not fixed in advance and
can be modified based on the results of the previous trials. In both cases, the
serial approach allows for a more accurate modeling of an attacker's behavior
than the parallel approach. In particular, the attacker can skip superfluous
elementary attacks and base his decisions on success or failure of the
previously executed elementary attacks.

In~\cite{WiJu}, an efficient algorithm for computing an attacker's expected
outcome assuming a given order of elementary attacks is provided. Taking
temporal dependencies into account allows the attacker to achieve better
expected outcome than when the parallel model
(Section~\ref{sec:multi-parameter_attack_trees}) is used. As remarked
in~\cite{JuWi2}, finding the best permutation of the elementary attacks in the
serial model for multi-parameter attack trees may turn computing the optimal
expected outcome into a super-exponential problem. In~\cite{Niit}, Niitsoo
proposed a decision-theoretical framework which makes possible to compute the
maximal expected outcome of a goal oriented attacker in linear time.
In~\cite{BuSt}, Buldas and Stepanenko propose a game theoretical framework to
compute upper bounds of the utility of fully-adaptive adversaries.

A prototype computer tool supporting the security analysis using the serial
model of multi-parameter attack trees has been implemented~\cite{Program4} and
described in~\cite{AndrusenkoMaster}.

A thorough comparison of the parallel and the serial model for multi-parameter
attack trees has been given in the Ph.D. thesis of J\"{u}rgenson~\cite{Jurg}.
Baca and Petersen mention that in order to use parametrized attack trees, the 
user needs to have a good understanding of the motivations of the 
attacker~\cite{BaPe}. To overcome this difficulty cumulative voting is used in 
countermeasure graphs (Section~\ref{sec:countermeausre_graphs}).

\subsubsection{Improved Attack Trees} 
\label{sec:improved_attack_trees}

\emph{Improved attack trees} are aim at dealing with security risks that arise
in space-based information systems. They were proposed by Wen-ping and Wei-min
in~$2011$~\cite{WeWe} to more precisely describe attack on the information
transmitting links, acquisitions systems and ground-based supporting and
application systems.

The formalism is based on attack trees and explicitly incorporates the use of
the sequential AND operator. It is not defined in a formal way. Improved
attack trees rely heavily on the description by Schneier and only
detail how to specifically compute the system risk.

Improved attack trees provide a specific formula to evaluate a risk value for
each leaf node. Starting from these risk values, the risk rate and the risk
possibility are computed and multiplied to compute the overall system risk.
The formulas distinguish between OR, AND and sequential AND nodes.

\subsection{Static Modeling of Attacks and Defenses}
\label{sec:both_static}

\subsubsection{Anti-Models} 
\label{sec:obstacle_trees}

\emph{Anti-models}~\cite{LaBrLaJa} have been introduced by Lamsweerde~et~al.
in~$2003$. They are closely related to AND-OR goal-refinement 
structures~\cite{LaLe} (sometimes called goal models) used for goal analysis in
requirements engineering. Anti-models extend such AND-OR goal-refinement
structures with the possibility to model malicious and intentional obstacles to
security goals, called anti-goals. They can be used to generate subtle attacks,
discard non-realizable or unlikely ones, and derive more effective customized
resolutions.

In~\cite{LaBrLaJa} and later in an extended version~\cite{Lams}, van
Lamsweerde~et~al. provide a six steps procedure for a systematic construction 
of anti-models. First, anti-goals, representing an attacker's goals, are 
obtained by negating confidentiality, privacy, integrity, availability,
authentication or non-repudiation requirements. For each anti-goal, the
questions ``who'' and ``why'' are asked to identify potential classes of 
attackers and their higher-level anti-goals. An AND-OR refinement process is
then applied to reach terminal anti-goals that are realizable by the attackers. 
The resulting AND-OR anti-models relate ``attackers, their anti-goals, 
referenced objects and anti-operations (necessary to achieve their anti-goals) 
to the attackees, their goals, objects, operations and vulnerabilities.'' The
construction of anti-models is only informally presented in~\cite{LaBrLaJa}.
Formal techniques developed for AND-OR goal-refinement structures (such as
refinement obstacle trees)~\cite{LaLe} can be used for the generation and
analysis of anti-models. In particular, real-time temporal logic can be 
employed to model anti-goals as sets of attack scenarios. After identifying
possible anti-goals, countermeasures expressed as epistemic extensions of
real-time temporal logic operators are selected based on severity or likelihood
of the corresponding threat and non-functional system goals that have been
identified earlier. Possible resolutions tactics, inspired by solutions 
proposed for analysis of non-functional requirements in software engineering, 
are described in~\cite{LaLe} and~\cite{Lams}. Applying resolution operators 
yields new security goals to be integrated in the model. These new goals are
then again refined with the help of AND-OR structures. These, in turn, may
require a new round of anti-model construction and analysis. 

The anti-models do not include quantitative analysis of security goals or
anti-goals. 

\subsubsection{Defense Trees} 
\label{sec:defense_trees}

\emph{Defense trees}\footnote{Papers by Bistarelli~et~al. use British English,
thus originally, the name of their formalism is \emph{defence trees}.} are
attack trees where leaf nodes are decorated with a set of countermeasures.
They have been introduced by Bistarelli~et~al. in~$2006$~\cite{BiFiPe}. The
approach combines qualitative and quantitative aspects and serves general
security modeling purposes.

The approach proposed by Bistarelli~et~al. was a first step towards integrating
a defender's behavior into models based on attack trees. The analysis
methodology for defense trees proposed in~\cite{BiFiPe} and~\cite{BiDaPe} uses
rigorous and formal techniques, such as calculation of economic indexes and game
theoretical solution concepts. However, the model itself is only introduced
verbally and a formal definition is not given.

In~\cite{BiFiPe}, the return on attack (ROA) and return on investment (ROI)
indexes are used for quantitative analysis of defense trees from the point of
view of an attacker and a defender, respectively. The calculation of ROI and ROA
is based on the following parameters: costs, impact, number of occurrences of a
threat and gain. The indexes provide a useful method to evaluate IT security
investments and to support the risk management process. In~\cite{BiDaPe}, game
theoretical reasoning was introduced to analyze attack--defense scenarios
modeled with the help of defense trees. In this paper, a defense tree represents
a game between two players: an attacker and a defender. The ROI and ROA indexes,
are used as utility functions and allow to evaluate the effectiveness and the
profitability of countermeasures. The authors of~\cite{BiDaPe} propose using
Nash equilibria to select the best strategy for the players.

In~\cite{BiPeTr}, defense trees have been extended to so called \emph{CP-defense
trees}, where modeling of preferences between countermeasures and actions is
possible. Transforming CP-defense trees into answer set optimization (ASO)
programs, allows to select the most suitable set of countermeasures, by
computing the optimal answer set of the corresponding ASO program. Formalisms
such as attack--defense trees  (Section~\ref{sec:attack_defense_trees}), and
attack countermeasure trees (Section~\ref{sec:attack_countermeasure_trees})
extended defense trees by allowing defensive actions to be placed at any node of
the tree and not only at the leaf nodes.

\subsubsection{Protection Trees} 
\label{sec:protection_trees}

\emph{Protection trees} are a tree-based formalism which allow a user to
allocate limited resources towards the appropriate defenses against specified 
attacks. The methodology was invented by Edge~et~al. in~$2006$, in order to
incorporate defenses in the attack tree methodology~\cite{EdDaRaMi}.

Protection tree are similar to attack trees since both decompose high level
goals into smaller manageable pieces by means of an AND-OR tree structure. The
difference is that in protection tree the nodes represent protections. A
protection tree is generated from an already established attack tree by finding
a protection against every leaf node of the attack tree. Then the attack tree is
traversed in a bottom-up way and new protection nodes are added to the
protection tree if the protection nodes do not already cover the parent attack
node.

The AND-OR structure of protection trees is enriched with three metrics, namely
probability of success, financial costs and performance costs on which the
standard bottom-up approach is applied~\cite{EdDaRaMi,EdRaGrBaBeRe,Edge}.
In~\cite{DaEdMiRa}, an additional  metrics, the impact, helps to further
prioritize where budget should be spent.

The formalism has been investigated in case studies on how the U.S. Department
of Homeland Security can allocate resources to protect their computer
networks~\cite{EdDaRaMi}, how an attack on an online banking system can be
mitigated cost-efficiently~\cite{EdRaGrBaBeRe}, how to cheaply protect against
an attack on computer and RFID networks~\cite{DaEdMiRa} as well as a mobile ad
hoc network~\cite{Edge}. When evaluating which defenses to install, the authors
propose to first prune the tree according to the attacker's assumed
capabilities. A larger, more applied case study to ``evaluate the effectiveness
of attack and protection trees in documenting the threats and vulnerabilities
present in a generic Unmanned Aerial Systems (UAS) architecture'' was performed
by Cowan~et~al.~\cite{CoGrPa}.

In~\cite{EdRaGrBaBeRe} a slightly different algorithm for creation of a
protection tree was proposed. Here a designer starts by finding defenses against
the root of an attack tree instead of the leaves, as in~\cite{EdDaRaMi,Edge}. An
approach similar to protection trees has been proposed in~\cite{RuHeCeMi} to
deal with the problem of threat modeling in software development. The paper uses
so called identification trees to identify threats in software design and
introduces the model of mitigation trees to describe countermeasures for
identified threats. Despite an obvious modeling analogy between protection 
trees and mitigation trees, no connection between the two models has been made 
explicit in the literature.

\subsubsection{Security Activity Graphs} 
\label{sec:security_activity_graphs}

In~$2006$, Ardi, Byers and Shahmehri introduced a formalism called
\emph{security activity graphs} (SAGs). The methodology was invented in order to
``improve security throughout the software development process''~\cite{ArBySh}.
SAGs depict possible vulnerability cause mitigations and are algorithmically
generated from vulnerability cause graphs
(Section~\ref{sec:vulnerability_cause_graphs}).

SAGs are a graphical representation of first order predicate calculus and are
based very loosely on ideas from fault tree analysis. In~\cite{ArBySh} the root
of a SAG is associated with a vulnerability, taken from a vulnerability cause
graph. The vulnerability mitigations are modeled with the help of activities
(leaf nodes). The syntax furthermore consists of AND-gates, OR-gates and split
gates. The AND and OR-gates strictly follow Boolean logic, whereas the split
gate allows one activity to be used in several parent activities, essentially
creating a DAGs structure. The syntax of SAGs was changed in~\cite{BySh2} for a
more concise illustration of the models. Split gates no longer appear in the
formalism. The functionality that simple activities can be distinguished from
compound activities (complex activities that may require further breakdown) was
added. Moreover cause references (possible attack points) serve as placeholders
for a different SAG associated with a particular cause.

In the SAG model, Boolean variables attached to the leaves of the SAG. A 
Boolean variable corresponding to an activity is true when it ``is implemented 
perfectly during software development'' otherwise, it is false. Then a value 
corresponding to the root of the SAG is deduced in a bottom-up fashion 
according to Boolean logic.

Visual representation of SAGs is supported by SeaMonster~\cite{MeSpHaBaKrVe} and
GOAT~\cite{GOAT}. Furthermore, SAGs have been used in~\cite{BySh2,BySh3} to 
model the vulnerability CVE-$2005$-$2558$ in MySQL that leads to ``denial of 
service or arbitrary code execution''.

Even though the model was devised in order to aid the software development
cycle, the authors explicitly state that SAGs ``lend themselves to other
applications such as process analysis.'' SAGs are the middle step of a
broader~$3$-steps approach for secure software development, with vulnerability 
cause graphs as a first step, and process component definition as a final step. 
In~$2010$ SAGs were replaced by security goal models 
(Section~\ref{sec:security_goal_models})

\subsubsection{Attack Countermeasure Trees} 
\hypertarget{ACT}{\label{sec:attack_countermeasure_trees}}

In~$2010$, Roy, Kim and Trivedi proposed \emph{attack countermeasure trees}
(ACT)~\cite{RoKiTr3,RoKiTr} as a methodology for attack and defense  modeling
which unifies analysis methods proposed for attack trees
(Section~\ref{sec:attack_trees}) with those introduced on defense trees
(Section~\ref{sec:defense_trees}). The main difference of ACTs with respect to
defense trees is that in ACTs defensive measures can be placed at any node of
the tree. Also, the quantitative analysis proposed for defense trees is extended
by incorporating probabilistic analysis into the model. ACTs were first
introduced in~\cite{RoKiTr} and then further developed in~\cite{RoKiTr2}.

ACTs may involve three distinct classes of events: attack events, detection
events and mitigation events. The set of classical AND and OR nodes, as defined
for attack trees, is extended with the possibility of using $k$-out-of-$n$
nodes. Generation and analysis of attack countermeasure scenarios is automated
using minimal cut sets (mincuts). Mincuts help to determine possible ways of
attacking and defending a system and to identify the system's most critical
components.

A rigorous mathematical framework is provided for quantitative analysis of ACTs 
in~\cite{RoKiTr} and~\cite{RoKiTr2}. The evaluation of the ROI and ROA
attributes, as proposed for defense trees (Section~\ref{sec:defense_trees}), has
been extended by adding the probability of attack, detection and mitigation
events. The authors of~\cite{RoKiTr2} provide algorithms for probability
computation on trees with an without repeated nodes. With the help of
probability parameters, further metrics, including cost, impact, Birnbaum's
importance measure and risk, are evaluated. The use of the Birnbaum's importance
measure (also called reliability importance measure, in the case of fault trees)
is used to prioritize defense mechanisms countering attack events. Furthermore,
in~\cite{RoKiTr2}, Roy~et~al. propose a cubic algorithm to select an optimal set
of countermeasures for an ACT. This addresses the problem of state-space
explosion that the intrusion response and recovery engine based on
attack-response trees (Section~\ref{sec:attack_response_trees}) suffers from.
Finally, in~\cite{RoKiTr4} the problem of selecting an optimal set of
countermeasures with and without having probability assignments has been
discussed.

The authors of~\cite{RoKiTr2} implemented a module for automatic description and
evaluation of ACTs in a modeling tool called Symbolic Hierarchical Automated
Reliability and Performance Evaluator~\cite{sharpe}. This implementation uses
already existing algorithms for analysis of fault trees and extends them with
algorithms to compute costs, impact and risk. Case studies concerning attacks on
the Border Gateway Protocol (BGP), SCADA systems and malicious insider attacks
have been performed using ACTs, as described in the Master thesis of
Roy~\cite{Roy}.

The model of attack countermeasure trees is very similar to 
attack--defense trees. The main differences between the two models 
are listed in Section~\ref{sec:attack_defense_trees}.

\subsubsection{Attack--Defense Trees} 
\label{sec:attack_defense_trees}

\emph{Attack--defense trees} (ADTrees) were proposed by Kordy~et~al. in
$2010$~\cite{KoMaRaSc}. They allow to illustrate security scenarios that involve
two opposing players: an attacker and a defender. Consequently it is  possible 
to model interleaving attacker and defender actions qualitatively and 
quantitatively. ADTrees can be seen as merging attack trees
(Section~\ref{sec:attack_trees}) and protection trees
(Section~\ref{sec:protection_trees}) into one formalism.

In ADTrees, both types of nodes, attacks and defenses, can be conjunctively as
well as disjunctively refined. Furthermore, the formalism allows for each node
to have one child of the opposite type. Children of opposite type represent
countermeasures. These countermeasures can be refined and countered again.  Two
sets of formal definitions build the basis of ADTrees: a graph-based definition
and an equivalent term-based definition. The graph-based definition ensures a
visual and intuitive handling of ADTrees models. The term-based representation
allows for formal reasoning about the models.  The formalism is enriched through
several semantics that allow to define  equivalent ADTree representations of a
scenario~\cite{KoMaRaSc2}. The necessity for multiple semantics is motivated by
diverse applications of ADTrees, in particular unification of other attack tree
related approaches and suitability for various kinds of computations.
In~\cite{KoPoSc}, the authors showed that, for a wide class of semantics (i.e.,
every semantics induced by a De Morgan lattice), ADTrees extend the modeling
capabilities of attack trees without increasing the computational complexity of
the model.  In~\cite{KoMaRaSc2} the most often used semantics for ADTrees have
been  characterized by finite axiom schemes, which provides an operational
method for  defining equivalent ADTree representations.  The authors
of~\cite{KoMaMeSc}, have established a connection between game theory and
graphical security assessment using ADTrees. More precisely, ADTrees under a
semantics derived from  propositional logics are shown to be  equally expressive
as two-player binary zero-sum extensive form games.

The standard bottom-up algorithm, formalized for attack trees in~\cite{MaOo}, 
has been extended to ADTrees in~\cite{KoMaRaSc2}. This required the introduction
of four new operators (two for conjunction and disjunction of defense nodes and
two for countermeasure links)~\cite{KoMaRaSc2}. Together with the two standard
operators (for conjunctions and disjunctions of attack nodes) and a set of
values, the six operators form an attribute domain. Specifying attribute domains
allows the user to quantify a variety of security relevant parameters, such as
time of attack, probability of defense, scenario satisfiability and
environmental costs. The authors of~\cite{KoMaRaSc2} show that every attribute
for which the attribute domain is based on a semi-ring can be evaluated on
ADTrees using the bottom-up algorithm. How to properly specify attribute domains
in terms of questions in natural language was presented in~\cite{KoMaSc}.

An extensive case study on a real-life RFID goods management system was
performed by academic and industrial researchers with different
backgrounds~\cite{BaKoMeSc}. The case study resulted in specific guidelines
about the use of attributes on ADTrees. A software tool, called the 
ADTool~\cite{ADTool}, supporting the attack-defense tree methodology, has been
developed as one of the outcomes of the ATREES project~\cite{ATREES}. The main
features of the tool are easy  creation, efficient editing, and quantitative
analysis of  ADTrees~\cite{KoKoMaSc}.  Since from a formal perspective, attack
trees  (Section~\ref{sec:attack_trees}), protection trees 
(Section~\ref{sec:protection_trees}), and defense trees 
(Section~\ref{sec:defense_trees}) are instances of attack--defense trees, the 
ADTool also supports all these formalisms.

Finally, ADTrees can be seen as natural extension of defense trees 
(Section~\ref{sec:defense_trees}) where defenses are only allowed as leaf 
nodes.  The ADTree formalism is quite similar to attack countermeasure trees 
(Section~\ref{sec:attack_countermeasure_trees}), however, there exist a couple
of fundamental differences between the two models. On the one hand, in ADTrees
defense nodes can be refined and countered, which is not possible in attack
countermeasure trees. On the other hand, attack countermeasure trees 
distinguish between detection and mitigation events which are both modeled with
defense nodes in ADTrees. Another difference is that attack countermeasure trees
are well suited to compute specific parameters, including probability,  return
on investment (ROI) and return on attack (ROA). ADTrees, in  turn, focus on
general methods for attribute  computation. A different formalism, also called
attack--defense trees, was used by Du~et~al. in~\cite{DuLiDuZh} to perform a
game-theoretic analysis of Vehicular ad-hoc network security by utilizing the
ROA and ROI utility functions. Despite  sharing the same name with the formalism
introduced in~\cite{KoMaRaSc}, the attack--defense tree approach used
in~\cite{DuLiDuZh} is built up on defense trees
(Section~\ref{sec:defense_trees}) and does not contain the possibility  to
refine countermeasures. Moreover it does not consider any formal semantics.

\subsubsection{Countermeasure Graphs} 
\label{sec:countermeausre_graphs}

\emph{Countermeasure graphs} provide a DAG-based structure for identification
and prioritization of countermeasures. They were introduced by Baca and
Petersen~\cite{BaPe} in~$2010$ as an integral part of the ``countermeasure
method for security'' which aims at simplifying countermeasure selection through
cumulative voting.

To build the graphical model, actors, goals, attacks and countermeasures are
identified. Goals explain why someone attacks a system, actors are the ones 
that attack the system, goal explain why actors attack a system, attacks detail 
how the system could get attacked and countermeasures describe how attacks 
could be prevented. When the representing events are related, edges are drawn 
between goals and actors, actors and attack as well as between attacks and 
countermeasures. More specifically, an edge is drawn between a goal and an 
actor if the actor pursues the goal. An edge is inserted between an actor
and an attack, if the actor is likely to be able to execute the attack. Finally 
an edge is drawn between an attack and a countermeasure if the countermeasure 
is able to prevent the attack. Priorities are assigned to goals, actors, 
attacks and countermeasures according to the rules of hierarchical cumulative 
voting~\cite{BeSv}. The higher the assigned priority is, the higher the threat 
level of the corresponding event is.

With the help of hierarchical cumulative voting~\cite{BeSv} the most effective
countermeasures can be deduced. Clever normalization and the fact that
countermeasures that prevent several attacks contribute more to the final
result than isolated countermeasures guarantee that the countermeasure with
the highest computed value is most efficient and should therefore be
implemented.

The methodology is demonstrated on an open source system, a first person 
shooter called Code~$43$~\cite{BaPe}.

\subsection{Sequential Modeling of Attacks and Defenses}
\label{sec:both_sequential}

\subsubsection{Insecurity Flows} 
\label{sec:insecurity_flows}

In~$1997$, Moskowitz and Kang described a model called \emph{insecurity flows}
to support risk assessment~\cite{MoKa}. It combines graph theory and discrete
probability, offering both graphical representation and quantification
capabilities to analyze how an ``invader can penetrate through security holes
to various protective security domains''. This analysis aims at identifying the
most vulnerable paths and the most appropriate security measures to eliminate
them.

From a high level perspective, insecurity flows are similar to reliability
block diagrams~\cite{Elec} used in reliability engineering, without however
mentioning such a similarity~\cite{MoKa}. The source corresponds to the
starting point of the attacker, the sink corresponds to the objective of the
attacker, and the asset under protection. An insecurity flow diagram is a
circuit connecting security measures, as serial or in parallel,
from the sink to the source. Serial nodes must be passed one after
the other by the attacker, whereas only one out of $n$ connected
in parallel must be passed to continue its path to the sink.
The graph is used to identify insecurity flows and quantify them
using probabilistic calculations. The paper provides a sound description of
the formalism and the associated quantifications.

Based on the circuit, the probability that the insecurity flow can pass
through the modeled security measures of a given system or architecture can be
computed. Probability computation formulas for simple serial and parallel
patterns are provided, whereas reduction formulas are proposed for more
elaborated circuits (decomposing them into the simple patterns).
Several defensive architectures can be compared along this metrics.

\subsubsection{Intrusion DAGs} 
\label{sec:i-dag}

\emph{Intrusion DAGs} (I-DAGs) have been introduced by 
Wu~et~al.~\cite{BagchiTech} as the underlying structure  for attack goals
representation in the Adaptive Intrusion Tolerant System, called ADEPTS
in~$2003$. The global goal of ADEPTS is to localize and automatically respond to
detected, possibly multiple and concurrent intrusions  on a distributed system. 

I-DAGs are directed acyclic  graphs representing intrusion goals in ADEPTS. 
I-DAGs are not necessarily rooted DAGs, i.e., they may have multiple roots.  The
 nodes of an I-DAG represent (sub-)goals  of an attack and can be  associated
with an alert from the intrusion detection framework described 
in~\cite{conf/acsac/WuFMB03}. A goal represented by a node can only be achieved 
if (some of) the goals of its children are achieved. To model the connection, 
I-DAGs use standard AND and OR refinement features similar to the refinements 
in attack trees. Each node stores two information sets: a cause service set 
(including all services that may be compromised in order to  achieve the goal)
and an effect service set (including all services that are  taken to be
compromised once the goal is achieved). The method presented
in~\cite{BagchiTech} allows to automatically trigger a response of appropriate
severity, based on a value which expresses the confidence that the goal 
corresponding to a node has been achieved. This provides dynamic aspects to the 
ADEPTS methodology. 

Three algorithms have been developed in order to support automated responses to
detected incidents. The goal of the first algorithm is to classify all nodes as
candidates for responses as follows. A bottom-up procedure assigns the 
compromised confidence index to each node situated on the paths between the 
node representing a detected incident and a root node. Then, a value called 
threshold is defined by the user and is  used by a top down procedure to label
the nodes as strong, weak, very  weak or  non-candidates for potential
responses. The second algorithm assigns the  response index to nodes. The
response index is a real number used to  determine the response to be taken for
a given node in the I-DAG. Finally,  the third algorithm is  based on so called
effectiveness index. It is responsible for dynamically  deciding which responses
are to be taken next. Intuitively, the effectiveness  index of a node is reduced
for every detected failure of a response action and  increased for every
successful deployment.

A lightweight distributed e-commerce system has been deployed to serve as a 
test bed for the  ADEPTS tool. The system contained~$6$ servers and has~$26$
nodes in the  corresponding I-DAG. The results of the experiments and analysis
are  described in~\cite{BagchiTech}. 

In~\cite{FoWuMaBaSp} and~\cite{FoWuMaBaSpTech}, the authors extend the model of 
intrusion DAGs to intrusion graphs (I-GRAPHs). The main difference is that, 
contrary to I-DAGs, I-GRAPHs may contain cycles. Nodes of an I-GRAPH do not 
need to be independent. All dependencies between the nodes are depicted by the 
edges between nodes. Additionally to AND and OR refinements, I-GRAPHs also make 
use of quorum edges. A value called minimum required quorum is assigned to 
quorum edges and represents the minimal number of children that need to be 
achieved in order to achieve the parent node.

\subsubsection{Bayesian Defense Graphs} 
\label{sec:bayesian_defense_graphs}

In a series of papers starting in~$2008$, Sommestad~et~al. construct a Bayesian
network for security (Section~\ref{sec:bayesian_networks}) that includes
defenses to perform enterprise architecture
analysis~\cite{FrSoEkJo,SoEkJo4,SoEkJo2,EkSo,SoEkNo}. Their model, explicitly
called \emph{Bayesian defense graphs}in~\cite{SoEkJo2}, is guided by the idea to
depict what exists in a system rather than what it is used for~\cite{SoEkJo2}.
This philosophy was adapted from~\cite{JoJoSoUl}. Bayesian defense graphs are
inspired by defense trees (Section~\ref{sec:defense_trees}) and therefore add
countermeasures to Bayesian networks. As a result, the formalism supports a
holistic system view including attack and defense components.

Bayesian defense graphs are built up on extended influence diagrams
(Section~\ref{sec:isolated_models}), including utility nodes, decisions nodes
chance nodes and arcs. Chance nodes and decision nodes are associated with
random variables that may assume one of several predefined and mutually
exclusive sates. The random variables are given as conditional probability
tables (or matrices). Utility nodes express the combination of states in chance
nodes and decision nodes. Countermeasures, which are controllable elements from
the perspective of the system owner, are represented as chance nodes with
adapted conditional probability tables. Finally, causal arcs (including an AND
or OR label) are drawn between the nodes indicating how the conditional
probabilities are related. A strength of Bayesian defense graphs is that they
allow to trade-off between collecting as much data as possible and the degree of
accuracy of the collected data. Through the use of iterative refinement, it is
possible to reduce the complexity of the model~\cite{SoEkJo2}. 

Like all formalisms that involve Bayesian statistics, Bayesian defense graphs 
use conditional probability tables to answer ``how do the security mechanisms
influence each other?'' and ``How do they contribute to enterprise-wide
security?''~\cite{SoEkJo4}. The authors of~\cite{SoEkJo4} exemplify how to 
compute the expected loss for both the current scenario and potential future 
scenarios. In~\cite{FrSoEkJo}, a suitable subset of a set of~$82$ security 
metrics known as Joint Quarterly Readiness Review (JQRR) metrics has been 
selected and adapted to Bayesian Defense graphs. The metrics serve as ``a 
posteriori indicators on the historical success rates of hostile attacks'' or 
``indicate the current state of countermeasures''. The formalism can handle 
causal and uncertainty measurements at the same time, by specifying how to 
combine the conditional probability tables.

With the help of a software tool for abstract models~\cite{JoJoSoUl}, Bayesian
defense graphs were applied by Sommestad~et~al. to analyze enterprise
architectures on numerous occasions.In~\cite{EkSo} ongoing efforts on Bayesian
defense graphs within the EU research project VIKING~\cite{Website_VIKING} are
summarized. The methodology is expanded in three follow-up papers that
illustrate security assessment based on an enterprise architecture
model~\cite{SoEkJo4,SoEkJo2} and information flow during a spoofing attack on a
server~\cite{FrSoEkJo}. In~\cite{SoEkNo}, a real case study was performed with a
power distribution operator to assess the security of wide-area networks (WANs)
used to operate electrical power systems. Since the results could not be
published the methodology was demonstrated on a fictitious example assessing the
security of two communication links with the help of conditional probability
tables~\cite{SoEkNo}.

A similar but less developed idea of using random variables, defenses and an
inference algorithm to compute the expected cost of an attack is presented by
Mirembe and Muyeba~\cite{MiMu}. 

\subsubsection{Security Goal Indicator Trees} 
\label{sec:security_goal_indicator_trees}

Peine, Jawurek and Mandel devised \emph{security goal indicator trees} in
$2008$, in order to support security inspections of software development and
documents~\cite{PeJaMa}.

A security goal indicator tree is a tree which combines negative and positive
security features that can be checked during an inspection, in order to see if a
security goal (e.g., secure password management) is met. With this objective in
mind, ``indicators'' can be linked in the resulting tree structure by three
types of relations: Conditional dependencies are represented by a special kind
of edge, Boolean combination are modeled by OR and AND gates, a
``specialization'' relation is represented by a UML-like inheritance symbol.
Moreover, a notion of ``polarity'' is defined for each node, attributing
positive or negative effect of a given property on security. The definition of
security goal indicator trees is semi-formal.

The formalism does not support quantitative evaluations.

Security goal indicator trees are implemented in a prototype tool, mentioned
in~\cite{PeJaMa}. They are used to formalize security inspection processes for
a distributed repository of digital cultural data in e-tourism application
in~\cite{JuElBaRa}. The formalism is extended to dependability inspection
in~\cite{KlElEs}.

\subsubsection{Attack-Response Trees} 
\label{sec:attack_response_trees}

In~$2009$, Zonouz, Khurana, Sanders, and Yardley introduced
\emph{attack-response trees} (ART) as a part of a methodology called response
and recovery engine (RRE), which was proposed to automate the intrusion response
process. The goal of the RRE is to provide an instantaneous response to
intrusions and thus eliminate the delay which occurs when the response process
is performed manually. The approach is modeled as a two-player Stackelberg
stochastic game between the leader (RRE) and the follower (attacker).
Attack-response trees have been used in~\cite{ZoKhSaYa}, for the first time.
This paper constitutes a part of the Ph.D. thesis of Zonouz~\cite{Zonouz}.

ARTs are an extension of attack trees (Section~\ref{sec:attack_trees}) that
incorporate possible response actions against attacks. They provide a formal way
to describe the system security based on possible intrusion and response
scenarios for the attacker and the response engine, respectively. An important
difference between attack trees and attack-response trees is that the former
represent all possible ways of achieving an attack goal and the latter are built
based on the attack consequences\footnote{A reader may notice that  what the
authors of~\cite{ZoKhSaYa} call ``sub-consequences'' are in fact the  causes of
the main consequence.}. In an attack-response tree, a violation of a security
property, e.g., integrity, confidentiality or availability, is assigned to the
root node (main consequence). Refining nodes represent sub-consequences whose
occurrence implies that the parent consequence will take place. Some consequence
nodes are then tagged by response nodes that represent response actions against
the consequence to which they are connected.

The goal of attack-response trees is to probabilistically verify whether the
security property specified by the root of an attack-response tree has been
violated, given the sequence of the received alerts and the successfully taken
response actions. First, a simple bottom-up procedure is applied in the case
when values~$0$ and~$1$ are assigned to the leaf nodes. More precisely, when a
response assigned to a node $v$ is activated (i.e., is assigned with~$1$), the
values in the subtree rooted in~$v$ are reset to~$0$. Second,~\cite{ZoKhSaYa}
also discusses the situation when uncertainties in intrusion detections and
alert notifications render the determination of Boolean values impossible. In
this case, satisfaction probabilities are assigned to the nodes of
attack-response trees and a game-theoretic algorithm is used to decide on the
optimal response action. In~\cite{ZoShRaKaPfAuIySaCo}, the RRE has been extended
to incorporate both IT system-level and business-level metrics to the model.
Here, the combined metrics are used to recommend optimal response actions to
security attacks.

The RRE has been implemented on top of the intrusion detection system (IDS)
Snort~$2.7$, as described in~\cite{Zonouz}. A validation of the approach on a
SCADA system use case~\cite{ZoKhSaYa} and a web-based retail  company
example~\cite{ZoShRaKaPfAuIySaCo} has shown that this dynamic method  performs
better than static response mechanisms based on lookup tables. The RRE allows to
recover the system with lower costs and is more helpful than static engines when
a large number of IDS alerts from different parts of the system are received.

As pointed out in~\cite{RoKiTr}, the approach described in this section suffers
from the state space explosion problem. To overcome this problem, attack
countermeasure trees (Section~\ref{sec:attack_countermeasure_trees}) have been
introduced. Their authors propose efficient algorithms for selecting an optimal
set of countermeasures.
\subsubsection{Boolean Logic Driven Markov Process} 
\label{sec:BDMP}

\emph{Boolean logic driven Markov processes} (BDMPs) are a general security
modeling formalism, which can also complete generic risk assessment procedures.
It was invented by Bouissou and Bon in~$2003$ in the safety and reliability
area~\cite{BoBo}, and was adapted to security modeling by
Pi\`{e}tre-Cambac\'{e}d\`{e}s and Bouissou in~$2010$~\cite{PiBo6,PiBo2}
\footnote{The original idea was introduced in a fast abstract by the same
authors in~$2009$~\cite{PiBo3}}. Its goal is to find a better trade-off between
readability, modeling power and quantification capabilities with respect to the
existing formalisms in general and attack trees in particular.

BDMPs combine the readability of classical attack trees with the modeling power
of Markov chains. They change the attack tree semantics by augmenting it with
links called triggers. In a first approach, triggers allow modeling of sequences
and simple dependencies by conditionally ``activating'' sub-trees of the global
structure. The root (top event) of an BDMP is the objective of the attacker. The
leaves correspond to attack steps or security events. They are associated to
Markov processes, dynamically selected in function of the states of some other
leaves. They can be connected by a wide choice of logical gates, including AND,
OR, and PAND gates, commonly used in dynamic fault trees 
(Section~\ref{sec:dynamic_fault_trees}). The overall approach allows for
sequential modeling in an attack tree-like structure, while enabling efficient
quantifications. BDMPs for security are well formalized~\cite{PiBo2}.

Success or realization parameters (mean time to success or to realization) are
associated to the leaves, depending on the basic event modeled. Defense-centric
attributes can also be added, reflecting detection and reaction capabilities
(the corresponding parameters are the probability or the mean-time to detection
for a given leaf, and the reduction of chance of success in case of detection). 
BDMPs for security allow for different types of quantification. These
quantifications include the computation of time-domain metrics (overall
mean-time to success, probability of success in a given time, ordered list of
attack sequences leading to the objectives), attack tree related metrics like
costs of attacks, handling of Boolean indicators (e.g., specific requirements),
and risk analysis oriented tools  like sensibility graphs by attack step or
event~\cite{PiDeBo}, etc.

The model construction and its analysis are supported by an industrial tool,
called KB3~\cite{KB3}. In~\cite{PiDeBo}, implementation issues and user feedback
are discussed and analyzed. BDMPs are used in~\cite{PiBo5,John} to integrate
safety and security analyses while~\cite{KriBoPi} develops a realistic use-case
based on the Stuxnet attack.

In several papers~\cite{PiBo6,PiBo2,PiDeBo}, the authors point out the intrinsic
limits of BDMPs to model cyclic behaviors and loops, as well as the difficulties
to assign relevant values for the leaves.

\subsubsection{Security Goal Models} 
\label{sec:security_goal_models}

In~$2010$, \emph{Security goal models} (SGMs) were formalized by Byers and
Shahmehri~\cite{BySh} in oder to identify the causes of software 
vulnerabilities and model their dependencies. They were introduced as a more
expressive replacement for attack trees (Section~\ref{sec:attack_trees}),
security goal indicator trees (Section~\ref{sec:security_goal_indicator_trees}),
vulnerability cause graphs (Section~\ref{sec:vulnerability_cause_graphs}), and
security activity graphs (Section~\ref{sec:security_activity_graphs}). The root
goal of a SGM corresponds to a vulnerability. ``Starting with the root, subgoals
are incrementally identified until a complete model has been
created''~\cite{ShMaOcByCaArJi}.

In SGMs, a goal can be anything that affects security or some other goal, e.g.,
it can be a vulnerability, a security functionality, a security-related software
development activity or an attack. SGMs have two types of goal refinements: one
type represents dependencies and one type modeling information flow. Dependency
nodes are connected with solid edges (dependence edge) and are depicted by white
nodes for contributing subgoals and by black nodes for countering subgoals.
Information edges are displayed with dashed edges. The formalism consists of a
syntactic domain (elements that make up the model), an abstract syntax (how
elements can be combined), a visual representation (used graphical symbols) and
a semantic transformation from the syntactic domain to the semantic domain. The
syntactic domain consists of the root, subgoals (contributing or counteracting),
dependency edges, operators AND and OR that express the connection of dependency
edges, annotation connected to nodes by annotation edges, stereotype (usually an
annotation about a dependency edge), ports that model information flow and
information edges that connect ports. The abstract syntax defined in a UML class
diagram~\cite{ShMaOcByCaArJi}.

It is possible to evaluate whether a security goal was successfully reached or
not. To do this, each cause is defined with a logical predicate (true/false).
Then the predicates are composed using Boolean logic and taking the
information from the information edges into account.

SGM were used in a case study about passive testing vulnerability detection,
i.e., examining the traces of a software system without the need for specific
test inputs. In a four step testing procedure vulnerabilities are first modeled
using SGMs. In the next step, causes are formally defined before SGMs are
converted into vulnerability detection conditions (VDC). In the final step
vulnerabilities are checked based on the VDCs. In~\cite{ShMaOcByCaArJi} this
procedure is performed on the xine media player~\cite{Website_xine} where an
older version contained the CVE-$2009$-$1274$ vulnerability. The case study is
executed with the help of ``TestInv-Code'', a program developed by Montimage
that can handle VCDs.

In~\cite{BySh}, the authors explicitly state that they have defined
transformations to and from attack trees VCGs, SAGs and SGITs so that SGMs can
be used with possibly familiar notation. (The transformations, however, were
omitted due to space restrictions.)

\subsubsection{Unified Parameterizable Attack Trees} 
\label{sec:unified_parametrizable_attack_trees}

In~$2011$, Wang, Whitley, Phan and Parish introduce \emph{unified 
parameterizable attack trees}\footnote{Wang~et~al. use British English, thus
originally, the name of their formalism is \emph{unified parametrizable attack
trees}.}~\cite{WaWhPhPa}. As the name suggests, the formalism was created as a
foundation to unify numerous existing extensions of attack trees
(Section~\ref{sec:attack_trees}). The formalism generalizes the notions of
connector types, edge augmentations and (node) attributes. With the help of
these generalizations it is possible to describe other extensions of attack
trees as structural extensions, computation extensions or hybrid extensions.

Unified parameterizable attack trees are defined as a~$5$-tuple, consisting of 
a set of nodes, a set of edges,  a set of allowed connectors (O-AND i.e., a time
or priority based AND, U-AND  i.e., an AND with a threshold condition and OR), a
set of attributes and a set  of edge augmentation structures that allows to
specify edge labels. Using this  definition, the authors of~\cite{WaWhPhPa}
identify defense trees (Section~\ref{sec:defense_trees}), attack countermeasure
trees (Section~\ref{sec:attack_countermeasure_trees}), attack-response trees
(Section~\ref{sec:attack_response_trees}), attack--defense trees
(Section~\ref{sec:attack_defense_trees}), protection trees
(Section~\ref{sec:protection_trees}), OWA trees (Section~\ref{sec:OWA_trees}),
and augmented attack trees (Section~\ref{sec:augmented_attack_trees}) as
structure-based extensions of attack tree that are covered by unified
parameterizable attack trees. They  classify multi-parameter attack  trees
(Section~\ref{sec:multi-parameter_attack_trees} and 
\ref{sec:multi-parameter_attack_trees-serial}) as a computational extension of
attack trees.

The formalism classifies attributes into the categories of ``attack
accomplishment attributes'', ``attack evaluation attributes'' and ``victim
system attributes'', but does not specify how to perform quantitative
evaluations.

Unified parameterizable attack trees are primarily built upon augmented attack 
trees (Section~\ref{sec:augmented_attack_trees}). In fact, the authors indicate
how to instantiate the node attributes, the edge augmentation and the connector 
type to obtain an augmented attack tree.

\section{Summary of the surveyed formalisms}
\label{sec:table}

In this section, we provide a consolidated view of all formalisms introduced in
Section~\ref{sec:main_survey}.
Tables~\ref{tab:comparison_table_1}--\ref{tab:comparison_table_3} characterize
the described methodologies (ordered alphabetically) according to the $13$ 
aspects presented in Table~\ref{tab:keywords_table2}. The aspects are grouped 
into formalism features and capabilities (Table~\ref{tab:comparison_table_1}),
formalism  characteristics (Table~\ref{tab:comparison_table_2}), and formalism
maturity and usability factors (Table~\ref{tab:comparison_table_3}). This
tabular view allows the reader to compare the features of the formalisms more
easily, it stresses their similarities and differences. Furthermore, the tables
support a user in selecting the most appropriate formalism(s) with respect to
specific modeling needs and requirements. We illustrate such a support on two
exemplary situations. 

\newcounter{exmpl}
\stepcounter{exmpl}

\paragraph{Example \theexmpl}

Let us assume that during a risk assessment, analysts want to investigate and
compare the efficiency of different defensive measures and controls, with
respect to several attack scenarios. Thereto, they need quantitative elements to
support the analysis technique they will choose. Furthermore, a software tool
and pre-existing use cases are required to facilitate their work. Using the
corresponding columns from 
Tables~\ref{tab:comparison_table_1}--\ref{tab:comparison_table_3} (i.e., attack 
or defensive, quantification, tool availability, case study) and choosing the
formalisms characterized by appropriate values (respectively: both, versatile or
specific, industrial or prototype, real(istic)), would help the analysts to
pre-selected attack countermeasure trees, attack--defense trees, BDMPs, 
intrusion DAGs, and security activity graphs, as potential modeling and analysis
techniques. The most suitable methodology could then be selected based on more
detailed information provided in Section~\ref{sec:main_survey}. For instance,
let us assume that the analysis requires the use of measures for probability of
success, the attacker's costs and the attacker's skills. Checking descriptions
of the pre-selected formalisms, given in Section~\ref{sec:main_survey}, would
convince the analysts that security activity graphs and intrusion DAGs would not
allow them to compute the expected quantitative elements. Therefore it would
reduce the choice to attack countermeasure trees, attack--defense trees, and
BDMPs. A more thorough investigation of the computational procedures and
algorithms described in the referred papers would help the analysts to make the
final decision on the formalism that best fits their needs. 

\stepcounter{exmpl}
\paragraph{Example \theexmpl}

Now, let us assume that a team of penetration testers wants to illustrate which
attack paths they have used to compromise different systems. Initially, this
does not significantly reduce the choice of possible formalisms, since they
could use all attack-oriented and all attack and defense oriented approaches.
However, to  keep the model as simple as possible, they start the selection
process by  looking at the attack-oriented methodologies only. Let us assume
further that the penetration testers also do not need to represent sequences of
actions.  With a similar reasoning as before, they first investigate a
possibility of using a static modeling technique. The team does not foresee to
perform any quantitative analysis. An important requirement, however, is to
employ a methodology which is already broadly used, with at least rudimentary
documented use cases on which they could rely to build their own models. Using
relevant columns from
Tables~\ref{tab:comparison_table_1}--\ref{tab:comparison_table_3}  (i.e., attack
or defense, sequential or static, paper count, use cases,  quantification) and
selecting formalisms characterized with appropriate values (respectively:  
attack or both, sequential or static, $>4$, real(istic) or toy, versatile,
specific or no), the team obtains a  large number of applicable formalisms. In
order to keep the formalism as simple  as possible, the analysts decide to
narrow the set of values they are  interested in to: attack, static,~$>4$,
real(istic) or toy, no. This strategy  yields the following most suitable
formalisms: attack trees, augmented attack  trees, and parallel model for
multi-parameter attack trees. The team would then be able to make a final choice
of the methodology based on complementary investigations starting from the
information and references provided by the corresponding textual descriptions
from Section~\ref{sec:main_survey}.

\begin{longtable}[c]{|m{0.25\textwidth}|m{0.14\textwidth}|m{0.13\textwidth}|
m{0.105\textwidth}|m{0.13\textwidth}|m{0.155\textwidth}|}
\caption{Aspects relating to the formalism's modeling capabilities}
\label{tab:comparison_table_1}
\\
\hline
\textbf{Name of formalism} & 
\multirow{2}{0.14\textwidth}{\textbf{Attack or defense}}\newline& 
\multirow{2}{0.13\textwidth}{\textbf{Sequential or static}}\newline&
\multirow{2}{0.105\textwidth}{\textbf{Quantifi\-cation}}\newline&
\multirow{2}{0.105\textwidth}{\textbf{Main Purpose}}\newline&
\textbf{Extension}
\\\hline
Anti-models\newline
(Section~\ref{sec:obstacle_trees})
& Both 
& Static
& No
& Req. eng.
& New formalism
\\
\hline
\multirow{2}{0.25\textwidth}{Attack countermeasure trees 
(Section~\ref{sec:attack_countermeasure_trees})} \newline
& Both 
& Static
& Specific
& Sec. mod.
& Structural \newline
Computational
\\\hline
Attack--defense trees
\newline
(Section~\ref{sec:attack_defense_trees})
& Both
& Static
& Versatile
& Sec. mod.
& Structural \newline
Computational
\\\hline
Attack-response trees
\newline
(Section~\ref{sec:attack_response_trees})
& Both
& Sequential
& Specific
& Int. det.
& Structural \newline
Quantitative
\\\hline
Attack trees
\newline
(Section~\ref{sec:attack_trees})
& Attack
& Static
& Versatile
& Sec. mod.
& New formalism
\\\hline
Augmented attack trees
\newline
(Section~\ref{sec:augmented_attack_trees})
& Attack
& Static
& Specific
& Sec. mod.
& Structural \newline
Computational
\\\hline
\multirow{2}{0.25\textwidth}{Augmented vulnerabil- \newline ity trees 
(Section~\ref{sec:vulnerability_trees})} \newline
& Attack
& Static
& Specific
& Risk
& Quantitative
\\\hline
Bayesian attack graphs
\newline
(Section~\ref{sec:bayesian_attack_graphs})
& Attack
& Sequential
& Specific
& Risk
& Structural \newline
Computational
\\\hline
Bayesian defense graphs
\newline
(Section~\ref{sec:bayesian_defense_graphs})
& Both
& Sequential
& Specific
& Risk
& Structural \newline
Computational
\\\hline
\multirow{2}{0.25\textwidth}{Bayesian networks for security 
(Section~\ref{sec:bayesian_networks})} \newline  
& Attack
& Sequential
& Specific
& Risk
& Structural \newline
Computational
\\\hline
BDMPs (Section~\ref{sec:BDMP})
& Both
& Sequential
& Versatile
& Sec. mod.
& Order \newline
Time
\\\hline
Compromise graphs
\newline
(Section~\ref{sec:compormise_graphs})
& Attack
& Sequential
& Specific
& Risk
& New formalism 
\\\hline
Countermeasure graphs
\newline
(Section~\ref{sec:countermeausre_graphs})
& Both
& Static
& Specific
& Sec. mod.
& Structural \newline
Computational
\\\hline
Cryptographic DAGs
\newline
(Section~\ref{sec:cryptographic_dag})
& Attack
& Sequential
& No
& Risk
& New formalism 
\\\hline
Defense trees
\newline
(Section~\ref{sec:defense_trees})
& Both
& Static
& Specific
& Sec. mod.
& Structural \newline
Computational
\\\hline
\multirow{2}{0.25\textwidth}{Dynamic fault trees for security 
(Section~\ref{sec:dynamic_fault_trees})} \newline
& Attack
& Sequential
& No
& Sec. mod.
& Order \newline
Time
\\\hline
Enhanced attack trees
\newline
(Section~\ref{sec:enhanced_attack_trees})
& Attack
& Sequential
& Specific
& Int. det.
& Order \newline
Time
\\\hline
Extended fault trees
\newline
(Section~\ref{sec:extended_fault_trees})
& Attack
& Static
& Specific
& Unification
& Structural
\\\hline
Fault trees for security
\newline
(Section~\ref{sec:fault_trees_for_sec})
& Attack
& Sequential
& No
& Sec. mod.
& Order
\\\hline
Improved attack trees
\newline
(Section~\ref{sec:improved_attack_trees})
& Attack
& Sequential
& Specific
& Risk
& Structural \newline
Computational
\\\hline
Insecurity flows
\newline
(Section~\ref{sec:insecurity_flows})
& Both
& Sequential
& Specific
& Risk
& New formalism
\\\hline
Intrusion DAGs
\newline
(Section~\ref{sec:i-dag})
& Both
& Sequential
& Specific
& Int. det.
& Structural \newline
Computational
\\\hline
OWA trees
\newline
(Section~\ref{sec:OWA_trees})
& Attack
& Static
& Specific
& 
Quantitative
& Structural \newline
Computational
\\\hline
\multirow{3}{0.25\textwidth}{Parallel model for multi-parameter attack 
trees (Section~\ref{sec:multi-parameter_attack_trees})}\newline $\phantom{x}$  
\newline
& Attack
& Static
& Specific
& 
Quantitative
& Quantitative \newline
Computational
\\\hline
Protection trees
\newline
(Section~\ref{sec:protection_trees})
& Defense
& Static
& Specific
& Sec. mod.
& New formalism 
\\\hline
Security activity graphs
\newline
(Section~\ref{sec:security_activity_graphs})
& Both
& Static
& Specific
& Soft. dev.
& New formalism
\\\hline
\multirow{2}{0.25\textwidth}{Security goal indicator trees 
(Section~\ref{sec:security_goal_indicator_trees})} \newline
& Defense
& Sequential
& No
& Soft. dev.
& New formalism 
\\\hline
\multirow{2}{0.25\textwidth}{Security goal models\newline  
(Section~\ref{sec:security_goal_models})} \newline
& Both
& Sequential
& Specific
& Unification
& Structural \newline
Computational
\\\hline
\multirow{3}{0.25\textwidth}{Serial model for multi-\newline parameter attack 
trees (Section~\ref{sec:multi-parameter_attack_trees-serial})} \newline 
$\phantom{x}$ \newline
& Attack
& Sequential
& Specific
& 
Quantitative
& Computational \newline
Order
\\\hline
\multirow{3}{0.25\textwidth}{Unified parameterizable attack trees 
(Section~\ref{sec:unified_parametrizable_attack_trees})} \newline $\phantom{x}$ 
\newline
& Both
& Sequential
& Versatile
& Unification
& Structural
\\\hline
\multirow{2}{0.25\textwidth}{Vulnerability cause graphs 
(Section~\ref{sec:vulnerability_cause_graphs})} \newline 
& Attack
& Sequential
& Specific
& Soft. dev.
& Structural \newline
Order
\\\hline
\end{longtable}

\begin{longtable}[c]{|m{0.25\textwidth}|m{0.12\textwidth}|m{0.40\textwidth}|
m{0.17\textwidth}|}
\caption{Aspects relating to the formalism's characteristics}
\label{tab:comparison_table_2}
\\
\hline
\textbf{Name of formalism} & 
\textbf{Structure} & 
\textbf{Connectors}&
\textbf{Formalization}
\\\hline
Anti-models
\newline
(Section~\ref{sec:obstacle_trees})
& Tree
& AND, OR
& Semi-formal
\\
\hline
\multirow{2}{0.25\textwidth}{Attack countermeasure trees 
(Section~\ref{sec:attack_countermeasure_trees})} \newline 
& Tree
& 
AND, OR,~$k$-out-of-$n$, counter leaves
& Formal
\\\hline
Attack--defense trees
\newline
(Section~\ref{sec:attack_defense_trees})
& Tree
& 
AND, OR, countermeasures
& Formal
\\\hline
Attack-response trees
\newline
(Section~\ref{sec:attack_response_trees})
& Tree
& AND, OR, responses
& Formal
\\\hline
Attack trees
\newline
(Section~\ref{sec:attack_trees})
& Tree
& AND, OR
& Formal
\\\hline
Augmented attack trees
\newline
(Section~\ref{sec:augmented_attack_trees})
& Tree
& AND, OR
& Formal
\\\hline
\multirow{2}{0.25\textwidth}{Augmented vulnerabil- \newline ity trees 
(Section~\ref{sec:vulnerability_trees})} \newline
& Tree
& AND, OR
& Informal
\\\hline
Bayesian attack graphs
\newline
(Section~\ref{sec:bayesian_attack_graphs})
& DAG
& AND, OR, conditional\newline probabilities
& Formal
\\\hline
Bayesian defense graphs
\newline
(Section~\ref{sec:bayesian_defense_graphs})
& DAG
& AND, OR, conditional\newline probabilities
& Formal
\\\hline
\multirow{2}{0.25\textwidth}{Bayesian networks for security 
(Section~\ref{sec:bayesian_networks})} \newline
& DAG
& AND, OR, conditional\newline probabilities
& Formal
\\\hline
BDMPs (Section~\ref{sec:bayesian_networks})
& DAG
& AND, OR, PAND, approx. OR, triggers
& Formal
\\\hline
Compromise graphs
\newline
(Section~\ref{sec:compormise_graphs})
& Unspecified
& None
& Formal
\\\hline
Countermeasure graphs
\newline
(Section~\ref{sec:countermeausre_graphs})
& DAG
& Countermeasures
& Informal
\\\hline
Cryptographic DAGs
\newline
(Section~\ref{sec:cryptographic_dag})
& DAG
& Dependence edges
& Informal
\\\hline
Defense trees
\newline
(Section~\ref{sec:defense_trees})
& Tree
& AND, OR, counter leaves
& Semi-formal
\\\hline
\multirow{2}{0.25\textwidth}{Dynamic fault trees for security 
(Section~\ref{sec:dynamic_fault_trees})} \newline
& Tree
& 
AND, OR, PAND, SEQ, FDEP, CSP
& Informal
\\\hline
Enhanced attack trees
\newline
(Section~\ref{sec:enhanced_attack_trees})
& Tree
& AND, OR, ordered-AND
& Formal
\\\hline
Extended fault trees
\newline
(Section~\ref{sec:extended_fault_trees})
& Tree
& AND, OR, merge gates
& Formal
\\\hline
Fault trees for security\newline
(Section~\ref{sec:fault_trees_for_sec})
& Tree
& AND, OR, PAND, XOR, inhibit
& Informal
\\\hline
Improved attack trees
\newline
(Section~\ref{sec:improved_attack_trees})
& Tree
& AND, OR, sequential AND
& Informal
\\\hline
Insecurity flows
\newline
(Section~\ref{sec:insecurity_flows})
& Unspecified
& None
& Formal
\\\hline
Intrusion DAGs
\newline
(Section~\ref{sec:i-dag})
& DAG
& AND, OR
& Semi-formal
\\\hline
\multirow{2}{0.25\textwidth}{OWA trees\newline (Section~\ref{sec:OWA_trees})} 
\newline
& Tree
& OWA operators
& Formal
\\\hline
\multirow{3}{0.25\textwidth}{Parallel model for multi-parameter attack 
trees (Section~\ref{sec:multi-parameter_attack_trees})}\newline $\phantom{x}$ 
\newline 
& Tree
& AND, OR
& Formal
\\\hline
Protection trees
\newline
(Section~\ref{sec:protection_trees})
& Tree
& AND, OR
& Informal
\\\hline
Security activity graphs
\newline
(Section~\ref{sec:security_activity_graphs})
& DAG
& AND, OR, split gate
& Semi-formal
\\\hline
\multirow{2}{0.25\textwidth}{Security goal indicator trees 
(Section~\ref{sec:security_goal_indicator_trees})} \newline 
& Tree
& AND, OR, dependence\newline edge, specialization edge
& Semi-formal
\\\hline
Security goal models
\newline
(Section~\ref{sec:security_goal_models})
& DAG
& 
AND, OR, dependence edge, information edge
& Formal
\\\hline
\multirow{3}{0.25\textwidth}{Serial model for multi-\newline parameter attack 
trees (Section~\ref{sec:multi-parameter_attack_trees-serial})} \newline 
$\phantom{x}$ \newline
& Tree
& AND, OR, ordered leaves
& Formal
\\\hline
\multirow{3}{0.25\textwidth}{Unified parameterizable attack trees 
(Section~\ref{sec:unified_parametrizable_attack_trees})} \newline $\phantom{x}$ 
\newline
& Tree
& 
AND, OR, PAND, time-based AND, threshold AND
& Formal
\\\hline
\multirow{2}{0.25\textwidth}{Vulnerability cause graphs 
(Section~\ref{sec:vulnerability_cause_graphs})} \newline 
& DAG
& AND, OR, sequential AND
& Informal
\\\hline
\end{longtable}

\begin{longtable}[c]{|m{0.25\textwidth}|m{0.15\textwidth}|m{0.16\textwidth}|
m{0.16\textwidth}|m{0.09\textwidth}|m{0.07\textwidth}|}
\caption{Aspects related to the formalism's maturity and usability}
\label{tab:comparison_table_3}
\\
\hline
\textbf{Name of formalism} & 
\multirow{2}{0.15\textwidth}{\textbf{Tool availability}}\newline & 
\textbf{Case study}
&
\textbf{External use}
&
\multirow{2}{0.09\textwidth}{\textbf{Paper count}}\newline
&
\textbf{Year}
\\
\hline
Anti-models
\newline
(Section~\ref{sec:obstacle_trees})
& No
& No
& No
& $3$
& $2006$
\\
\hline
\multirow{2}{0.25\textwidth}{Attack countermeasure trees 
(Section~\ref{sec:attack_countermeasure_trees})} \newline 
& Prototype
& Real(istic)
& No
& $4$
& $2010$
\\\hline
Attack--defense trees
\newline
(Section~\ref{sec:attack_defense_trees})
& Prototype
& Real(istic)
& Collaboration
& $6$
& $2010$
\\\hline
Attack-response trees
\newline
(Section~\ref{sec:attack_response_trees})
& Prototype
& Toy case study
& No
& $3$
& $2009$
\\\hline
Attack trees
\newline
(Section~\ref{sec:attack_trees})
& Commercial 
& Real(istic)
& Independent
& $>100$
& $1991$
\\\hline
Augmented attack trees
\newline
(Section~\ref{sec:augmented_attack_trees})
& No
& Real(istic)
& Independent
& $6$
& $2005$
\\\hline
Augmented vulnerabil-\newline ity trees
(Section~\ref{sec:vulnerability_trees})
& No
& Real(istic)
& Independent
& $3$
& $2003$
\\\hline
Bayesian attack graphs
\newline
(Section~\ref{sec:bayesian_attack_graphs})
& Commercial 
& Toy case study
& Independent
& $10$
& $2005$
\\\hline
Bayesian defense graphs
\newline
(Section~\ref{sec:bayesian_defense_graphs})
& Prototype
& Real(istic)
& No
& $5$
& $2008$
\\\hline
\multirow{2}{0.25\textwidth}{Bayesian networks for security
(Section~\ref{sec:bayesian_networks})} \newline 
& Commercial 
& Real(istic)
& Independent
& $14$
& $2004$
\\\hline
BDMPs (Section~\ref{sec:BDMP})
& Commercial 
& Real(istic)
& Independent
& $5$
& $2010$
\\\hline
Compromise graphs\newline
(Section~\ref{sec:compormise_graphs})
& No
& Real(istic)
& Collaboration
& $3$
& $2006$
\\\hline
Countermeasure graphs\newline
(Section~\ref{sec:countermeausre_graphs})
& No
& Toy case study
& No
& $1$
& $2010$
\\\hline
Cryptographic DAGs\newline
(Section~\ref{sec:cryptographic_dag})
& No
& No
& No
& $1$
& $1996$
\\\hline
Defense trees\newline
(Section~\ref{sec:defense_trees})
& No
& No
& No
& $3$
& $2006$
\\\hline
\multirow{2}{0.25\textwidth}{Dynamic fault trees for security
(Section~\ref{sec:dynamic_fault_trees})} \newline
& No
& No
& No
& $1$
& $2009$
\\\hline 
Enhanced attack trees\newline
(Section~\ref{sec:enhanced_attack_trees})
& No
& No
& No
& $1$
& $2007$
\\\hline
Extended fault trees\newline
(Section~\ref{sec:extended_fault_trees})
& No
& No
& No
& $1$
& $2007$
\\\hline
Fault trees for security\newline
(Section~\ref{sec:fault_trees_for_sec})
& Commercial 
& Real(istic)
& Independent
& $3$
& $2003$
\\\hline
Improved attack trees\newline
(Section~\ref{sec:improved_attack_trees})
& No
& No
& No
& $1$
& $2011$
\\\hline
Insecurity flows\newline
(Section~\ref{sec:insecurity_flows})
& No
& No
& No
& $1$
& $1997$
\\\hline
Intrusion DAGs\newline
(Section~\ref{sec:i-dag})
& Prototype
& Real(istic)
& No
& $2$
& $2003$
\\\hline
OWA trees\newline
(Section~\ref{sec:OWA_trees})
& No
& No
& No
& $2$
& $2005$
\\\hline
\multirow{3}{0.25\textwidth}{Parallel model for multi-parameter attack 
trees (Section~\ref{sec:multi-parameter_attack_trees})}\newline $\phantom{x}$ 
\newline
& Prototype
& Real(istic)
& Collaboration
& $5$
& $2006$
\\\hline
Protection trees\newline
(Section~\ref{sec:protection_trees})
& No
& Toy case study
& No
& $4$
& $2006$
\\\hline
Security activity graphs\newline
(Section~\ref{sec:security_activity_graphs})
& Prototype
& Real(istic)
& No
& $2$
& $2006$
\\\hline
\multirow{2}{0.25\textwidth}{Security goal indicator trees 
(Section~\ref{sec:security_goal_indicator_trees})} \newline 
& Prototype
& Real(istic)
& No
& $3$
& $2008$
\\\hline
Security goal models\newline
(Section~\ref{sec:security_goal_models})
& No
& Real(istic)
& No
& $2$
& $2010$
\\\hline
\multirow{3}{0.25\textwidth}{Serial model for multi-\newline parameter attack 
trees (Section~\ref{sec:multi-parameter_attack_trees-serial})} \newline 
$\phantom{x}$ 
\newline 
& Prototype
& No
& No
& $3$
& $2010$
\\\hline
\multirow{3}{0.25\textwidth}{Unified parameterizable attack trees 
(Section~\ref{sec:unified_parametrizable_attack_trees})} \newline $\phantom{x}$ 
\newline 
& No
& No
& No
& $1$
& $2011$
\\\hline
\multirow{2}{0.25\textwidth}{Vulnerability cause graphs 
(Section~\ref{sec:vulnerability_cause_graphs})} \newline 
& Commercial 
& Real(istic)
& Independent
& $4$
& $2006$
\\\hline
\end{longtable}

\section{Alternative Methodologies}
\label{sec:alternative}

We close this survey with a short overview of alternative methodologies for
security modeling and analysis.  The formalisms described here are outside the
main scope of this paper, because they  were not originally introduced for the
purposes of attack and defense  modeling or they are not based on the DAG
structure.  However, for the sake of completeness, we find important to  briefly
present those approaches as well.   The objective of this section is to give
pointers to other existing methodological tools for security assessment,  rather
than to perform a thorough overview of all related formalisms. This explains why
the description of the formalisms given here is less complete and structured
than the information provided in  Section~\ref{sec:main_survey}. 

\subsection{Petri Nets for Security}

In the mid~$1990$s, models based on Petri nets have been applied for security 
analysis~\cite{KuSp,Dacier1994}. In~$1994$, Kumar and Spafford~\cite{KuSp}
adopted colored Petri nets for  security  modeling. They illustrate how to model
reference scenarios for an intrusion  detection device. Also in~$1994$,
Dacier~\cite{Dacier1994} used Petri nets in  his  Ph.D. thesis as part of a
larger quantification model that describes the progress of an attacker taking
over a system. A useful property of Petri nets  is their great modeling
capability and in particular their ability to take into  account the sequential
aspect of attacks, the modeling of concurrent action and  various forms of
dependency. Petri nets are widely used and have various  specific extensions. To
corroborate this statement,  we list a few existing ones. Kumar and Spafford's
work relies on \emph{colored Petri nets}~\cite{KuSp}, Dacier's on
\emph{stochastic Petri nets}~\cite{Dacier1994}, McDermott's on \emph{disjunctive
Petri  nets}~\cite{McDe}, Horvath and D\"{o}rges' on \emph{reference
nets}~\cite{HoDo}, Dalton II~et~al.'s on \emph{generalized stochastic Petri 
nets}~\cite{DaMiCoRa}, Pudar~et~al.'s on \emph{deterministic time transition 
Petri nets}~\cite{PuMaLi} and Xu and Nygard's on \emph{aspect-oriented Petri 
nets}~\cite{XuNy}. Several articles on Petri nets merge the formalism with 
other approaches. Horvath and D\"{o}rges combine Petri nets with the concept of 
\emph{security patterns}~\cite{HoDo} while Dalton II~et~al.~\cite{DaMiCoRa}, and
more thoroughly Pudar~et~al.~\cite{PuMaLi}, combine Petri nets and attack 
trees. 

In~$1994$, Dacier embedded Petri nets into a higher level formalism called
\emph{privilege graphs}. They model an attacker's progress in obtaining access
rights for a desired target~\cite{Dacier1994,DaDe}.  In a privilege graph, a
node represents a set of privileges and an edge a method for transferring these
privileges to the attacker. This corresponds to the exploitation of a
vulnerability. The model includes an attacker's ``memory''  which forbids him to
go through privilege states that he has already acquired.  In addition, an
attacker's ``good sense'' is modeled which prevents him from  regressing.
In~\cite{DaDeKa}, Dacier~et~al. proposed to transform a privilege  graph into
Markov chain corresponding to all possible successful  attack scenarios. The
method has been applied to help  system administrators to monitor the security
of their systems.

In~\cite{ZaFe}, Zakzewska and Ferragut presented a model extending Petri nets 
in order to model real-time cyber conflicts. This formalism is able to 
represent situational awareness, concurrent actions, incomplete information and 
objective functions. Since it makes use of stochastic transitions, it is well 
suited to reason about stochastic non-controlled events. The formalism is used 
to run simulations of cyber attacks in order to experimentally analyze 
cyber conflicts. The authors also performed a 
comparison of their \emph{extended Petri nets} model with other security 
modeling techniques. In particular, they showed that extended Petri 
nets are more readable and more expressive than attack graphs, especially with 
respect to the completeness of the models. 

\subsection{Attack Graphs}
\label{sec:attack_graphs}

The term \emph{attack graph} has been first introduced by Phillips and 
Swiler~\cite{PhSw,SwPhElCh} in~$1998$, and has extensively been used ever 
since. The nodes of an attack graph represent possible states of a system during
the attack. The edges correspond to changes of states due to an attacker's 
actions. An attack graph is generated automatically based on three types of
inputs: attack templates (generic representations of attacks including required
conditions), a detailed description of the system to be attacked  (topology,
configurations of components, etc.), and the attacker's profile (his 
capability, his tools, etc.). Quantifications, such as average probabilities or 
time to success, can be deduced by giving  weight to the edges and by finding
shortest paths in the graph.

Starting in~$2002$, Sheyner~\cite{ShHaJhLiWi,Shey} made extensive  contributions
 to popularize attack graphs by associating them with model checking techniques.
To limit the risk of combinatorial explosion, a large number of methods were 
developed. Ammann~et~al.~\cite{AmWiKa} restricted the graphs by exploiting a 
monotony property, thereby eliminating backtracking in terms of privilege 
escalation. Noel and Jajodia~\cite{NoJaObJa,JaNoOb} took configuration aspects 
into account.  A complete state of the art  concerning the contributions to the
field between~$2002$ and~$2005$ can be  found in~\cite{LiIn}. In $2006$, Wang et
al. introduced a relational model for attack  graphs~\cite{WaYaSiJa}.  The
approach facilitates interactive analysis of the models and improves its 
performance.  Ou~et~al.~\cite{OuBoQu} optimized the  generation and
representation of attack graphs by transforming them into  \emph{logical attack
graphs} of polynomial size with respect to the number of  components of the
computer network analyzed.  During the same year,  Ingols~et~al.~\cite{InLiPi}
proposed \emph{multiple-prerequisite graphs}, which  also severely reduce the
complexity of the graphs.   In~\cite{MeBaZhClWi}, Mehta~et~al. proposed an
algorithm  for classification of states in order to identify the most relevant 
parts of an attack graph. In~$2008$, Malhotra~et~al.~\cite{MaBaGh} did the same
based on the notion of \emph{attack surface} described in \cite{Mana}. The vast 
majority of the authors mentioned have also worked on visualization 
aspects~\cite{NoJa,NoJaKaJa,WiLiIn,HoVaOuQu}. Kotenko and 
Stepashkin~\cite{KoSt} described a complete software platform for implementing 
concepts and metrics of attack graphs. On a theoretical level, Braynov and 
Jadliwala~\cite{BrJa} extended the model with several attackers.

Starting in~$2003$, the problem of quantitative  assessment of the security of
networked systems using attack  graphs has been extensively 
studied~\cite{NoJaObJa,WaNoJa,WaSiJa2,WaSiJa,WaIsLoSiJa}.  The work presented
in~\cite{NoJaObJa} and \cite{WaNoJa}  focuses on minimal cost of removing
vulnerabilities in hardening a network.  In~\cite{WaSiJa2}, the authors
introduced a metric, called attack resistance, which is used to compare the
security of different network configurations.   The approach was then extended
in~\cite{WaSiJa} into a general abstract framework  for measuring various
aspects of network security.  In~\cite{WaIsLoSiJa}, Wang et al. introduced a
metric incorporating probabilities  of the existence of the vulnerabilities
considered in the graph.

In his master thesis, Louthan IV~\cite{Lout} proposed extensions to the attack 
graph modeling framework to permit modeling continuous, in addition to 
discrete, system elements and their interactions. In~\cite{WaIsLoSiJa},
Wang~et~al. addressed the problem of likelihood  quantification of potential
multi-step attacks on networked environments, that  combine multiple
vulnerabilities. They developed an attack graph-based  probabilistic metric for
network security and proposed heuristics for efficient  computation.
In~\cite{NoJaWaSi}, Noel~et~al. used attack graphs to understand  how different
vulnerabilities can be combined to form an attack on a network.  They simulated
incremental network penetration and assessed the overall  security of a network
system by propagating attack likelihoods. The method  allows to give scores to
risk mitigation options in terms of  maximizing security and minimizing cost. It
can be used to study cost/benefit  trade-offs for analyzing return on security
investment. 

Dawkins and Hale~\cite{DaHa} developed a concept similar to attack graphs called
\emph{attack chains}. The model is based on a deductive tree 
structure approach but also allows for inductive reasoning using 
\emph{goal-inducing attack chains}, to extract scenarios leading to a given 
aim. These models are also capable of generating attack trees, which may be 
quantified by conventional methods. Aspects concerning software implementation 
are described in~\cite{ClTyDaHa}.

\subsection{Approaches Derived from UML Diagrams}

We start this section with a short description of two formalisms derived from 
UML diagrams, namely \emph{abuse cases} of McDermott and Fox~\cite{DeFo} and 
\emph{misuse cases} of Sindre and 
Opdahl~\cite{SiOpLo2,SiOpLo,SiOpLoBr,Alex,SiOp} which were later extended by 
R\o{}stad in~\cite{Rost}. These techniques are not specifically intended to
model attacks but rather to  capture threats and abusive behavior which have to
be taken into account when  eliciting security requirements (for misuse cases)
as well as for design and  testing (for abuse cases).  The flexibility of misuse
and abuse cases allows for expressive graphical  modeling of attack scenarios
without mathematical formalization that supports  quantification.

In~\cite{Fire}, Firesmith argues that  misuse and abuse cases are ``highly
effective ways of analyzing security threats but are inappropriate for  the
analysis and specification of security requirements''. The reasoning is  that
misuse cases focus on how misusers can successfully attack the system.  Thus
they often model specific architectural mechanisms and solutions, e.g.,  the use
of passwords, rather than actual security requirements, e.g.,  authentication
mechanisms. To specify security requirements, he suggested to  use
\emph{security use cases}. Security use cases focus on how an application 
achieves its goals. According to Firesmith, they provide ``a highly-reusable 
way of organizing, analyzing, and specifying security 
requirements''~\cite{Fire}. 

Diallo et al. presented a comparative  evaluation of the common
criteria~\cite{CC},  misuse cases, and attack  trees~\cite{DiRoSiAlRi}. Opdahl
and Sindre~\cite{OpSi} compared usability aspects and  modeling features of
misuse cases and attack trees.   UML-based approaches can be combined with other
types of models.  The combination of misuse cases and attack trees  appears not
only to be simple but also useful and  relevant~\cite{ToJeRo,MeToJe}. 
In~\cite{KaSiOp}, K\'{a}rp\'{a}ti~et~al.  adapted use case maps to security as
\emph{misuse case maps}. Katta~et~al.~\cite{Katta2010} combined \emph{UML
sequence diagrams} with misuse cases in  a new formalism called \emph{misuse
sequence diagrams}.  A misuse sequence diagram represents a sequence of attacker
interactions with  system components and depicts how the components were misused
over time by  exploiting their vulnerabilities. The authors of~\cite{Katta2010}
performed  usability and performance comparison of misuse sequence diagrams and
misuse case  maps. In~\cite{KaSiOp2},  K\'{a}rp\'{a}ti~et~al. integrated five
different representation techniques  in a method called \emph{hacker attack
representation method} (HARM).  The methodologies used in HARM are: attack
sequence descriptions (summarizing  attacks in natural language), misuse case
maps (depicting the system  architecture targeted by the attack and visualizing
the traces of the  exploits), misuse case diagrams (showing threats in relation
to the wanted  functionality) attack trees (representing the hierarchical
relation between  attacks) and attack patterns (describing an attack in detail
by adding  information about context and solutions). Combining such diverse
representation  techniques has two goals. First, it provides ``an integrated
view of security attacks  and system architecture''. Second, the HARM method is
especially well  suited in involving different stakeholders, including
non-technical people  preferring informal representations. 

In~\cite{Sind}, Sindre adapted UML activity diagrams to security. The resulting 
\emph{mal-activity diagrams} constitute an alternative for misuse cases when 
the author considers the latter to be unsuitable. This is for instance the case 
in situations where a large numbers of interactions need to be specified within 
a or outside a system. Case studies mainly concern social engineering 
attacks~\cite{KarpatiIJSSE12}. 

\subsection{Isolated models}
\label{sec:isolated_models}

In this section we gather a number of isolated models. Most of them contain 
cycles and therefore are outside of the main scope of this paper. However, we
mention them because they build upon a formalism described in 
Section~\ref{sec:main_survey}.

The \emph{stratified node topology} was proposed by Daley~et~al.~\cite{DaLaDa} 
as an extension of attack trees, in~$2002$. The  formalism consists of a
directed graph which is aimed at providing a context  sensitive attack modeling
framework. It supports incident correlation, analysis  and prediction and
extends attack trees by separating the nodes into three  distinct classes based
on their functionality: event-level nodes,  state-level nodes and top-level
nodes. The directed edges between the nodes are  classified into implicit and
explicit links. Implicit links allow individual  nodes to imply other nodes in
the tree; explicit links are created when an  attack provides a capability to
execute additional nodes, but does not actually  invoke a new instance of a
node. As in attack trees, the set of linked nodes  can be connected
disjunctively as well as conjunctively. In comparison with  attack trees, the
authors drop the requirement of a designated root node, along  with the
requirement that the graphs have to be acyclic. Due to the functional 
distinction of the nodes, the stratified node topology can keep the vertical 
ordering, even if the modeled scenario is cyclic.

In~$2010$, Abdulla~et~al.~\cite{AbCeKa} described a model called \emph{attack 
jungles}. When trying to use attack trees as formalized by Mauw and Oostdijk 
in~\cite{MaOo} to illustrate the security of a GSM radio network, the authors 
of~\cite{AbCeKa} encountered modeling problems related to the presence of 
cycles as well as analysis problems related to reusability of nodes in real 
life scenarios. This led them to propose attack jungles, which extend attack 
trees with multiple roots, reusable nodes and cycles that allow for modeling  of
attacks which depend on each other. Attack jungles are formalized as 
multigraphs and their formal semantics extend the semantics based on multisets 
proposed in~\cite{MaOo}. In order to find possible ways of attacking a system, 
a backwards reachability algorithm for analysis of attack jungles was 
described. Moreover, the notion of an attribute domain for quantitative 
analysis, as proposed for attack trees in~\cite{MaOo}, is extended to fit the 
new structure of attack jungles. By dividing attack components (nodes)  into
reusable and not reusable ones, it is possible to reason about and better 
analyze realistic scenarios. For instance, in an attack jungle a component  used
once can be reused multiple times without inducing any extra cost. 

\emph{Extended influence diagrams}~\cite{JoLaNaSi} form another related
formalism which is not based on a DAG structure. Extended influence diagrams are
built upon influence diagrams, introduced by Matheson and Howard in the
1960s~\cite{MaHo}, which, in turn, are an extension of Bayesian networks. 
Influence diagrams are also known as relevance diagrams, decision diagrams or
decision networks and are used to provide a high-level visualization of decision
problems under uncertainty~\cite{EzBe2010}. Extended influence diagrams allow to
model the relationships between decisions, events and outcomes of an enterprise
architecture. They employ the following three types of node: ellipses which
represent events (also known as chance nodes), rectangles which depict decision
nodes and diamonds which represent utility nodes (or outcomes). In addition the
formalism allows to specify how a node is defined, how well it can be controlled
and how the nodes relate to each other. The latter is achieved with different
types of edges. Moreover, transformation rules between graphs govern switching
between different levels of abstraction of a scenario (expanding and
collapsing). The rules also ensure that graphs do not contradict each other.
In~\cite{LaJoNa}, the authors show how to elicit knowledge from scientific
texts, generating extended influence diagrams and in~\cite{EkSo} the authors
outline how extended influence diagrams can be used for cyber security
management.

\section{Conclusion} \label{sec:conclusion}

This work presents a complete and methodical overview of DAG-based techniques
for modeling attack and defense scenarios. Some of the described methodologies
have extensively been studied and are widely used to support security and risk
assessment processes. Others emerged from specific, practical developments and 
have remained isolated methods. This survey provides a structured description of
the existing formalisms, gives pointers to related papers, tools and projects,
and proposes a general classification of the presented approaches. To classify
the formalisms, we have used $13$ aspects concerning graphical, formal and
usability characteristics of the analyzed models.

Two general trends can be observed in the field of graphical security modeling:
\emph{unification} and \emph{specification}. The objective of the methodologies
developed within the first trend is to unify existing approaches and propose
general solutions that can be used for analysis of a broad spectrum of security 
scenarios. The corresponding formalisms are well suited for reasoning about
situations involving diversified aspects, such as digital, physical and social
components, simultaneously. Such models usually have sound formal foundations
and are extensively studied from a theoretical point of view. They are augmented
with formal semantics and a general mathematical framework for quantitative
analysis. Examples of such models developed within the unification trend are
attack--defense trees, unified parameterizable attack trees, multi-parameter
attack trees, OWA trees, Bayesian attack graphs, and Bayesian defense graphs.

The second observed trend, i.e., the specification trend, aims at developing 
methodologies for addressing domain specific security problems. Studied domains
include intrusion detection (e.g., attack-response trees, intrusion DAGs),
secure software development (e.g., security activity graphs, security goal
indicator trees), and security requirements engineering (e.g., anti-models).
Formalisms developed within this trend are often based on empirical studies and
practical needs. They concentrate on domain specific  metrics, such as the
\emph{response index}, which is used for the analysis of intrusion DAGs. These
approaches often remain isolated and seldom relate to or build upon other
existing approaches.

The multitude of methodologies presented in this survey shows that graphical
security modeling is a young but very rapidly growing area. Thus, further
development is necessary and new directions need to be explored before security
assessment can fully benefit from graphical models. One of the research
questions which has not yet received enough attention is building graphical
models from pre-existing attack templates and patterns. Addressing this question
would make automatic model creation possible and replace the tedious,
error-prone, manual construction process. It would therefore strongly relieve
the industrial sector when building large-scale practical models. 

The idea of reusing attack patterns is not new. It has already been mentioned
in~$2001$ by Moore et al.~\cite{MoElLi}. An excellent initiative was taken by
the FP7 project SHIELDS~\cite{Website_SHIELDS}, in which the Security
Vulnerability Repository Service (SVRS) has been developed. The SVRS is an
on-line library of various security models including attack trees~\cite{SVRS}. 
A natural follow-up step would be to propose methods for automatic or
semi-automatic construction of complex, specific models from general attack or
vulnerability patterns. This would require developing algorithms for correct
composition and comparison of models, standardizing employed node labels and
introducing an agent-based view into the formalisms. 

Using security patterns makes threat analysis more efficient and accurate. First
generating a general model from existing libraries constitutes a good starting
point for further model refinement and analysis. Furthermore, although new
technological opportunities arise every day, empirical studies show that most
attackers reuse the same attack vectors with little or no modification. Often
the same company is attacked several times by an intruder exploiting the same
already known vulnerability.

There still exists a gap between theoretical research and practical employment
of graphical security models. Tighter interaction between the scientific and
industrial security communities would be very beneficial for the future of the
field. Setting up dedicated events, such as workshops, conferences or panel
discussions, would provide a platform for the exchange of ideas, closer
collaboration and a faster dissemination of results. This would allow
practitioners to better understand the capabilities of theoretical models and
give scientists an opportunity to learn what the practical and industrial needs
are. Once a bridge between the two communities is built, a natural next step
will be to include graphical models into standardized and commonly used auditing
and risk assessment tools and practices. Due to the sound formal foundations of 
the graphical models as well as their user friendliness, this would greatly
improve the quality and usability of the currently used, mostly table-based,
practical risk and auditing methodologies.

\section*{Acknowledgments}

The authors would like to thank Sjouke Mauw and Pieter Hartel for their 
comments on a preliminary version of this survey, which helped them to improve 
the paper.

\bibliographystyle{plain} 
\bibliography{arxiv}

\begin{thebibliography}{100}

\bibitem{CC}
ISO/IEC 15408.
\newblock Common criteria for information technology security evaluation
  (version 3.1, revision 4), 2012.

\bibitem{AbCeKa}
Parosh~Aziz Abdulla, Jonathan Cederberg, and Lisa Kaati.
\newblock {Analyzing the Security in the GSM Radio Network Using Attack
  Jungles}.
\newblock In Tiziana Margaria and Bernhard Steffen, editors, {\em ISoLA (1)},
  volume 6415 of {\em LNCS}, pages 60--74. Springer, 2010.

\bibitem{Accurate_Annual_Report}
{ACCURATE}.
\newblock {A Center for Correct Usable Reliable Auditable and Transparent
  Elections: Annual Report 2006}.
\newblock
  \url{http://accurate-voting.org/wp-content/uploads/2007/02/AR.2007.pdf},
  2007.
\newblock Accessed November 12, 2012.

\bibitem{AiBoDoFeGeKrLe}
Amer Aijaz, Bernd Bochow, Florian D\"{o}tzer, Andreas Festag, Matthias Gerlach,
  Rainer Kroh, and Tim Leinm\"{u}ller.
\newblock {Attacks on Inter Vehicle Communication Systems - an Analysis}.
\newblock In {\em 3rd International Workshop on Intelligent Transportation},
  pages 189--194, 2006.

\bibitem{Alex}
Ian Alexander.
\newblock {Misuse cases: Use cases with hostile intent}.
\newblock {\em IEEE software}, 20(1):58--66, 2003.

\bibitem{AlPa}
Qutaibah Althebyan and Brajendra Panda.
\newblock {A Knowledge-Based Bayesian Model for Analyzing a System after an
  Insider Attack}.
\newblock In Sushil Jajodia, Pierangela Samarati, and Stelvio Cimato, editors,
  {\em Proceedings of The Ifip Tc 11 23rd International Information Security
  Conference}, volume 278 of {\em IFIP International Federation for Information
  Processing}, pages 557--571. Springer Boston, 2008.
\newblock \url{10.1007/978-0-387-09699-5_36}.

\bibitem{Program1}
Amenaza.
\newblock {SecurITree}.
\newblock \url{http://www.amenaza.com/}, 2001--2013.
\newblock Accessed October 5, 2012.

\bibitem{AmWiKa}
Paul Ammann, Duminda Wijesekera, and Saket Kaushik.
\newblock {Scalable, graph-based network vulnerability analysis}.
\newblock In {\em Proceedings of the 9th {ACM Conference on Computer and
  ommunications Security (CCS'02)}}, pages 217--224, {Washington, DC, USA},
  November 2002.

\bibitem{Amor}
Edward~G. Amoroso.
\newblock {\em {Fundamentals of Computer Security Technology}}.
\newblock Prentice-Hall, Inc., Upper Saddle River, NJ, USA, 1994.

\bibitem{AnJuCe}
Xiangdong An, Dan Jutla, and Nick Cercone.
\newblock {Privacy intrusion detection using dynamic Bayesian networks}.
\newblock In {\em Proceedings of the 8th {International Conference for
  Electronic Commerce (ICEC'06)}}, pages 208--215, Fredericton, Canada, August
  2006.

\bibitem{Andefirst}
Ross~J. Anderson.
\newblock {\em {Security engineering - a guide to building dependable
  distributed systems}}.
\newblock Wiley, 1st edition, 2001.

\bibitem{Program4}
Alexander Andrusenko.
\newblock {AForest}.
\newblock \url{http://research.cyber.ee/~alexander/}, 2008.
\newblock Accessed October 5, 2012.

\bibitem{AndrusenkoMaster}
Alexander Andrusenko.
\newblock {R\"{u}ndepuude Metoodika Ja Seda Toetav Tarkvaraline Raamistik}.
\newblock Master's thesis, Tallinn University, 2010.
\newblock
  \url{http://www.cyber.ee/publikatsioonid/20-magistri-ja-doktoritood/loputoeoede-failid/Andrusenko-MA.pdf}.

\bibitem{aniketos}
{ANIKETOS}.
\newblock {ANIKETOS: Ensuring Trustworthiness and Security in Service
  Composition, FP7 project, grant agreement 257930}.
\newblock \url{http://www.aniketos.eu/}, 2010--2014.
\newblock Accessed January 19, 2013.

\bibitem{ArBySh}
Shanai Ardi, David Byers, and Nahid Shahmehri.
\newblock {Towards a structured unified process for software security}.
\newblock In {\em Proceedings of the 2006 international workshop on Software
  engineering for secure systems}, SESS '06, pages 3--10, New York, NY, USA,
  2006. ACM.

\bibitem{Arnb}
Stefan Arnborg.
\newblock {Efficient algorithms for combinatorial problems on graphs with
  bounded decomposability -- A survey}.
\newblock {\em BIT Numerical Mathematics}, 25:1--23, 1985.
\newblock 10.1007/BF01934985.

\bibitem{ATREES}
{ATREES}.
\newblock {Attack Trees, project funded by the Fonds National de la Recherche,
  Luxembourg under grants C08/IS/26 and PHD-09-167}.
\newblock \url{http://satoss.uni.lu/projects/atrees/}, 2009--2012.
\newblock Accessed March 4, 2013.

\bibitem{BaPe}
Dejan Baca and Kai Petersen.
\newblock {Prioritizing Countermeasures through the Countermeasure Method for
  Software Security (CM-Sec)}.
\newblock In Muhammad~Ali Babar, Matias Vierimaa, and Markku Oivo, editors,
  {\em PROFES}, volume 6156 of {\em LNIBP}, pages 176--190. Springer, 2010.

\bibitem{BaKoMeSc}
Alessandra Bagnato, Barbara Kordy, Per~H\aa{}kon Meland, and Patrick
  Schweitzer.
\newblock {Attribute Decoration of Attack--Defense Trees}.
\newblock {\em International Journal of Secure Software Engineering, Special
  Issue on Security Modeling}, 3(2):1--35, 2012.

\bibitem{BeSv}
Patrik Berander and Mikael Svahnberg.
\newblock {Evaluating two ways of calculating priorities in requirements
  hierarchies - An experiment on hierarchical cumulative voting}.
\newblock {\em Journal of Systems and Software}, 82(5):836--850, May 2009.

\bibitem{BiDaPe}
Stefano Bistarelli, Marco Dall'Aglio, and Pamela Peretti.
\newblock {Strategic Games on Defense Trees}.
\newblock In Theodosis Dimitrakos, Fabio Martinelli, Peter Y.~A. Ryan, and
  Steve~A. Schneider, editors, {\em FAST}, volume 4691 of {\em LNCS}, pages
  1--15. Springer, 2006.

\bibitem{BiFiPe}
Stefano Bistarelli, Fabio Fioravanti, and Pamela Peretti.
\newblock {Defense Trees for Economic Evaluation of Security Investments}.
\newblock In {\em ARES}, pages 416--423. IEEE Computer Society, 2006.

\bibitem{BiPeTr}
Stefano Bistarelli, Pamela Peretti, and Irina Trubitsyna.
\newblock {Analyzing Security Scenarios Using Defence Trees and Answer Set
  Programming}.
\newblock {\em Electron. Notes Theor. Comput. Sci.}, 197(2):121--129, 2008.

\bibitem{Bodl}
Hans~L. Bodlaender.
\newblock A linear time algorithm for finding tree-decompositions of small
  treewidth.
\newblock In {\em Proceedings of the twenty-fifth annual ACM symposium on
  Theory of computing}, STOC '93, pages 226--234, New York, NY, USA, 1993. ACM.

\bibitem{BoFeGi}
Silvia Bortot, Mario Fedrizzi, and Silvio Giove.
\newblock {Modelling fraud detection by attack trees and Choquet integral}.
\newblock DISA Working Papers 2011/09, Department of Computer and Management
  Sciences, University of Trento, Italy, August 2011.

\bibitem{Boui}
Marc Bouissou.
\newblock {A Generalization of Dynamic Fault Trees through Boolean logic Driven
  Markov Processes (BDMP)}.
\newblock In {\em Proceedings of the 16th European Safety and Reliability
  Conference {(ESREL'07)}}, Stavanger, Norway, June 2007.

\bibitem{BoBo}
Marc Bouissou and Jean-Louis Bon.
\newblock {A new formalism that combines advantages of fault-trees and Markov
  models: Boolean logic driven Markov processes}.
\newblock {\em Reliability Engineering \& System Safety}, 82(2):149--163,
  November 2003.

\bibitem{BrJa}
Sviatoslav Braynov and Murtuza Jadliwala.
\newblock {Representation and analysis of coordinated attacks}.
\newblock In {\em Proceedings of the 2003 {ACM} Workshop on {Formal Methods in
  Security Engineering (FMSE'03)}}, pages 43--51, {Washington, D.C., USA},
  2003.

\bibitem{BrPa}
Phillip~J. Brooke and Richard~F. Paige.
\newblock {Fault trees for security system design and analysis}.
\newblock {\em Computers \& Security}, 22(3):256--264, 2003.

\bibitem{BuPaUnPaWaSa}
Donald~L. Buckshaw, Gregory~S. Parnell, Willard~L. Unkenholz, Donald~L. Parks,
  James~M. Wallner, and O.~Sami Saydjari.
\newblock {Mission Oriented Risk and Design Analysis of Critical Information
  Systems}.
\newblock {\em Military Operations Research}, 10(2):19--38, 2005.

\bibitem{BuLaPrSaWi}
Ahto Buldas, Peeter Laud, Jaan Priisalu, M\"{a}rt Saarepera, and Jan Willemson.
\newblock {Rational Choice of Security Measures Via Multi-Parameter Attack
  Trees}.
\newblock In Javier L{\'o}pez, editor, {\em CRITIS}, volume 4347 of {\em LNCS},
  pages 235--248. Springer, 2006.

\bibitem{BuTr}
Ahto Buldas and Triinu M\"{a}gi.
\newblock {Practical Security Analysis of E-Voting Systems}.
\newblock In Miyaji et~al. \cite{DBLP:conf/iwsec/2007}, pages 320--335.

\bibitem{BuSt}
Ahto Buldas and Roman Stepanenko.
\newblock {Upper Bounds for Adversaries' Utility in Attack Trees}.
\newblock In Jens Grossklags and Jean~C. Walrand, editors, {\em GameSec},
  volume 7638 of {\em LNCS}, pages 98--117. Springer, 2012.

\bibitem{Buon2}
Alessandro Buoni.
\newblock {Fraud Detection: From Basic Techniques to a Multi-Agent Approach}.
\newblock In {\em Management and Service Science (MASS), 2010 International
  Conference on}, pages 1--4, August 2010.

\bibitem{Buon}
Alessandro Buoni.
\newblock {\em {Fraud Detection in the Banking Sector}}.
\newblock PhD thesis, {\AA{}bo Akademi University}, Finland, 2012.

\bibitem{BuFe}
Alessandro Buoni and Mario Fedrizzi.
\newblock {Consensual Dynamics and Choquet Integral in an Attack Tree-based
  Fraud Detection System}.
\newblock In Joaquim Filipe and Ana L.~N. Fred, editors, {\em ICAART (1)},
  pages 283--288. SciTePress, 2012.

\bibitem{BuFeMe}
Alessandro Buoni, Mario Fedrizzi, and J\'{o}zsef Mezei.
\newblock {A Delphi-Based Approach to Fraud Detection Using Attack Trees and
  Fuzzy Numbers}.
\newblock In {\em Proceeding of the IASK International Conferences}, pages
  21--28, 2010.

\bibitem{BuFeMe2}
Alessandro Buoni, Mario Fedrizzi, and J\'{o}zsef Mezei.
\newblock {Combining Attack Trees and Fuzzy Numbers in a Multi-Agent Approach
  to Fraud Detection}.
\newblock {\em International Journal of Electronic Business}, 9(3):186--202,
  2011.

\bibitem{ByArShDu}
David Byers, Shanai Ardi, Nahid Shahmehri, and Claudiu Duma.
\newblock {Modeling software vulnerabilities with vulnerability cause graphs}.
\newblock In {\em Proceedings of the International Conference on Software
  Maintenance (ICSM'06)}, pages 411--422, September 2006.

\bibitem{BySh3}
David Byers and Nahid Shahmehri.
\newblock {Design of a Process for Software Security}.
\newblock In {\em Second International Conference on Availability, Reliability
  and Security (ARES'07)}, pages 301--309, April 2007.

\bibitem{BySh2}
David Byers and Nahid Shahmehri.
\newblock {A Cause-Based Approach to Preventing Software Vulnerabilities}.
\newblock In {\em Proceedings of the Third International Conference on
  Availability, Reliability and Security (ARES'08)}, pages 276--283,
  Washington, DC, USA, 2008. IEEE Computer Society.

\bibitem{BySh}
David Byers and Nahid Shahmehri.
\newblock {Unified modeling of attacks, vulnerabilities and security
  activities}.
\newblock In {\em SESS '10: Proceedings of the 2010 ICSE Workshop on Software
  Engineering for Secure Systems}, pages 36--42, New York, NY, USA, 2010. ACM.

\bibitem{ByFrMi}
Eric~J. Byres, Matthew Franz, and Darrin Miller.
\newblock {The Use of Attack Trees in Assessing Vulnerabilities in SCADA
  Systems}.
\newblock {\em International Infrastructure Survivability Workshop}, {}:{},
  December 2004.

\bibitem{CaKiKi}
Giovanni Cagalaban, Taihoon Kim, and Seoksoo Kim.
\newblock {Improving SCADA control systems security with software vulnerability
  analysis}.
\newblock In {\em Proceedings of the 12th WSEAS international conference on
  Automatic control, modelling \&\#38; simulation}, ACMOS'10, pages 409--414,
  Stevens Point, Wisconsin, USA, 2010. World Scientific and Engineering Academy
  and Society (WSEAS).

\bibitem{Camtepe2006}
Seyit~Ahmet \c{C}amtepe and B\"{u}lent Yener.
\newblock {A Formal Method for Attack Modeling and Detection}.
\newblock Technical Report {TR-06-01}, Rensselaer Polytechnic Institute, {Troy,
  NY, USA}, 2006.

\bibitem{CaYe}
Seyit~Ahmet \c{C}amtepe and B\"{u}lent Yener.
\newblock {Modeling and detection of complex attacks}.
\newblock In {\em {Proceedings of the 3rd International Conference on Security
  and Privacy in Communications Networks (SecureComm 2007)}}, pages 234--243,
  Nice, France, September 2007.

\bibitem{ChHa}
Nicolas Chaufette and Tommie Haag.
\newblock {Vulnerability Cause Graphs: A Case of Study}.
\newblock \url{http://www.ida.liu.se/~TDDD17/oldprojects/2007/projects/3.pdf},
  2007.
\newblock Accessed October 5, 2012.

\bibitem{ClSiTyHa}
Kevin Clark, Ethan Singleton, Stephen Tyree, and John Hale.
\newblock {Strata-Gem: risk assessment through mission modeling}.
\newblock In {\em Proceedings of the 4th {ACM Workshop on Quality of Protection
  (QoP'08)}}, pages 51--58, {Alexandria, Virginia, USA}, October 2008.

\bibitem{ClTyDaHa}
Kevin Clark, Stephen Tyree, Jerald Dawkins, and John Hale.
\newblock {Qualitative and quantitative analytical techniques for network
  security assessment}.
\newblock In {\em Proceedings of the {5th IEEE Systems, Man and Cybernetics
  Information Assurance Workshop (IAW'04)}}, pages 321--328, {West Point, USA},
  June 2004.

\bibitem{CoCoFr}
Sean Convery, David Cook, and Matt Franz.
\newblock {An Attack Tree for the Border Gateway Protocol}.
\newblock {IETF Internet Draft:}
  \url{http://tools.ietf.org/html/draft-ietf-rpsec-bgpattack-00}, February
  2004.
\newblock Accessed October 5, 2012.

\bibitem{CoGrPa}
Robert Cowan, Michael Grimaila, and Raju Patel.
\newblock {Using Attack and Protection Trees to Evaluate Risk in an Embedded
  Weapon System}.
\newblock In {\em Proceedings of the 3rd International Conference on
  Information Warfare and Security ({ICIW 2008})}, pages 97--108, Omaha,
  Nebraska, USA, April 2008.

\bibitem{Dacier1994}
Marc Dacier.
\newblock {\em {Vers une \'evaluation quantitative de la s\'ecurit\'e
  informatique}}.
\newblock PhD thesis, {Laboratoire d'Analyse et d'Architecture des Syst\`emes
  du CNRS (LAAS)}, 1994.

\bibitem{DaDe}
Marc Dacier and Yves Deswarte.
\newblock {Privilege graph: An extension to the typed access matrix model}.
\newblock In Dieter Gollmann, editor, {\em ESORICS'1994}, volume 875 of {\em
  LNCS}, pages 319--334. Springer, 1994.

\bibitem{DaDeKa}
Marc Dacier, Yves Deswarte, and Mohamed Ka\^{a}niche.
\newblock {Models and tools for quantitative assessment of operational
  security}.
\newblock In Sokratis~K. Katsikas and Dimitris Gritzalis, editors, {\em SEC},
  volume~54 of {\em IFIP Conference Proceedings}, pages 177--186. Chapman {\&}
  Hall, 1996.

\bibitem{DaLaDa}
Kristopher Daley, Ryan Larson, and Jerald Dawkins.
\newblock {A Structural Framework for Modeling Multi-Stage Network Attacks}.
\newblock In {\em ICPP Workshops}, pages 5--10. IEEE Computer Society, 2002.

\bibitem{DaEdMiRa}
George~C. {Dalton II}, Kenneth~S. Edge, Robert~F. Mills, and Richard~A. Raines.
\newblock {Analysing security risks in computer and Radio Frequency
  Identification (RFID) networks using attack and protection trees}.
\newblock {\em International Journal of Security and Networks}, 5(2):87--95,
  2010.

\bibitem{DaMiCoRa}
George~C. {Dalton II}, Robert~F. Mills, John~M. Colombi, and Richard~A. Raines.
\newblock {Analyzing Attack Trees using Generalized Stochastic Petri Nets}.
\newblock In {\em Information Assurance Workshop, 2006 {IEEE}}, pages 116--123,
  West Point, NY, 2006.

\bibitem{DaKo}
Ram Dantu and Prakash Kolan.
\newblock {Risk Management Using Behavior Based Bayesian Networks}.
\newblock In Paul~B. Kantor, Gheorghe Muresan, Fred Roberts, Daniel~Dajun Zeng,
  Fei-Yue Wang, Hsinchun Chen, and Ralph~C. Merkle, editors, {\em ISI}, volume
  3495 of {\em LNCS}, pages 115--126. Springer, 2005.

\bibitem{DaKoAkLo}
Ram Dantu, Prakash Kolan, Robert Akl, and Kall Loper.
\newblock Classification of attributes and behavior in risk management using
  bayesian networks.
\newblock In {\em IEEE Intelligence and Security Informatics}, pages 71--74,
  2007.

\bibitem{DaKoCa}
Ram Dantu, Prakash Kolan, and Jo\ {a}o W.~Cangussu.
\newblock {Network risk management using attacker profiling}.
\newblock {\em Security and Communication Networks}, 2(1):83--96, 2009.

\bibitem{DaKaKo}
Ram Dantu, Kall Loper, and Prakash Kolan.
\newblock {Risk management using behavior based attack graphs}.
\newblock In {\em International Conference on Information Technology: Coding
  and Computing (ITCC'04)}, volume~1, pages 445--449, april 2004.

\bibitem{DaHa}
Jerald Dawkins and John Hale.
\newblock {A systematic approach to multi-stage network attack analysis}.
\newblock In {\em Proceedings of the 2nd {IEEE International Information
  Assurance Workshop (IAWA'04)}}, pages 48--56, {Charlotte, NC, USA}, April
  2004.

\bibitem{Website_Genie}
{Decision Systems Laboratory, University of Pittsburgh}.
\newblock {GeNIe \& SMILE}.
\newblock \url{http://genie.sis.pitt.edu/}, 1996--2013.
\newblock Accessed November 6, 2012.

\bibitem{Website_FTAP}
{Department of Engineering, University of Maryland}.
\newblock {Fault Tree Analysis Programs}.
\newblock \url{http://www.enre.umd.edu/tools/ftap.htm}, ca. 2004.
\newblock Accessed October 5, 2012.

\bibitem{DePoRaWh}
Rinku Dewri, Nayot Poolsappasit, Indrajit Ray, and Darrell Whitley.
\newblock {Optimal security hardening using multi-objective optimization on
  attack tree models of networks}.
\newblock In {\em Proceedings of the 14th ACM Conference on Computer and
  Communications Security}, CCS '07, pages 204--213, New York, NY, USA, 2007.
  ACM.

\bibitem{DeRaPoWh}
Rinku Dewri, Indrajit Ray, Nayot Poolsappasit, and Darrell Whitley.
\newblock {Optimal security hardening on attack tree models of networks: a
  cost-benefit analysis}.
\newblock {\em Int. J. Inf. Secur.}, 11(3):167--188, June 2012.

\bibitem{DiRoSiAlRi}
Mamadou~H. Diallo, Jose Romero-Mariona, Susan~E. Sim, Thomas~A. Alspaugh, and
  Debra~J. Richardson.
\newblock {A Comparative Evaluation of Three Approaches to Specifying Security
  Requirements}.
\newblock In {\em Proceedings of the 12th International Working Conference on
  {Requirements Engineering: Foundation for Software Quality (REFSQ 2006)}},
  Luxembourg, {Grand-Duchy of Luxembourg}, June 2006.

\bibitem{DuLiDuZh}
Suguo Du, Xiaolong Li, Junbo Du, and Haojin Zhu.
\newblock {An attack-and-defence game for security assessment in vehicular ad
  hoc networks}.
\newblock {\em Peer-to-Peer Networking and Applications}, {}:1--14, 2012.
\newblock 10.1007/s12083-012-0127-9.

\bibitem{Dugan1990}
Joanne~Bechta Dugan, Salvatore~J. Bavuso, and Mark~A. Boyd.
\newblock {Fault Trees and Sequence Dependencies}.
\newblock In {\em Proceedings of the Reliability and Maintainability Annual
  Symposium ({RAMS'90})}, pages 286--293, {Los Angeles, CA, USA}, January 1990.

\bibitem{DuBaBo}
Joanne~Bechta Dugan, Salvatore~J. Bavuso, and Mark~A. Boyd.
\newblock {Dynamic fault tree models for fault tolerant computer systems}.
\newblock {\em IEEE Transactions on Reliability}, 41(3):363--377, 1992.

\bibitem{DuSuCo}
Joanne~Bechta Dugan, Kevin~J. Sullivan, and David Coppit.
\newblock {Developing a Low-Cost, High-Quality Software Tool for Dynamic Fault
  Tree Analysis}.
\newblock {\em IEEE Transactions on Reliability}, 49(1):49--59, 2000.

\bibitem{KB3}
{EDF R \& D}.
\newblock {KB3 Platform tools}.
\newblock
  \url{http://research.edf.com/research-and-the-scientific-community/software/kb3-44337.html},
  2011--2012.
\newblock Accessed October 19, 2012.

\bibitem{Edge}
Kenneth~S. Edge.
\newblock {\em {A Framework for Analyzing and Mitigating the Vulnerabilities of
  Complex Systems via Attack and Protection Trees}}.
\newblock PhD thesis, Air Force Institute of Technology, Wright Patterson AFB,
  OH, USA, July 2007.

\bibitem{EdDaRaMi}
Kenneth~S. Edge, George~C. {Dalton II}, Richard~A. Raines, and Robert~F. Mills.
\newblock {Using Attack and Protection Trees to Analyze Threats and Defenses to
  Homeland Security}.
\newblock In {\em MILCOM}, pages 1--7. IEEE, 2006.

\bibitem{EdRaGrBaBeRe}
Kenneth~S. Edge, Richard~A. Raines, Michael Grimaila, Rusty Baldwin, Robert
  Bennington, and Christopher Reuter.
\newblock {The Use of Attack and Protection Trees to Analyze Security for an
  Online Banking System}.
\newblock In {\em 40th Annual Hawaii International Conference on System
  Sciences, 2007. (HICSS 2007)}, page 144b, January 2007.

\bibitem{EkSo}
Mathias Ekstedt and Teodor Sommestad.
\newblock {Enterprise architecture models for cyber security analysis}.
\newblock In {\em Proceedings of the {IEEE/PES Power System Conference and
  Exposition (PSCE'09)}}, pages 1--6, {Seattle, USA}, March 2009.

\bibitem{EoPaPaCh}
Jung-Ho Eom, Min-Woo Park, Seon-Ho Park, and Tai-Myoung Chung.
\newblock {A framework of defense system for prevention of insider's malicious
  behaviors}.
\newblock In {\em 13th International Conference on Advanced Communication
  Technology (ICACT'11)}, pages 982--987, February 2011.

\bibitem{Eric}
Clifton~A. {Ericson II}.
\newblock {Fault Tree Analysis - A History}.
\newblock In {\em Proceedings of the 17th International System Safety
  Conference {(ISSC'99)}}, {Orlando, FL, USA}, August 1999.

\bibitem{Espe}
Jeanne~H. Espedalen.
\newblock {Attack Trees Describing Security in Distributed Internet-Enabled
  Metrology}.
\newblock Master's thesis, Gj\o{}vik University, 2007.

\bibitem{EvHeKyPiWa}
Shelby Evans, David Heinbuch, Elizabeth Kyule, John Piorkowski, and James
  Wallner.
\newblock {Risk-based systems security engineering: stopping attacks with
  intention}.
\newblock {\em IEEE Security and Privacy}, 2(6):59--62, 2004.

\bibitem{Website_EVITA}
{EVITA}.
\newblock {E-safety vehicle intrusion protected applications: FP7 project,
  grant agreement 224275}.
\newblock \url{http://www.evita-project.org/}, 2008--2011.
\newblock Accessed October 5, 2012.

\bibitem{EzBe2010}
Barry~Charles Ezell, Steven~P. Bennett, Detlof {von Winterfeldt}, John
  Sokolowski, and Andrew~J. Collins.
\newblock {Probabilistic risk analysis and terrorism risk}.
\newblock {\em Risk analysis an official publication of the Society for Risk
  Analysis}, 30(4):575--589, 2010.

\bibitem{FeXi}
Nan Feng and Jing Xie.
\newblock {A Bayesian networks-based security risk analysis model for
  information systems integrating the observed cases with expert experience}.
\newblock {\em Scientific Research and Essays}, 7(10):1103--1112, 2012.

\bibitem{FeBaMoJi}
{Pl\'{\i}nio C\'{e}sar Sim\~{o}es} Fernandes, Tania Basso, Regina Moraes, and
  Mario Jino.
\newblock {Attack Trees Modeling for Security Tests in Web Applications}.
\newblock In {\em Brazilian Workshop on Systematic and Automated Software
  Testing}, pages 3--12, November 2010.

\bibitem{Fire}
Donald~J. Firesmith.
\newblock {Security Use Cases}.
\newblock {\em Journal of Object Technology}, 2(3):53--64, May 2003.

\bibitem{FoWuMaBaSp}
Bingrui Foo, Yu-Sung Wu, Yu-Chun Mao, Saurabh Bagchi, and Eugene Spafford.
\newblock {ADEPTS: adaptive intrusion response using attack graphs in an
  e-commerce environment}.
\newblock In {\em International Conference on Dependable Systems and Networks
  (DSN'05)}, pages 508--517, June-1 July 2005.

\bibitem{Fost}
Nathalie~L. Foster.
\newblock {\em {The application of software and safety engineering techniques
  to security protocol development}}.
\newblock PhD thesis, University of York, 2002.

\bibitem{FoMaCi}
Igor~Nai Fovino, Marcelo Masera, and Alessio~De Cian.
\newblock {Integrating cyber attacks within fault trees}.
\newblock {\em Reliability Engineering \& System Safety}, 94(9):1394--1402,
  September 2009.

\bibitem{FrSoEkJo}
Ulrik Franke, Teodor Sommestad, Mathias Ekstedt, and Pontus Johnson.
\newblock {Defense Graphs and Enterprise Architecture for Information Assurance
  Analysis}.
\newblock In {\em Proceedings of the 26th {Army Science Conference}}, {Orlando,
  Florida, USA}, December 2008.

\bibitem{FrWa}
Marcel Frigault and Lingyu Wang.
\newblock {Measuring Network Security Using Bayesian Network-Based Attack
  Graphs}.
\newblock In {\em Computer Software and Applications, 2008. COMPSAC '08. 32nd
  Annual IEEE International}, pages 698--703, 28 2008-aug. 1 2008.

\bibitem{FrWaSiJa}
Marcel Frigault, Lingyu Wang, Anoop Singhal, and Sushuil Jajodia.
\newblock {Measuring network security using dynamic Bayesian network}.
\newblock In {\em Proceedings of the 4th {ACM Workshop on Quality of Protection
  (QoP'08)}}, pages 23--30, {Alexandria, Virginia, USA}, October 2008.

\bibitem{FuChWaLeTaAnLi}
Case Fung, Yi-Liang Chen, Xinyu Wang, Joseph Lee, Richard Tarquini, Mark
  Anderson, and Richard Linger.
\newblock {Survivability Analysis of Distributed Systems Using Attack Tree
  Methodology}.
\newblock In {\em MILCOM}, pages 583--589, Atlantic City, NJ, 2005.

\bibitem{GrJo}
Lars Grunske and David Joyce.
\newblock {Quantitative risk-based security prediction for component-based
  systems with explicitly modeled attack profiles}.
\newblock {\em Journal of Systems and Software}, 81(8):1327--1345, 2008.

\bibitem{Harr}
Patrick~D. Harrington.
\newblock {\em {Noncooperative potential Games to improve network security}}.
\newblock PhD thesis, {Oklahoma State University}, USA, 2010.

\bibitem{HeWoSlHoMiLu}
Guy Helmer, Johnny Wong, Mark Slagell, Vasant Honavar, Les Miller, and Robyn
  Lutz.
\newblock {A Software Fault Tree Approach to Requirements Analysis of an
  Intrusion Detection System}.
\newblock {\em Journal of Requirements Engineering}, 7(4):207--220, December
  2002.

\bibitem{HeWoSlHoMiWaWaSt}
Guy Helmer, Johnny Wong, Mark Slagell, Vasant Honavar, Les Miller, Yanxin Wang,
  Xia Wang, and Natalia Stakhanova.
\newblock {Software fault tree and coloured Petri net-based specification,
  design and implementation of agent-based intrusion detection systems}.
\newblock {\em International Journal of Information and Computer Security},
  1(1/2):109--142, 2007.

\bibitem{HeApFuRoRuWe}
Olaf Henniger, Ludovic Apvrille, Andreas Fuchs, Yves Roudier, Alastair Ruddle,
  and Benjamin Weyl.
\newblock {Security requirements for automotive on-board networks}.
\newblock In {\em 9th International Conference on Intelligent Transport Systems
  Telecommunications,({ITST})}, pages 641--646, Lille, 2009.

\bibitem{HiUnJaSaLu}
Victoria Higuero, Juan~Jos\'e Unzilla, Eduardo Jacob, Purificaci\'on S\'aiz,
  and David Luengo.
\newblock {Application of `Attack Trees' Technique to Copyright Protection
  Protocols Using Watermarking and Definition of a New Transactions Protocol
  SecDP (Secure Distribution Protocol)}.
\newblock In {\em Proceedings of the 2nd International Workshop on Multimedia
  Interactive Protocols and Systems {(MIPS'04), LNCS 3311}}, pages 264--275,
  Grenoble, France, September 2004.

\bibitem{Hogg}
Ida Hogganvik.
\newblock {\em {A graphical approach to security risk analysis}}.
\newblock PhD thesis, Faculty of Mathematics and Natural Sciences, University
  of Oslo, 2007.

\bibitem{HoVaOuQu}
John Homer, Ashok Varikuti, Xinming Ou, and Miles~A. McQueen.
\newblock {Improving Attack Graph Visualization through Data Reduction and
  Attack Grouping}.
\newblock In {\em Proceedings of the 5th international {Workshop on
  Visualization For Computer Security (VizSEC'08)}}, pages 68--79, Cambridge,
  MA, USA, September 2008.

\bibitem{HoDo}
Viktor Horvath and Till D\"{o}rges.
\newblock {From security patterns to implementation using petri nets}.
\newblock In {\em Proceedings of the fourth international workshop on Software
  engineering for secure systems}, SESS '08, pages 17--24, New York, NY, USA,
  2008. ACM.

\bibitem{HoFrEn}
Siv~Hilde Houmb, Virginia N.~L. Franqueira, and Erlend~A. Engum.
\newblock {Quantifying security risk level from CVSS estimates of frequency and
  impact}.
\newblock {\em Journal of Systems and Software}, 83(9):1662--1634, 2009.

\bibitem{HoLe}
Michael Howard and David LeBlanc.
\newblock {\em {Writing Secure Code}}.
\newblock Microsoft Press, 2nd edition, 2002.

\bibitem{Ingo}
Terrance~R. Ingoldsby.
\newblock {Understanding Risk Through Attack Tree Analysis}.
\newblock {\em Computer Security Journal}, 20(2):33--59, 2004.

\bibitem{InLiPi}
Kyle~W. Ingols, Richard Lippmann, and Keith Piwowarski.
\newblock {Practical Attack Graph Generation for Network Defense}.
\newblock In {\em Proceedings of the 22nd {Annual Computer Security
  Applications Conference (ACSAC'06)}}, pages 121--130, {Washington, DC, USA},
  December 2006.

\bibitem{Elec}
{International Electrotechnical Commission (IEC)}.
\newblock {Analysis techniques for dependability -- Reliability block diagram
  and boolean methods}.
\newblock {IEC 61078}, January 2006.
\newblock 2nd Ed.

\bibitem{Elec2}
{International Electrotechnical Commission (IEC)}.
\newblock {Fault tree analysis}.
\newblock {IEC 61025}, December 2006.
\newblock 2nd Ed.

\bibitem{Program2}
Isograph.
\newblock {AttackTree+}.
\newblock \url{http://www.isograph-software.com/atpover.htm}, 2004--2005.
\newblock Accessed October 5, 2012.

\bibitem{Lout}
George Robert~Louthan IV.
\newblock {Hybrid Attack Graphs for Modeling Cyber-physical Systems}.
\newblock Master's thesis, University of Tulsa, USA, 2011.

\bibitem{JaNoOb}
Sushil Jajodia, Steven Noel, and Brian O'Berry.
\newblock {\em {Managing Cyber Threats: Issues, Approaches, and Challenges}},
  chapter {Topological Analysis of Network Attack Vulnerability}, pages
  247--266.
\newblock Springer US, 2005.
\newblock editor = {Kumar, Vipin and Srivastava, Jaideep and Lazarevic,
  Aleksandar}.

\bibitem{JeNi}
Finn~V. Jensen and Thomas~D. Nielsen.
\newblock {\em {Bayesian Networks and Decision Graphs}}.
\newblock Springer Publishing Company, Incorporated, 2nd edition, 2007.

\bibitem{John}
Chris~W. Johnson.
\newblock {Using Assurance Cases and {Boolean logic Driven Markov Processes} to
  Formalise Cyber Security Concerns for Safety-Critical Interaction with Global
  Navigation Satellite Systems}.
\newblock {\em ECEASST}, 45:1--18, 2011.

\bibitem{JoJoSoUl}
Pontus Johnson, Erik Johansson, Teodor Sommestad, and Johan Ullberg.
\newblock {A Tool for Enterprise Architecture Analysis}.
\newblock In {\em EDOC}, pages 142--156. IEEE Computer Society, 2007.

\bibitem{JoLaNaSi}
Pontus Johnson, Robert Lagerstr\"{o}m, Per N\"{a}rman, and M\aa{}rten
  Simonsson.
\newblock {Enterprise architecture analysis with extended influence diagrams}.
\newblock {\em Information Systems Frontiers}, 9(2-3):163--180, July 2007.

\bibitem{JuElBaRa}
Christian Jung, Frank Elberzhager, Alessandra Bagnato, and Fabio Raiteri.
\newblock {Practical Experience Gained from Modeling Security Goals: Using
  SGITs in an Industrial Project}.
\newblock In {\em International Conference on Availability, Reliability, and
  Security (ARES'10)}, pages 531--536, February 2010.

\bibitem{Jurg}
Aivo J\"{u}rgenson.
\newblock {\em {Efficient Semantics of Parallel and Serial Models of Attack
  Trees}}.
\newblock PhD thesis, {Tallinn University of Technology, Faculty of Information
  Technology, Department of Informatics}, 2010.
\newblock Available at \url{http://digi.lib.ttu.ee/i/?496}.

\bibitem{JuWi}
Aivo J\"{u}rgenson and Jan Willemson.
\newblock {Processing Multi-Parameter Attacktrees with Estimated Parameter
  Values}.
\newblock In Miyaji et~al. \cite{DBLP:conf/iwsec/2007}, pages 308--319.

\bibitem{JuWi3}
Aivo J\"{u}rgenson and Jan Willemson.
\newblock {Computing Exact Outcomes of Multi-parameter Attack Trees}.
\newblock In Robert Meersman and Zahir Tari, editors, {\em OTM Conferences
  (2)}, volume 5332 of {\em LNCS}, pages 1036--1051. Springer, 2008.

\bibitem{JuWi2}
Aivo J\"{u}rgenson and Jan Willemson.
\newblock {On Fast and Approximate Attack Tree Computations}.
\newblock In {\em Proceedings of the 6th international conference on
  Information Security Practice and Experience}, ISPEC'10, pages 56--66,
  Berlin, Heidelberg, 2010. Springer-Verlag.

\bibitem{KaSiOp}
P\'{e}ter K\'{a}p\'{a}rti, Guttorm Sindre, and Andreas~L. Opdahl.
\newblock {Visualizing cyber attacks with misuse case maps}.
\newblock In {\em Proceedings of the 16th International Working Conference on
  Requirements Engineering: Foundation for Software Quality ({REFSQ 2010})},
  pages 262--275, Essen, Germany, June 2010.

\bibitem{KarpatiIJSSE12}
P{\'e}ter K{\'a}rp{\'a}ti, Guttorm Sindre, and Raimundas Matulevicius.
\newblock {Comparing Misuse Case and Mal-Activity Diagrams for Modelling Social
  Engineering Attacks}.
\newblock {\em IJSSE}, 3(2):54--73, 2012.

\bibitem{Karp}
Kaarina Karppinen.
\newblock {Security Measurement Based on Attack Trees in a Mobile Ad Hoc
  Network Environment}.
\newblock Master's thesis, {VTT and University of Oulu}, 2005.
\newblock Available at
  \url{http://www.vtt.fi/inf/pdf/publications/2005/P580.pdf}.

\bibitem{Katta2010}
Vikash Katta, P{\'e}ter K{\'a}rp{\'a}ti, Andreas~L. Opdahl, Christian
  Raspotnig, and Guttorm Sindre.
\newblock {Comparing Two Techniques for Intrusion Visualization}.
\newblock In Patrick van Bommel, Stijn Hoppenbrouwers, Sietse Overbeek, Erik
  Proper, and Joseph Barjis, editors, {\em PoEM}, volume~68 of {\em Lecture
  Notes in Business Information Processing}, pages 1--15. Springer, 2010.

\bibitem{Khan}
Parvaiz~Ahmed Khand.
\newblock {System level security modeling using attack trees}.
\newblock In {\em Proceedings of the 2nd {International Conference on Computer,
  Control and Communication (IC4)}}, pages 1--6, Karachi, Pakistan, February
  2009.

\bibitem{KhSe}
Parvaiz~Ahmed Khand and Poong~Hyun Seong.
\newblock {An Attack model development process for the Cyber Security of Safety
  Related Nuclear Digital {I\&C} Systems}.
\newblock In {\em Proceedings of the {Korean Nucleary Society (KNS) Fall
  meeting}}, Korea, October 2007.

\bibitem{Kien}
Darrell~M. Kienzle.
\newblock {\em {Practical Computer Security Analysis}}.
\newblock PhD thesis, School of Engineering and Applied Science, University of
  Virginia, USA, 1998.

\bibitem{KiWu}
Darrell~M. Kienzle and William~A. Wulf.
\newblock {A Practical Approach to Security Assessment}.
\newblock In {\em Proceedings of the 1997 New Security Paradigms Workshop},
  NSPW '97, pages 5--16, New York, NY, USA, 1997. ACM.

\bibitem{KlElEs}
Johannes Kloos, Frank Elberzhager, and Robert Eschbach.
\newblock {Systematic Construction of Goal Indicator Trees for Indicator-Based
  Dependability Inspections}.
\newblock In {\em 36th EUROMICRO Conference on Software Engineering and
  Advanced Applications (SEAA'10)}, pages 279--282, September 2010.

\bibitem{Koot}
Laurens Koot.
\newblock {\em {Security of mobile TAN on smartphones}}.
\newblock PhD thesis, Radboud University Nijmegen, Faculty of Science, The
  Netherlands, 2012.

\bibitem{KoKoMaSc}
Barbara Kordy, Piotr Kordy, Sjouke Mauw, and Patrick Schweitzer.
\newblock {ADTool: Security Analysis Using Attack--Defense Trees}, 2013.
\newblock submitted.

\bibitem{KoMaMeSc}
Barbara Kordy, Sjouke Mauw, Matthijs Melissen, and Patrick Schweitzer.
\newblock {Attack--Defense Trees and Two-Player Binary Zero-Sum Extensive Form
  Games Are Equivalent}.
\newblock In Tansu Alpcan, Levente Butty{\'a}n, and John~S. Baras, editors,
  {\em GameSec}, volume 6442 of {\em LNCS}, pages 245--256. Springer, 2010.

\bibitem{KoMaRaSc}
Barbara Kordy, Sjouke Mauw, Sa\v{s}a Radomirovi\'c, and Patrick Schweitzer.
\newblock {Foundations of Attack--Defense Trees}.
\newblock In Pierpaolo Degano, Sandro Etalle, and Joshua~D. Guttman, editors,
  {\em FAST}, volume 6561 of {\em LNCS}, pages 80--95. Springer, 2010.

\bibitem{KoMaRaSc2}
Barbara Kordy, Sjouke Mauw, Sa\v{s}a Radomirovi\'c, and Patrick Schweitzer.
\newblock {Attack--Defense Trees}.
\newblock {\em Journal of Logic and Computation}, 2012.
\newblock Available at
  \url{http://logcom.oxfordjournals.org/content/early/2012/06/21/logcom.exs029
  }.

\bibitem{KoMaSc}
Barbara Kordy, Sjouke Mauw, and Patrick Schweitzer.
\newblock {Quantitative Questions on Attack--Defense Trees}.
\newblock In {\em ICISC}, volume 7839 of {\em LNCS}, pages 49--64. Springer,
  2012.

\bibitem{KoPoSc}
Barbara Kordy, Marc Pouly, and Patrick Schweitzer.
\newblock {Computational Aspects of Attack--Defense Trees}.
\newblock In {\em Security \& Intelligent Information Systems}, volume 7053 of
  {\em LNCS}, pages 103--116. Springer, 2011.

\bibitem{ADTool}
Piotr Kordy and Patrick Schweitzer.
\newblock {ADTool}.
\newblock \url{http://satoss.uni.lu/projects/atrees/adtool}, 2012.
\newblock Accessed March 1st, 2013.

\bibitem{KoSt}
Igor Kotenko and Mikhail Stepashkin.
\newblock {Analyzing Network Security using Malefactor Action Graphs}.
\newblock {\em International Journal of Computer Science and Network Security},
  6(6):226--235, 2006.

\bibitem{KriBoPi}
Siwar Kriaa, Marc Bouissou, and Ludovic Pi\`{e}tre-Cambac\'{e}d\`{e}s.
\newblock {Modeling the Stuxnet Attack with BDMP: Towards More Formal Risk
  Assessments}.
\newblock In {\em {Proceedings of the 7th International Conference on Risks and
  Security of Internet and Systems (CRiSIS 2012)}}, Cork, Ireland, October
  2012.

\bibitem{KuSp}
Sandeep Kumar and Eugene~H. Spafford.
\newblock {A Pattern-Matching Model for Misuse Intrusion Detection}.
\newblock In {\em Proceedings of the 17th National Computer Security Conference
  {(NCSC'94)}}, pages 11--21, {Baltimore, USA}, October 1994.

\bibitem{LaJoNa}
Robert Lagerstr{\"o}m, Pontus Johnson, and Per N{\"a}rman.
\newblock {Extended Influence Diagram Generation}.
\newblock In Ricardo Jardim-Gon\c{c}alves, J{\"o}rg~P. M{\"u}ller, Kai Mertins,
  and Martin Zelm, editors, {\em IESA}, pages 599--602. Springer, 2007.

\bibitem{Program3}
Eric~L. Lazarus.
\newblock {AttackDog}.
\newblock \url{https://decisionsmith.com/doc/adog}, 2010--2011.
\newblock Accessed July 21, 2010.

\bibitem{LaDiEpHa}
Eric~L. Lazarus, David~L. Dill, Jeremy Epstein, and Joseph~Lorenzo Hall.
\newblock {Applying a Reusable Election Threat Model at the County Level}.
\newblock In {\em Proceedings of the 2011 Conference on Electronic voting
  Technology/Workshop on Trustworthy Elections}, EVT/WOTE'11, pages 1--14,
  Berkeley, CA, USA, August 2011. USENIX Association.

\bibitem{LeBy2}
David~John Leversage and Eric~James Byres.
\newblock {Comparing Electronic Battlefields: Using Mean Time-To-Compromise as
  a Comparative Security Metric}.
\newblock In {\em Proceedings of the 4th International Conference on Methods,
  Models, and Architectures for Network Security ({MMM-ACNS'07}), {CCIS 1}},
  pages 213--227, St Petersburg, Russia, September 2007.

\bibitem{LeBy}
David~John Leversage and Eric~James Byres.
\newblock {Estimating a System's Mean Time-to-Compromise}.
\newblock {\em IEEE Security and Privacy}, 6:52--60, January 2008.

\bibitem{Leve}
Nancy~G. Leveson.
\newblock {\em {Safeware: System Safety and Computers}}.
\newblock Addison-Wesley Professional, April 1995.

\bibitem{LeHa}
Nancy~G. Leveson and Peter~R. Harvey.
\newblock Software fault tree analysis.
\newblock {\em Journal of Systems and Software}, 3(2):173--181, 1983.

\bibitem{LiLiFeHe}
Xiaohong Li, Ran Liu, Zhiyong Feng, and Ke~He.
\newblock {Threat modeling-oriented attack path evaluating algorithm}.
\newblock {\em Transactions of Tianjin University}, 15(3):162--167, 2009.

\bibitem{LiZaRuLi}
Xiaoli Lin, Pavol Zavarsky, Ron Ruhl, and Dale Lindskog.
\newblock {Threat Modeling for CSRF Attacks}.
\newblock In {\em International Conference on Computational Science and
  Engineering (CSE'09)}, volume~3, pages 486--491, August 2009.

\bibitem{LiMo}
Richard~C. Linger and Andrew~P. Moore.
\newblock {Foundations for Survivable System Development: Service Traces,
  Intrusion Traces, and Evaluation Models}, 2001.
\newblock Accessed October 5, 2012.

\bibitem{LiIn}
Richard Lippmann and Kyle~W. Ingols.
\newblock {An annotated review of past papers on attack graphs}.
\newblock Project Report {ESC-TR-2005-054}, {Massachusetts Institute of
  Technology (MIT), Lincoln Laboratory}, March 2005.

\bibitem{LiMa}
Yuan Liu and Hong Man.
\newblock {Network vulnerability assessment using Bayesian networks}.
\newblock In {\em Proceedings of {SPIE} Data Mining, Intrusion Detection,
  Information Assurance, and Data Networks Security 2005}, volume 5812, pages
  61--71, Orlando, FL, USA, March 2005.

\bibitem{Magi}
Triinu M\"{a}gi.
\newblock {Practical Security Analysis of E-voting Systems}.
\newblock Master's thesis, Tallin University of Technology, Faculty of
  Information Technology, Department of Informatics, Estonia, 2007.
\newblock Available at \url{http://triinu.net/e-voting/}.

\bibitem{MaBaGh}
Samresh Malhotra, Somak Bhattacharya, and S.~K. Ghosh.
\newblock {A Vulnerability and Exploit Independent Approach for Attack Path
  Prediction}.
\newblock In {\em Proceedings of the IEEE 8th International Conference on
  Computer and Information Technology Workshops}, pages 282--287, Sydney,
  Australia, July 2008.

\bibitem{MaCaMoArBySh}
Amel Mammar, Ana Cavalli, Edgardo {Montes de Oca}, Shanai Ardi, David Byers,
  and Nahid Shahmehri.
\newblock Mod\'{e}lisation et d\'{e}tection formelles de vuln\'{e}rabilit\'{e}s
  logicielles par le test passif.
\newblock In {\em 4\`{e}me Conf\'{e}rence sur la S\'{e}curit\'{e} des
  Architectures R\'{e}seaux et des Syst\`{e}mes d'Information (SAR-SSI)}, page
  12pp, June 2009.

\bibitem{Mana}
Pratyusa~K. Manadhata.
\newblock {\em {An Attack Surface Metric}}.
\newblock PhD thesis, Carnegie Mellon University, December 2008.

\bibitem{MaThFe}
Theodore~W. Manikas, Mitchell~A. Thornton, and David~Y. Feinstein.
\newblock {Using Multiple-Valued Logic Decision Diagrams to Model System Threat
  Probabilities}.
\newblock {\em IEEE International Symposium on Multiple-Valued Logic},
  0:263--267, 2011.

\bibitem{MaDoHeKoXu}
Aaron Marback, Do~Hyunsook, Ke~He, Samuel Kondamarri, and Dianxiang Xu.
\newblock {Security test generation using threat trees}.
\newblock In {\em Automation of Software Test, 2009. AST '09. ICSE Workshop
  on}, pages 62--69, May 2009.

\bibitem{Mars}
Charles Marshall.
\newblock {Attack Trees and Their Uses in BGP and SMTP Analysis}.
\newblock
  \url{http://citeseerx.ist.psu.edu/viewdoc/summary?doi=10.1.1.122.3609}, 2008.
\newblock Accessed October 5, 2012.

\bibitem{MaFoCi}
Marcelo Masera, Igor~Nai Fovino, and Alessio~De Cian.
\newblock {Integrating cyber attacks within fault trees}.
\newblock In Terje Aven and Jan~Erik Vinnem, editors, {\em Risk, Reliability
  and Societal Safety (Proceedings of the 16th European Safety and Reliability
  Conference (ESREL'07)}, pages 1--8, London, 2007. Taylor \& Francis Group.

\bibitem{MaHo}
Jim~E. Matheson and Ron~A. Howard.
\newblock {\em {An Introduction to Decision Analysis}}.
\newblock Strategic Decisions Group, Menlo Park, CA, 1968.

\bibitem{MaOo}
Sjouke Mauw and Martijn Oostdijk.
\newblock {Foundations of Attack Trees}.
\newblock In Dongho Won and Seungjoo Kim, editors, {\em ICISC}, volume 3935 of
  {\em LNCS}, pages 186--198. Springer, 2005.

\bibitem{McDe}
John~P. McDermott.
\newblock {Attack net penetration testing}.
\newblock In {\em Proceedings of the 2000 {Workshop on New Security Paradigms
  (NSPW'00)}}, pages 15--21, Cork, Ireland, September 2000.

\bibitem{DeFo}
John~P. McDermott and Chris Fox.
\newblock {Using abuse case models for security requirements analysis}.
\newblock In {\em Proceedings of the 15th {Annual Computer Security
  Applications Conference (ACSAC'99)}}, pages 55--64, {Phoenix, USA}, December
  1999.

\bibitem{LaPoDa}
Stephen McLaughlin, Dmitry Podkuiko, and Patrick McDaniel.
\newblock {Energy theft in the advanced metering infrastructure}.
\newblock In {\em Proceedings of the 4th International Conference on Critical
  Information Infrastructures Security}, CRITIS'09, pages 176--187, Berlin,
  Heidelberg, 2010. Springer-Verlag.

\bibitem{LaPoMiDeDa}
Stephen McLaughlin, Dmitry Podkuiko, Sergei Miadzvezhanka, Adam Delozier, and
  Patrick McDaniel.
\newblock {Multi-vendor penetration testing in the advanced metering
  infrastructure}.
\newblock In {\em {Proceedings of the 26th Annual Computer Security
  Applications Conference (ACSAC'10)}}, pages 107--116, Austin, Texas, USA,,
  December 2010.

\bibitem{McQueen2005}
Miles~A. McQueen, Wayne~F. Boyer, Mark~A. Flynn, and George~A. Beitel.
\newblock Time-to-compromise model for cyber risk reduction estimation.
\newblock In {\em Proceedings of the 1st Workshop on {Quality of Protection
  (QoP'05)}}, pages 49--64, Milan, Italy, September 2005.

\bibitem{QuBoFlBe}
Miles~A. McQueen, Wayne~F. Boyer, Mark~A. Flynn, and George~A. Beitel.
\newblock {Quantitative Cyber Risk Reduction Estimation Methodology for a Small
  SCADA Control System}.
\newblock In {\em Proceedings of the 39th Annual {Hawaii International
  Conference on System Sciences (HICSS-39)}}, volume~9, pages 226--237,
  {Hawaii, USA}, January 2006.

\bibitem{MeHoSt}
Nancy~R. Mead, Eric~D. Hough, and Theodore~R. {Stehney II}.
\newblock {Security Quality Requirements Engineering (SQUARE) Methodology}.
\newblock Technical Report {CMU/SEI-2005-TR-009}, Carnegie Mellon University,
  2005.

\bibitem{Mead2}
Catherine Meadows.
\newblock {A representation of Protocol Attacks for Risk Assessment}.
\newblock In {\em Proceedings of the {DIMACS Workshop on Network Threats}},
  pages 1--10, {New Brunswick, NJ, USA}, December 1996.

\bibitem{MeBaZhClWi}
Vaibhav Mehta, Constantinos Bartzis, Haifeng Zhu, Edmund Clarke, and Jeannette
  Wing.
\newblock {Ranking Attack Graphs}.
\newblock In {\em Proceedings of the 9th International Symposium on {Recent
  Advances in Intrusion Detection (RAID'06), LNCS 4219}}, pages 127--144,
  Hamburg, Germany, September 2006.

\bibitem{Program6}
Per~H\aa{}kon Meland.
\newblock {SeaMonster}, 2007--2010.
\newblock Accessed October 5, 2012.

\bibitem{MeToJe}
Per~H\aa{}kon Meland, Inger~Anne T{\o}ndel, and Jostein Jensen.
\newblock {Idea: Reusability of Threat Models - Two Approaches with an
  Experimental Evaluation}.
\newblock In {\em International Symposium on {Engineering Secure Software and
  Systems(ESSoS)}}, pages 114--122, Pisa, Italy, February 2010.

\bibitem{MeSpHaBaKrVe}
Per~H{\aa}kon Meland, Daniele~Giuseppe Spampinato, Eilev Hagen, Egil~Trygve
  Baadshaug, Kris-Mikael Krister, and Ketil~Sandanger Velle.
\newblock {SeaMonster: Providing tool support for security modeling}.
\newblock In {\em Norsk Informasjonssikkerhetskonferanse (NISK'08)}, 2008.

\bibitem{MiMu}
Drake~Patrick Mirembe and Maybin Muyeba.
\newblock {Threat Modeling Revisited: Improving Expressiveness of Attack}.
\newblock In {\em EMS '08: Proceedings of the 2008 Second UKSIM European
  Symposium on Computer Modeling and Simulation}, pages 93--98, Washington, DC,
  USA, 2008. IEEE Computer Society.

\bibitem{MiKaYa}
Shivani Mishra, Krishna Kant, and R.~S. Yadav.
\newblock Multi tree view of complex attack -- stuxnet.
\newblock In {\em Proceedings of the {ACITY 2012} Conference}, pages 171--188,
  Chennai, India, July 2012.

\bibitem{DBLP:conf/iwsec/2007}
Atsuko Miyaji, Hiroaki Kikuchi, and Kai Rannenberg, editors.
\newblock {\em Advances in Information and Computer Security, Second
  International Workshop on Security, IWSEC 2007, Nara, Japan, October 29-31,
  2007, Proceedings}, volume 4752 of {\em LNCS}. Springer, 2007.

\bibitem{Mobe}
Fredrik Moberg.
\newblock {Security Analysis of an Information System Using an Attack
  Tree-based Methodology}.
\newblock Master's thesis, Chalmers University of Technology, 2000.

\bibitem{MoElLi}
Andrew~P. Moore, Robert~J. Ellison, and Richard~C. Linger.
\newblock {Attack Modeling for Information Security and Survivability}.
\newblock Technical Note {CMU/SEI-2001-TN-001}, Carnegie Mellon University,
  March 2001.

\bibitem{MoCaMa}
Anderson Morais, Ana Cavalli, and Eliane Martins.
\newblock {A Model-Based Attack Injection Approach for Security Validation}.
\newblock In {\em Proceedings of the 4th International Conference on Security
  of Information and Networks}, SIN '11, pages 103--110, New York, NY, USA,
  2011. ACM.

\bibitem{MoMaCaJi}
Anderson Nunes~Paiva Morais, Eliane Martins, Ana~R. Cavalli, and Willy Jimenez.
\newblock {Security Protocol Testing Using Attack Trees}.
\newblock In {\em CSE (2)}, pages 690--697. IEEE Computer Society, 2009.

\bibitem{MoYa}
Ikuya Morikawa and Yuji Yamaoka.
\newblock {Threat Tree Templates to Ease Difficulties in Threat Modeling}.
\newblock In {\em 14th International Conference on Network-Based Information
  Systems (NBiS'11)}, pages 673--678, September 2011.

\bibitem{MoKa}
Ira~S. Moskowitz and Myong~H. Kang.
\newblock {An insecurity flow model}.
\newblock In {\em Proceedings of the 1997 {Workshop on New Security Paradigms
  (NSPW'97)}}, pages 61--74, {Langdale, Cumbria, UK}, September 1997.

\bibitem{NaJoLaFrEk}
Per N\"{a}rman, Pontus Johnson, Robert Lagerstr\"{o}m, Ulrik Franke, and
  Mathias Ekstedt.
\newblock {Data Collection Prioritization for System Quality Analysis}.
\newblock {\em Electron. Notes Theor. Comput. Sci.}, 233:29--42, March 2009.

\bibitem{Neap}
Richard~E. Neapolitan.
\newblock {\em {Learning Bayesian Networks}}.
\newblock Prentice Hall, 2003.

\bibitem{Niel}
Jason~R. Nielsen.
\newblock {Evaluating Information Assurance Control Effectiveness on An Air
  Force Supervisory Control And Data Acquisition ({SCADA}) System}.
\newblock Master's thesis, US Air Force Institute of Technology, March 2011.
\newblock Available at \url{
  http://oai.dtic.mil/oai/oai?verb=getRecord&metadataPrefix=html&identifier=ADA541615}.

\bibitem{Niit}
Margus Niitsoo.
\newblock {Optimal adversary behavior for the serial model of financial attack
  trees}.
\newblock In {\em Proceedings of the 5th International Conference on Advances
  in Information and Computer Security}, IWSEC'10, pages 354--370, Berlin,
  Heidelberg, 2010. Springer-Verlag.

\bibitem{NiXiYoSi}
Zhu Ning, Chen Xin-yuan, Zhang Yong-fu, and Xin Si-yuan.
\newblock {Design and Application of Penetration Attack Tree Model Oriented to
  Attack Resistance Test}.
\newblock In {\em International Conference on Computer Science and Software
  Engineering}, volume~3, pages 622--626, December 2008.

\bibitem{NoElJaKaOhPr}
Steven Noel, Matthew Elder, Sushil Jajodia, Pramod Kalapa, Scott O'Hare, and
  Kenneth Prole.
\newblock {Advances in Topological Vulnerability Analysis}.
\newblock In {\em Proceedings of the 2009 Cybersecurity Applications \&
  Technology Conference for Homeland Security}, CATCH '09, pages 124--129,
  Washington, DC, USA, 2009. IEEE Computer Society.

\bibitem{NoJaKaJa}
Steven Noel, Michael Jacobs, Pramod Kalapa, and Sushil Jajodia.
\newblock {Multiple coordinated views for network attack graphs}.
\newblock In {\em Proceedings of the 2005 IEEE Workshop on Visualization for
  Computer Security (VizSEC 05)}, pages 99--106, {Minneapolis, USA}, October
  2005.

\bibitem{NoJa}
Steven Noel and Sushil Jajodia.
\newblock {Managing attack graph complexity through visual hierarchical
  aggregation}.
\newblock In {\em Proceedings of the 2004 {ACM workshop on Visualization and
  data mining for computer security (VizSEC'04)}}, pages 109--118, {George
  Mason University, Fairfax, VA, USA}, October 2004.

\bibitem{NoJaObJa}
Steven Noel, Sushil Jajodia, Brian O'Berry, and Michael Jacobs.
\newblock {Efficient Minimum-cost Network Hardening via Exploit Dependency
  Graphs}.
\newblock In {\em Proceedings of the 19th {Annual Computer Security
  Applications Conference (ACSAC'03)}}, pages 86--95, {Las Vegas, NV, USA},
  December 2003.

\bibitem{NoJaWaSi}
Steven Noel, Sushil Jajodia, Lingyu Wang, and Anoop Singhal.
\newblock {Measuring Security Risk of Networks Using Attack Graphs}.
\newblock {\em IJNGC}, 1(1):135--147, 2010.

\bibitem{OnTuThNaSzMa}
Poramate Ongsakorn, Kyle Turney, Mitchell~A. Thornton, Suku Nair, Stephen~A.
  Szygenda, and Theodore Manikas.
\newblock {Cyber threat trees for large system threat cataloging and analysis}.
\newblock In {\em 4th Annual IEEE Systems Conference}, pages 610--615, April
  2010.

\bibitem{OpSi}
Andreas~L. Opdahl and Guttorm Sindre.
\newblock {Experimental comparison of attack trees and misuse cases for
  security threat identification}.
\newblock {\em Information \& Software Technology}, 51(5):916--932, 2009.

\bibitem{Opel}
Alexander Opel.
\newblock {Design and Implementation of a Support Tool for Attack Trees}.
\newblock Master's thesis, Technische Universiteit Eindhoven, Otto-von-Guericke
  University, Magdeburg, Germany, March 2005.

\bibitem{Ostl}
Ryan~T. Ostler.
\newblock {Defensive Cyber Battle Damage Assessment through Attack Methodology
  Modeling}.
\newblock Master's thesis, Air Force Institute of Technology, Department of
  Electrical and Computer Engineering, USA, 2011.

\bibitem{OuBoQu}
Xinming Ou, Wayne~F. Boyer, and Miles~A. McQueen.
\newblock {A scalable approach to attack graph generation}.
\newblock In {\em Proceedings of the 13th {ACM} conference on Computer and
  Communications Security ({CCS'06})}, pages 336--345, Alexandria, Virginia,
  {USA}, November 2006.

\bibitem{OuGoAp}
Xinming Ou, Sudhakar Govindavajhala, and Andrew~W. Appel.
\newblock {MulVAL: A logic-based network security analyzer}.
\newblock In {\em 14th USENIX Security Symposium}, pages 113--128, 2005.

\bibitem{PaLeChKiLeKw}
Gee-Yong Park, Cheol~Kwon Lee, Jong~Gyun Choi, Dong~Hoon Choi, Young~Jun Lee,
  and Kee-Choon Kwon.
\newblock {Cyber Security Analysis by Attack Trees for a Reactor Protection
  System}.
\newblock In {\em Proceedings of the Korean Nuclear Society (KNS) Fall
  Meeting}, Pyeong Chang, Korea, October 2008.

\bibitem{PaGrRa}
Sandip~C. Patel, James~H. Graham, and Patricia A.~S. Ralston.
\newblock {Quantitatively assessing the vulnerability of critical information
  systems: A new method for evaluating security enhancements}.
\newblock {\em International Journal of Information Management},
  28(6):483--491, December 2008.

\bibitem{Pear2}
Judea Pearl.
\newblock {Fusion, propagation, and structuring in belief networks}.
\newblock {\em Artificial Intelligence}, 29(3):241--288, 1986.

\bibitem{Pear}
Judea Pearl.
\newblock {\em {Probabilistic Reasoning in Intelligent Systems: Networks of
  Plausible Inference}}.
\newblock Morgan Kaufmann, 1988.

\bibitem{PeJaMa}
Holger Peine, Marek Jawurek, and Stefan Mandel.
\newblock {Security Goal Indicator Trees: A Model of Software Features that
  Supports Efficient Security Inspection}.
\newblock In {\em HASE '08: Proceedings of the 2008 11th IEEE High Assurance
  Systems Engineering Symposium}, pages 9--18, Washington, DC, USA, 2008. IEEE
  Computer Society.

\bibitem{KaSiOp2}
{P\'{e}ter K\'{a}p\'{a}rti and Guttorm Sindre and Andreas L. Opdahl}.
\newblock {Towards a Hacker Attack Representation Method}.
\newblock In {\em Proceedings of the 5th {ICSOFT} Conference}, pages 92--101,
  2010.

\bibitem{PhSw}
Cynthia Phillips and Laura~Painton Swiler.
\newblock {A graph-based system for network-vulnerability analysis}.
\newblock In {\em Proceedings of the 1998 {Workshop on New Security Paradigms
  (NSPW'98)}}, pages 71--79, {Charlottesville, Virginia, USA}, September 1998.

\bibitem{Piet}
Ludovic Pi\`{e}tre-Cambac\'{e}d\`{e}s.
\newblock {\em {Des relations entre s\^{u}ret\'e et s\'{e}curit\'{e}}}.
\newblock PhD thesis, {T\'el\'ecom ParisTech}, 2010.

\bibitem{PiBo3}
Ludovic Pi\`{e}tre-Cambac\'{e}d\`{e}s and Marc Bouissou.
\newblock {The promising potential of the BDMP formalism for security
  modeling}.
\newblock In {\em Proceedings of the 39th Annual {IEEE/IFIP} International
  Conference on Dependable Systems and Networks ({DSN 2009}), Supplemental
  Volume}, Estoril, Portugal, June 2009.
\newblock Fast Abstract track.

\bibitem{PiBo2}
Ludovic Pi\`{e}tre-Cambac\'{e}d\`{e}s and Marc Bouissou.
\newblock {Attack and Defense Modeling with BDMP}.
\newblock In Igor Kotenko and Victor Skormin, editors, {\em Computer Network
  Security}, volume 6258 of {\em LNCS}, pages 86--101. Springer, 2010.

\bibitem{PiBo6}
Ludovic Pi\`{e}tre-Cambac\'{e}d\`{e}s and Marc Bouissou.
\newblock {Beyond attack trees: dynamic security modeling with Boolean logic
  Driven Markov Processes (BDMP)}.
\newblock In {\em Proceedings of the {8th European Dependable Computing
  Conference (EDCC-8)}}, pages 199--208, Valencia, Spain, April 2010.

\bibitem{PiBo5}
Ludovic Pi\`{e}tre-Cambac\'{e}d\`{e}s and Marc Bouissou.
\newblock {Modeling safety and security interdepedencies with BDMP (Boolean
  logic Driven Markov Processes)}.
\newblock In {\em {IEEE} International Conference on Systems, Man, and
  Cybernetics ({SMC 2010})}, pages 2852--2861, Istanbul, Turkey, October 2010.

\bibitem{PiDeBo}
Ludovic Pi\`{e}tre-Cambac\'{e}d\`{e}s, Yann Deflesselle, and Marc Bouissou.
\newblock {Security modeling with BDMP: from theory to implementation}.
\newblock In {\em 6th {IEEE} International Conference on Network and
  Information Systems Security ({SAR-SSI 2011})}, pages 1--8, La Rochelle,
  France, May 2011.

\bibitem{PoRa}
Nayot Poolsapassit and Indrajit Ray.
\newblock {Investigating Computer Attacks Using Attack Trees}.
\newblock In Philip Craiger and Sujeet Shenoi, editors, {\em Advances in
  Digital Forensics III}, volume 242 of {\em IFIP International Federation for
  Information Processing}, pages 331--343. Springer Boston, 2007.

\bibitem{PoDeRa}
Nayot Poolsappasit, Rinku Dewri, and Indrajit Ray.
\newblock {Dynamic Security Risk Management Using Bayesian Attack Graphs}.
\newblock {\em IEEE Transactions on Dependable and Secure Computing},
  9(1):61--74, Jan-Feb 2012.

\bibitem{Pose}
Simona Posea.
\newblock {Renewal Periods for Cryptographic Keys}.
\newblock Master's thesis, Eindhoven University of Technology, Department of
  Mathematics and Computer Science, Eindhoven, The Netherlands, August 2012.

\bibitem{PoKo}
Marc Pouly and J\"{u}rg Kohlas.
\newblock {\em {Generic Inference: A Unifying Theory for Automated Reasoning}}.
\newblock John Wiley \& Sons, Inc., 2011.

\bibitem{PuMaLi}
Srdjan Pudar, Govindarasu Manimaran, and Chen-Ching Liu.
\newblock {PENET:} a practical method and tool for integrated modeling of
  security attacks and countermeasures.
\newblock {\em Computers \& Security}, 28(8):754--771, May 2010.

\bibitem{Pumf}
David Pumfrey.
\newblock {\em {The Principled Design of Computer System Safety Analyses}}.
\newblock PhD thesis, Department of Computer Science, University of York, York,
  UK, September 1999.

\bibitem{QiLe}
Xinzhou Qin and Wenke Lee.
\newblock Attack plan recognition and prediction using causal networks.
\newblock In {\em 20th Annual Computer Security Applications Conference}, pages
  370--379, December 2004.

\bibitem{RaPo}
Indrajit Ray and Nayot Poolsapassit.
\newblock {Using Attack Trees to Identify Malicious Attacks from Authorized
  Insiders}.
\newblock In Sabrina di~Vimercati, Paul Syverson, and Dieter Gollmann, editors,
  {\em ESORICS'2005}, volume 3679 of {\em LNCS}, pages 231--246. Springer
  Berlin / Heidelberg, 2005.

\bibitem{ReVeOlCu}
Kamil Reddy, Hein~S. Venter, Martin Olivier, and Iain Currie.
\newblock {Towards Privacy Taxonomy-Based Attack Tree Analysis for the
  Protection of Consumer Information Privacy}.
\newblock In {\em Proceedings of the 6th Annual Conference on {Privacy,
  Security and Trust (PST '08)}}, pages 56--64, New Brunswick, Canada, October
  2008.

\bibitem{ReSeKoStHo}
Andreas Reinhardt, Daniel Seither, Andr{\'e} K{\"o}nig, Ralf Steinmetz, and
  Matthias Hollick.
\newblock {Protecting IEEE 802.11s Wireless Mesh Networks Against Insider
  Attacks}.
\newblock In {\em LCN}, pages 224--227. IEEE, 2012.

\bibitem{Rost}
Lillian R\o{}stad.
\newblock {An extended misuse case notation: Including vulnerabilities and the
  insider threat}.
\newblock In {\em Proceedings of the 12th International Working Conference on
  {Requirements Engineering: Foundation for Software Quality (REFSQ 2006)}},
  pages 33--43, Luxembourg, {Grand-Duchy of Luxembourg}, June 2006.

\bibitem{Roy}
Arpan Roy.
\newblock {Attack Countermeasure Trees: A Non-state-space Approach Towards
  Analyzing Security and Finding Optimal Countermeasure Sets}.
\newblock Master's thesis, Duke University, Department of Electrical and
  Computer Engineering, USA, 2010.

\bibitem{RoKiTr3}
Arpan Roy, Dong~Seong Kim, and Kishor~S. Trivedi.
\newblock {ACT: Attack Countermeasure Trees for Information Assurance
  Analysis}.
\newblock In {\em Proceedings of {INFOCOM IEEE Conference on Computer
  Communications Workshops}}, pages 1--2, {San Diego, CA, USA}, March 2010.

\bibitem{RoKiTr}
Arpan Roy, Dong~Seong Kim, and Kishor~S. Trivedi.
\newblock {Cyber security analysis using attack countermeasure trees}.
\newblock In {\em Proceedings of the Sixth Annual Workshop on Cyber Security
  and Information Intelligence Research}, CSIIRW '10, pages 28:1--28:4, New
  York, NY, USA, 2010. ACM.

\bibitem{RoKiTr2}
Arpan Roy, Dong~Seong Kim, and Kishor~S. Trivedi.
\newblock {Attack Countermeasure Trees (ACT): towards unifying the constructs
  of attack and defense trees}.
\newblock {\em Security and Communication Networks}, 5(8):929--943, 2012.

\bibitem{RoKiTr4}
Arpan Roy, Dong~Seong Kim, and Kishor~S. Trivedi.
\newblock {Scalable optimal countermeasure selection using implicit enumeration
  on attack countermeasure trees}.
\newblock In Robert~S. Swarz, Philip Koopman, and Michel Cukier, editors, {\em
  DSN}, pages 1--12. IEEE Computer Society, 2012.

\bibitem{RuHeCeMi}
Guifr\'{e} Ruiz, Elisa Heymann, Eduardo C\'{e}sar, and Barton~P. Miller.
\newblock {Automating Threat Modeling through the Software Development
  Life-Cycle}, September 2012.

\bibitem{SaDuPa}
Vineet Saini, Qiang Duan, and Vamsi Paruchuri.
\newblock {Threat Modeling Using Attack Trees}.
\newblock {\em Journal of Computing Small Colleges}, 23(4):124--131, 2008.

\bibitem{SaSaScWa}
Chris Salter, O.~Sami Saydjari, Bruce Schneier, and Jirn Wallner.
\newblock {Toward a secure system engineering methodology}.
\newblock In {\em Proceedings of the 1998 {Workshop on New Security Paradigms
  (NSPW '98)}}, pages 2--10, Charlottesville, Virginia, United States,
  September 1998.

\bibitem{Sameer}
K.~C. Sameer.
\newblock {Attack Generation From System Models}.
\newblock Master's thesis, Technical University of Denmark, Denmark, 2011.

\bibitem{SaWoXu}
Michael Sanford, Daniel Woodraska, and Dianxiang Xu.
\newblock {Security Analysis of FileZilla Server Using Threat Models}.
\newblock In {\em SEKE}, pages 678--682. Knowledge Systems Institute Graduate
  School, 2011.

\bibitem{Sche}
Stuart~Edward Schechter.
\newblock {\em {Computer Security Strength and Risk - A Quantitative
  Approach}}.
\newblock PhD thesis, Harvard University, Cambridge, Massachusetts, May 2004.

\bibitem{Schn}
Bruce Schneier.
\newblock {Attack Trees: Modeling Security Threats}.
\newblock {\em Dr. Dobb's Journal of Software Tools}, 24(12):21--29, 1999.

\bibitem{Schn2}
Bruce Schneier.
\newblock {\em {Secrets \& Lies: Digital Security in a Networked World}}.
\newblock Wiley, Indianapolis, Ind., 2004.

\bibitem{Scut}
Marco Scutari.
\newblock {Learning Bayesian Networks with the bnlearn R Package}.
\newblock {\em Journal of Statistical Software}, 35(3):1--22, July 2010.

\bibitem{ShMaOcByCaArJi}
Nahid Shahmehri, Amel Mammar, Edgardo~Montes de~Oca, David Byers, Ana Cavalli,
  Shanai Ardi, and Willy Jimenez.
\newblock {An advanced approach for modeling and detecting software
  vulnerabilities}.
\newblock {\em Information and Software Technology}, 54(9):997--1013, 2012.

\bibitem{Shey}
Oleg Sheyner.
\newblock {\em {Scenario Graphs and Attack Graphs}}.
\newblock PhD thesis, {Carnegie Mellon University (CMU)}, Pittsburgh, PA, 2004.

\bibitem{ShHaJhLiWi}
Oleg Sheyner, Joshua Haines, Somesh Jha, Richard Lippmann, and Jeannette~M.
  Wing.
\newblock {Automated generation and analysis of attack graphs}.
\newblock In {\em Proceedings of the {IEEE} Symposium on Security and Privacy
  ({S\&P'02})}, pages 273--284, {Oakland, California, USA}, May 2002.

\bibitem{GOAT}
SHIELDS.
\newblock {GOAT}.
\newblock \url{https://www.ida.liu.se/divisions/adit/security/goat/},
  2008--2010.
\newblock Accessed October 5, 2012.

\bibitem{Website_SHIELDS}
{SHIELDS}.
\newblock {SHIELDS: Detecting known security vulnerabilities from within design
  and development tools, FP7 project, grant agreement 215995}.
\newblock \url{http://www.shields-project.eu/}, 2008--2010.
\newblock Accessed October 5, 2012.

\bibitem{SVRS}
{SHIELDS}.
\newblock {Final SHIELDS approach guide - Deliverable D1.4}.
\newblock
  \url{http://www.shields-project.eu/files/docs/D1.4%20Final%20SHIELDS%20Approach%20Guide.pdf},
  2010.
\newblock Accessed February 28, 2013.

\bibitem{Sind}
Guttorm Sindre.
\newblock {Mal-Activity Diagrams for Capturing Attacks on Business Processes}.
\newblock In {\em Proceedings of the 13th International Working Conference on
  {Requirements Engineering: Foundation for Software Quality (REFSQ 2007), LNCS
  4542}}, pages 355--366, Trondheim, Norway, June 2007.

\bibitem{SiOpLo2}
Guttorm Sindre and Andreas~L. Opdahl.
\newblock {Eliciting Security Requirements by Misuse Cases}.
\newblock In {\em Proceedings of 37th International Conference on Technology of
  Object-Oriented Languages and Systems {(TOOLS-PACIFIC 2000)}}, pages
  120--131, Sydney, Australia, November 2000.

\bibitem{SiOpLo}
Guttorm Sindre and Andreas~L. Opdahl.
\newblock {Templates for misuse case description}.
\newblock In {\em Proceedings of the 7th International Working Conference on
  {Requirements Engineering: Foundation for Software Quality (REFSQ 2001)}},
  pages 125--136, Interlaken, Switzerland, June 2001.

\bibitem{SiOp}
Guttorm Sindre and Andreas~L. Opdahl.
\newblock {Eliciting security requirements with misuse cases}.
\newblock {\em Journal of Requirements Engineering}, 10:34--44, 2005.
\newblock 10.1007/s00766-004-0194-4.

\bibitem{SiOpLoBr}
Guttorm Sindre, Andreas~L. Opdahl, and G\o{}ran~F. Brevik.
\newblock {Generalization/specialization as a structuring mechanism for misuse
  cases}.
\newblock In {\em Proceedings of the 2nd {Symposium on Requirements Engineering
  for Information Security (SREIS'02)}}, Raleigh, NC, USA, October 2002.

\bibitem{Website_Square_Tool}
{Software Engineering Institute -- Carnegie Mellon University}.
\newblock {SQUARE: System Quality Requirements Engineering}.
\newblock \url{https://www.cert.org/sse/square-tool.html}, 2004--2009.
\newblock Accessed November 12, 2012.

\bibitem{SoEkJo4}
Teodor Sommestad, Mathias Ekstedt, and Pontus Johnson.
\newblock {Combining defense graphs and enterprise architecture models for
  security analysis}.
\newblock In {\em Proceedings of the 12th {IEEE} International Conference on
  {Enterprise Distributed Object Computing (EDOC'08)}}, pages 349--355,
  M\"{u}nchen, Germany, September 2008.

\bibitem{SoEkJo2}
Teodor Sommestad, Mathias Ekstedt, and Pontus Johnson.
\newblock {Cyber Security Risks Assessment with Bayesian Defense Graphs and
  Architectural Models}.
\newblock In {\em Proceedings of the 42nd Annual Hawaii International
  Conference on System Sciences ({HICSS-42})}, pages 1--10, {Hawaii, USA},
  January 2009.

\bibitem{SoEkNo}
Teodor Sommestad, Mathias Ekstedt, and Lars Nordstr{\"o}m.
\newblock {Modeling security of power communication systems using defense
  graphs and influence diagrams}.
\newblock {\em IEEE Transactions on Power Delivery}, 24(4):1801--1808, October
  2009.

\bibitem{StVeDuFrMiRa}
Michael Stamatelatos, William Vesely, Joanne Dugan, Joseph Fragola, Joseph
  {Minarick III}, and Jan Railsback.
\newblock {Fault Tree Handbook with Aerospace Applications}.
\newblock {U.S. National Aeronautics and Space Administration (NASA) Handbook:}
  \url{http://www.hq.nasa.gov/office/codeq/doctree/fthb.pdf}, August 2002.
\newblock Version 1.1.

\bibitem{StSc}
Jan Steffan and Markus Schumacher.
\newblock {Collaborative attack modeling}.
\newblock In {\em Proceedings of the 2002 {ACM Symposium on Applied Computing
  (SAC'02)}}, pages 253--259, Madrid, Spain, March 2002.

\bibitem{SuSv}
Husam Suleiman and Davor Svetinovic.
\newblock {Evaluating the effectiveness of the security quality requirements
  engineering (SQUARE) method: a case study using smart grid advanced metering
  infrastructure}.
\newblock {\em Requirements Engineering}, {}:1--29, 2012.
\newblock 10.1007/s00766-012-0153-4.

\bibitem{SwSn}
Frank Swiderski and Window Snyder.
\newblock {\em {Threat modeling}}.
\newblock Microsoft Press, Redmond, 2004.

\bibitem{SwPhElCh}
Laura~Painton Swiler, Cynthia Phillips, David Ellis, and Stefan Chakerian.
\newblock {Computer-attack graph generation tool}.
\newblock {\em DARPA Information Survivability Conference and Exposition II
  (DISCEX'01)}, 2:307--321, 2001.

\bibitem{TaJo}
Eedee Tanu and Johnnes Arreymbi.
\newblock {An examination of the security implications of the supervisory
  control and data acquisition (SCADA) system in a mobile networked
  environment: An augmented vulnerability tree approach.}
\newblock In {\em Proceedings of Advances in Computing and Technology, (AC\&T)
  The School of Computing and Technology 5th Annual Conference}, pages
  228--242. University of East London, School of Computing, Information
  Technology and Engineering, 2010.

\bibitem{TeLiGo}
Chee-Wooi Ten, Chen-Ching Liu, and Govindarasu Manimaran.
\newblock {Vulnerability Assessment of Cybersecurity for SCADA Systems Using
  Attack Trees}.
\newblock In {\em Proceedings of the {IEEE} Power Engineering Society General
  Meeting}, pages 1--8, Tampa, {USA}, June 2007.

\bibitem{TeMaLi}
Chee-Wooi Ten, Govindarasu Manimaran, and Chen-Ching Liu.
\newblock {Cybersecurity for Critical Infrastructures: Attack and Defense
  Modeling}.
\newblock {\em IEEE Transactions on Systems, Man and Cybernetics, Part A:
  Systems and Humans}, 40(4):853--865, July 2010.

\bibitem{Website_xine}
the~xine project.
\newblock {xine multimedia engine}.
\newblock \url{http://www.xine-project.org/home}, 2002--2012.
\newblock Accessed October 26, 2012.

\bibitem{TiLaFiHa}
Terry Tidwell, Ryan Larson, Kenneth Fitch, and John Hale.
\newblock {Modeling Internet Attacks}.
\newblock In {\em Proceedings of the {2nd IEEE Systems, Man and Cybernetics
  Information Assurance Workshop (IAW '01)}}, pages 54--59, {West Point, USA},
  June 2001.

\bibitem{ToJeRo}
Inger~Anne T\o{}ndel, Jostein Jensen, and Lillian R\o{}stad.
\newblock {Combining Misuse Cases with Attack Trees and Security Activity
  Models}.
\newblock In {\em International Conference on Availability, Reliability and
  Security}, pages 438--445, Los Alamitos, CA, USA, 2010. IEEE Computer
  Society.

\bibitem{trespass}
{TREsPASS}.
\newblock {Technology-supported Risk Estimation by Predictive Assessment of
  Socio-technical Security, FP7 project, grant agreement 318003}.
\newblock \url{http://www.trespass-project.eu/}, 2012--2016.
\newblock Accessed January 16, 2013.

\bibitem{sharpe}
Kishor~S. Trivedi and Robin Sahner.
\newblock {SHARPE at the age of twenty two}.
\newblock {\em SIGMETRICS Perform. Eval. Rev.}, 36(4):52--57, March 2009.

\bibitem{DepD}
{U.S. Department of Defense (DoD)}.
\newblock {Standard Practice For System Safety}.
\newblock {MIL-STD-882D}, June 1988.

\bibitem{Nucl}
{U.S. Nuclear Regulatory Commission (NRC)}.
\newblock {Cyber Security Programs For Nuclear Facilities}.
\newblock Regulatory Guide {5.71}, January 2010.

\bibitem{Lams}
Axel {van Lamsweerde}.
\newblock {Elaborating security requirements by construction of intentional
  anti-models}.
\newblock In {\em 26th International Conference on Software Engineering
  (ICSE'04)}, pages 148--157, May 2004.

\bibitem{LaBrLaJa}
Axel {van Lamsweerde}, Simon Brohez, Renaud~De Landtsheer, and David Janssens.
\newblock {From System Goals to Intruder Anti-Goals: Attack Generation and
  Resolution for Security Requirements Engineering}.
\newblock In {\em Proceedings of RHAS'03}, pages 49--56, 2003.

\bibitem{LaLe}
Axel {van Lamsweerde} and Emmanuel Letier.
\newblock {Handling Obstacles in Goal-Oriented Requirements Engineering}.
\newblock {\em IEEE Trans. Softw. Eng.}, 26:978--1005, October 2000.

\bibitem{VeGoRoHa}
William~E. Vesely, Francine~F. Goldberg, Norman~H. Roberts, and David~F. Haasl.
\newblock {Fault Tree Handbook}.
\newblock Technical Report NUREG-0492, U.S. Regulatory Commission, 1981.

\bibitem{ViJo}
Stilianos Vidalis and Andy Jones.
\newblock {Using vulnerability trees for decision making in threat assessment}.
\newblock Technical Report CS-03-02, School of Computing, University of
  Glamorgan, Pontypridd, Wales, UK, 2003.

\bibitem{Website_VIKING}
{VIKING}.
\newblock {FP7 project, grant agreement 225643}.
\newblock \url{http://www.vikingproject.eu}, 2008--2011.
\newblock Accessed November 6, 2012.

\bibitem{VISPER}
{VISPER}.
\newblock {VISPER: The VIrtual Security PERimeter for digital, physical, and
  organisational security, project funded by the Sentinels programme}.
\newblock \url{http://www.sentinels.nl/en/content/visper}, 2007--2011.
\newblock Accessed February 27, 2013.

\bibitem{WaLiZh}
Hui Wang, Shufen Liu, and Xinjia Zhang.
\newblock {An improved model of attack probability prediction system}.
\newblock {\em Wuhan University Journal of Natural Sciences}, 11:1498--1502,
  2006.

\bibitem{WaPhWhPa3}
Jie Wang, Raphael C.-W. Phan, John~N. Whitley, and David~J. Parish.
\newblock {Augmented Attack Tree Modeling of Distributed Denial of Services and
  Tree Based Attack Detection Method}.
\newblock In {\em Proceedings of the {10th IEEE International Conference on
  Computer and Information Technology (CIT 2010)}}, pages 1009--1014,
  {Bradford, UK}, June 2010.

\bibitem{WaPhWhPa}
Jie Wang, Raphael C.-W. Phan, John~N. Whitley, and David~J. Parish.
\newblock {Augmented attack tree modeling of SQL injection attacks}.
\newblock In {\em Proceedings of the 2nd {IEEE International Conference on
  Information Management and Engineering (ICIME)}}, pages 182--186, Chengdu,
  China, April 2010.

\bibitem{WaPhWhPa2}
Jie Wang, Raphael C.-W. Phan, John~N. Whitley, and David~J. Parish.
\newblock {Quality of detectability (QoD) and QoD-aware AAT-based attack
  detection}.
\newblock In {\em Proceedings of the 2010 {International Conference for
  Internet Technology and Secured Transactions (ICITST)}}, pages 1--6,
  November, London, UK 2010.

\bibitem{WaWhPhPa}
Jie Wang, John~N. Whitley, Raphael C.-W. Phan, and David~J. Parish.
\newblock {Unified Parametrizable Attack Tree}.
\newblock {\em International Journal for Information Security Research},
  1(1):20--26, 2011.

\bibitem{WaIsLoSiJa}
Lingyu Wang, Tania Islam, Tao Long, Anoop Singhal, and Sushil Jajodia.
\newblock {An Attack Graph-Based Probabilistic Security Metric}.
\newblock In {\em Proceeedings of the 22nd Annual {IFIP WG 11.3 Working
  Conference on Data and Applications Security (DAS'2008), LNCS 5094}}, pages
  283--296, {London, UK}, July 2008.

\bibitem{WaNoJa}
Lingyu Wang, Steven Noel, and Sushil Jajodia.
\newblock {Minimum-cost network hardening using attack graphs}.
\newblock {\em Comput. Commun.}, 29(18):3812--3824, November 2006.

\bibitem{WaSiJa2}
Lingyu Wang, Anoop Singhal, and Sushil Jajodia.
\newblock {Measuring the Overall Security of Network Configurations Using
  Attack Graphs}.
\newblock In Steve Barker and Gail-Joon Ahn, editors, {\em Data and
  Applications Security XXI}, volume 4602 of {\em LNCS}, pages 98--112.
  Springer Berlin / Heidelberg, 2007.
\newblock \url{10.1007/978-3-540-73538-0_9}.

\bibitem{WaSiJa}
Lingyu Wang, Anoop Singhal, and Sushil Jajodia.
\newblock {Toward measuring network security using attack graphs}.
\newblock In {\em Proceedings of the 2007 ACM workshop on Quality of
  protection}, QoP '07, pages 49--54, New York, NY, USA, 2007. ACM.

\bibitem{WaYaSiJa}
Lingyu Wang, Chao Yao, Anoop Singhal, and Sushil Jajodia.
\newblock {Interactive Analysis of Attack Graphs Using Relational Queries}.
\newblock In Ernesto Damiani and Peng Liu, editors, {\em Data and Applications
  Security XX}, volume 4127 of {\em LNCS}, pages 119--132. Springer Berlin
  Heidelberg, 2006.

\bibitem{WaLeRo}
Mathew Warren, Shona Leitch, and Ian Rosewall.
\newblock {Attack vectors against social networking systems : the Facebook
  example }.
\newblock In {\em Proceedings of The 9th Australian Information Security
  Management Conference}. SECAU - Security Research Centre, 2011.

\bibitem{Wats}
H.~A. Watson.
\newblock {\em {Launch Control Safety Study}}, volume~1.
\newblock Bell Labs, Murray Hill, NJ, 1961.

\bibitem{Weis}
Jonathan~D. Weiss.
\newblock {A system security engineering process}.
\newblock In {\em 14th Nat. Comp. Sec. Conf.}, pages 572--581, 1991.

\bibitem{WeWe}
Lv~Wen-ping and Li~Wei-min.
\newblock {Space Based Information System Security Risk Evaluation Based on
  Improved Attack Trees}.
\newblock In {\em Third International Conference on Multimedia Information
  Networking and Security (MINES'11)}, pages 480--483, November 2011.

\bibitem{WhPhWaPa}
John~N. Whitley, Raphael C.-W. Phan, Jie Wang, and David~J. Parish.
\newblock {Attribution of attack trees}.
\newblock {\em Computers \& Electrical Engineering}, 37(4):624--628, 2011.

\bibitem{WiJu}
Jan Willemson and Aivo J\"{u}rgenson.
\newblock {Serial Model for Attack Tree Computations}.
\newblock In D.~Lee and S.~Hong, editors, {\em ICISC}, volume 5984 of {\em
  LNCS}, pages 118--128. Springer, 2010.

\bibitem{WiLiIn}
Leevar Williams, Richard Lippmann, and Kyle~W. Ingols.
\newblock {An interactive attack graph cascade and reachability display}.
\newblock In {\em Proceedings of the 2007 Workshop on Visualization for
  Computer Security {(VizSEC'07)}}, pages 221--236, {Sacramento, CA, USA},
  October 2007.

\bibitem{FoWuMaBaSpTech}
Yu-Sung Wu, Bingrui Foo, Yu-Chun Mao, Saurabh Bagchi, and Eugene Spafford.
\newblock {Automated Aaptive Intrusion Containment in Systems of Interacting
  Services}.
\newblock Technical Report Paper 68, Purdue University, School of Electrical
  and Computer Engineering, West Lafayette, IN 47907-2035, 2005.

\bibitem{BagchiTech}
Yu-Sung Wu, Bingrui Foo, Blake Matheny, Tyler Olsen, and Saurabh Bagchi.
\newblock {ADEPTS: Adaptive Intrusion Containment and Response using Attack
  Graphs in an E-commerce Environment}.
\newblock Technical report, Purdue University, School of Electrical and
  Computer Engineering, December 2003.

\bibitem{conf/acsac/WuFMB03}
Yu-Sung Wu, Bingrui Foo, Yongguo Mei, and Saurabh Bagchi.
\newblock {Collaborative Intrusion Detection System (CIDS): A Framework for
  Accurate and Efficient IDS.}
\newblock In {\em ACSAC}, pages 234--244. IEEE Computer Society, 2003.

\bibitem{XiLiOuLiLe}
Peng Xie, Jason~H. Li, Xinming Ou, Peng Liu, and Renato Levy.
\newblock {Using Bayesian networks for cyber security analysis}.
\newblock In {\em IEEE/IFIP International Conference on Dependable Systems and
  Networks (DSN'10)}, pages 211--220, 28 2010-july 1 2010.

\bibitem{XuNy}
Dianxiang Xu and Kendall~E. Nygard.
\newblock {Threat-driven modeling and verification of secure software using
  aspect-oriented Petri nets}.
\newblock {\em IEEE Transactions on Software Engineering}, 32(4):265--278,
  2006.

\bibitem{Yage}
Ronald~R. Yager.
\newblock {OWA trees and their role in security modeling using attack trees}.
\newblock {\em Inf. Sci.}, 176(20):2933--2959, 2006.

\bibitem{ZaFe}
Anita~N. Zakrzewska and Erik~M. Ferragut.
\newblock {Modeling cyber conflicts using an extended Petri Net formalism}.
\newblock In {\em Computational Intelligence in Cyber Security (CICS), 2011
  IEEE Symposium on}, pages 60--67, april 2011.

\bibitem{ZhYu}
Chengli Zhao and Zhiheng Yu.
\newblock {Quantitative Analysis of Survivability Based on Intrusion
  Scenarios}.
\newblock In David Jin and Sally Lin, editors, {\em Advances in Electronic
  Engineering, Communication and Management Vol.2}, volume 140 of {\em LNEE},
  pages 701--705. Springer Berlin Heidelberg, 2012.
\newblock \url{10.1007/978-3-642-27296-7_105}.

\bibitem{Zonouz}
Saman~Aliari Zonouz.
\newblock {\em {Game-theoretic Intrusion Response and Recovery}}.
\newblock PhD thesis, {University of Illinois at Urbana-Champaign}, USA, 2011.
\newblock Available at
  \url{https://www.ideals.illinois.edu/bitstream/handle/2142/29667/AliariZonouz_Saman.pdf?sequence=1}.

\bibitem{ZoKhSaYa}
Saman~Aliari Zonouz, Himanshu Khurana, William~H. Sanders, and Timothy~M.
  Yardley.
\newblock {RRE: A game-theoretic intrusion Response and Recovery Engine}.
\newblock In {\em IEEE/IFIP International Conference on Dependable Systems
  Networks (DSN'09)}, pages 439--448, July 2009.

\bibitem{ZoShRaKaPfAuIySaCo}
Saman~Aliari Zonouz, Aashish Sharma, HariGovind~V. Ramasamy, Zbigniew~T.
  Kalbarczyk, Birgit Pfitzmann, Kevin McAuliffe, Ravishankar~K. Iyer,
  William~H. Sanders, and Eric Cope.
\newblock {Managing business health in the presence of malicious attacks}.
\newblock In {\em IEEE/IFIP 41st International Conference on Dependable Systems
  and Networks Workshops (DSN-W'11)}, pages 9--14, June 2011.

\end{thebibliography}

\end{document}